%% file: main_dns.tex
\documentclass[preprint]{aastex}

\usepackage[dvipdfmx]{color}
\usepackage{amsmath}
\usepackage{booktabs}
\usepackage{natbib}
\newcommand{\myemail}{kitagawa@solar.mtk.nao.ac.jp}
\newcommand{\kmpers}{\mathrm{km} \, \mathrm{s}^{-1}}
\newcommand{\logt}{\log T \, [\mathrm{K}]}

\slugcomment{}

\shorttitle{Electron density of active region outflows measured by \textit{Hinode}/EIS}
\shortauthors{Kitagawa \& Yokoyama}

\begin{document}


\title{Electron density of active region outflows measured by the EUV Imaging Spectrometer onboard \textit{Hinode}}


\author{N.\ Kitagawa\altaffilmark{1}}

\and

\author{T.\ Yokoyama\altaffilmark{2}}

\affil{\altaffilmark{1} National Astronomical Observatory of Japan, 2-21-1 Osawa, Mitaka, Tokyo, Japan}
\affil{\altaffilmark{2} The University of Tokyo, 7-3-1 Hongo, Bunkyo, Tokyo, Japan}


\altaffiltext{1}{E-mail: \myemail}


\begin{abstract}
	\input{abstract_1.tex}
\end{abstract}





\section{Introduction}
  \input{itdn_aroutflow.tex}
  \input{dns_itdn.tex}

\section{Observation and calibration}
    \label{sect:obs}
    \input{dns_raster_scans.tex}

\section{Density diagnostics of upflows}
    \label{subsect:dns_diag}
    \input{dns_analysis_preface.tex}
  \subsection{Integration of observational pixels}
    \input{dns_analysis_integration_pixels.tex}
  \subsection{De-blending of Si \textsc{vii} from Fe \textsc{xiv} 274.20{\AA}}
    \label{sec:de-blend}
    \input{dns_analysis_de-blending.tex}
  \subsection{Simultaneous fitting of the two Fe \textsc{xiv} emission lines}
    \input{dns_analysis_simul_fit.tex}
  \subsection{Density inversion}
    \label{sec:dns_inv}
    \input{dns_analysis_dens_inv.tex}

\section{Density derived from Fe \textsc{xiv} 264.78{\AA}/274.20{\AA}}
  \label{sect:dns_results}
  \subsection{Results from single Gaussian fitting}
    \input{dns_results_1G.tex}
  \subsection{Density of the upflows}
    \input{dns_results_n.tex}
  \subsection{Column depth}
    \input{dns_results_h.tex}

\section{$\lambda$-$n_{\mathrm{e}}$ diagram}
  \label{chap:ndv}
  \input{ndv_itdn.tex}
  \subsection{Method}
    \input{ndv_method.tex}
    \input{ndv_method_2.tex}
    \input{ndv_method_3.tex}
  \subsection{Verification of the method}
    \input{ndv_test.tex}
    \subsubsection{Dependence on electron density}
      \input{ndv_test_density.tex}
    \subsubsection{Dependence on velocity}
      \input{ndv_test_velocity.tex}
  \subsection{$\lambda$-$n_\mathrm{e}$ diagram in AR10978}
    \input{ndv_sv.tex}

\section{Discussion}
  \label{sect:dns_dis}
  \subsection{Theoretical estimation of electron density}
    \input{dns_discuss_n.tex}
  \subsection{Mass transport by the outflow}
    \label{sect:dis_mass}
    \input{dis_mass.tex}
  \subsection{Eastern and western outflow regions}
    \label{sect:dis_ew}
    \input{dis_ew.tex}

\section{Summary}
  \label{sect:dns_sum}
  \input{dns_sum.tex}
  \input{ndv_sum.tex}
\acknowledgments

Hinode is a Japanese mission developed and launched by ISAS/JAXA, with NAOJ as domestic partner and NASA and STFC (UK) as international partners.  It is operated by these agencies in co-operation with ESA and NSC (Norway).  We extend our gratitude to Hirohisa Hara and Toshifumi Shimizu who made a large number of insightful comments with their expertise in spectroscopy.  We are also thankful to the GCOE program for proofreading/editing assistance.  



{\it Facilities:} \facility{Hinode}.



\appendix

\section{Fe \textsc{xiv} 264.78{\AA} intensity and electron density}
  \label{sect:dns_app_in}
  \input{dns_app_in.tex}
\section{Uncertainty in Si \textsc{vii} density}
  \label{sect:dns_results_sivii}
  \input{dns_results_sivii.tex}

\input{main_dns.bbl}
\end{document}

%% file: abstract_1.tex
In order to better understand the nature of active region outflows, the electron density was measured 
by using a density-sensitive line pair Fe \textsc{xiv} $264.78${\AA}/$274.20${\AA}.  
Since coronal line profiles of the outflow region are composed of a major 
component 
with a Doppler shift of $\leq 10 \, \kmpers$ 
and a 
minor component (enhanced blue wing: EBW) 
blueshifted by up to $100 \, \kmpers$, 
we extracted EBW from the line profiles through double-Gaussian fitting.  
We tried applying the simultaneous fitting to those two Fe \textsc{xiv} lines with several physical restrictions.  
Electron density for both components ($n_{\mathrm{Major}}$ and $n_{\mathrm{EBW}}$, respectively) was calculated 
by referring to 
the theoretical intensity ratio as a function of electron density as per 
the CHIANTI database.  
We studied six locations in the outflow regions around NOAA AR10978.  The average electron density was $n_{\mathrm{Major}} = 10^{9.16 \pm 0.16} \, \mathrm{cm}^{-3}$ and $n_{\mathrm{EBW}} = 10^{8.74 \pm 0.29} \, \mathrm{cm}^{-3}$.  The magnitude relationship between $n_{\mathrm{Major}}$ and $n_{\mathrm{EBW}}$ was opposite in the eastern and western outflow regions.  The column depth was also calculated 
for each component, which leads to the result that the outflows possess only a small fraction ($\sim 0.1$) 
in the eastern region, while they dominate over 
the major component in the line profiles 
by a factor of five in the western region.  
When taking into account the extending coronal structures, the western region can be thought to represent the mass leakage.  In contrast, we suggest a possibility that the eastern region actually contributes to the mass supply to coronal loops. 


%% file: itdn_aroutflow.tex

Spectral coverage sensitive to the coronal temperature and unprecedented high signal-to-noise ratio of \textit{Hinode}/EIS enabled us to reveal the existence of upflows at the edge of active regions \citep{doschek2008,harra2008}.  These upflows have been called ``active region (AR) outflows'', and are considered to be ejected from the bottom of the corona.  It has previously been confirmed that these outflows persist for several days in the images taken by X-Ray Telescope (XRT) onboard \textit{Hinode} \citep{sakao2007}.  Some authors interpreted AR outflows as the source of the solar wind \citep{harra2008,baker2009,brooks2011}. 

\citet{doschek2008} analyzed emission line profiles of Fe \textsc{xii} $195.12${\AA} and revealed that the outflows are observed at the dark region outside an active region core.  A preliminary result from EIS has shown that there is a clear boundary between closed hot loops in the AR core ($\sim 3 \times 10^6 \, \mathrm{K}$) and extended cool loops ($\lesssim 1 \times 10^6 \, \mathrm{K}$) where the blueshift was observed \citep{delzanna2008}.  The upflows were seen in the low density and low radiance area.  Meanwhile, redshift was observed in the AR core for all emission lines (Fe \textsc{viii}--\textsc{xv}).  This apparent lack of signatures of any upflows at active region cores was explained as a strong 
major 
component 
closer to the rest wavelength 
in line profiles hinders the signal of upflows \citep{doschek2012}, but it has not been proved yet.  
The magnetic configuration of the outflow region has been modeled by magnetic field extrapolation from the photospheric magnetogram \citep{harra2008,baker2009}, and it was revealed that AR outflows emanate from the footpoints of extremely long coronal loops in the edge of an active region \citep{harra2008}.  Close investigation revealed that AR outflows are located near the footpoints of quasi separatrix layers (QSLs), which forms the changes of the connectivity of the magnetic fields from closed coronal loops into open regions \citep{baker2009,delzanna2011}. 

The velocity of the outflow lies within the range of a few tens up to $\sim 100 \, \kmpers$.  These velocities were derived by subtracting the fitted single-Gaussian from raw line profiles \citep{hara2008}, and by double-Gaussian fitting \citep{bryans2010}.  By using extrapolated magnetic fields, the actual velocity was derived from the Doppler measurement and found to have a speed of $60$--$125 \, \kmpers$ \citep{harra2008}.  The upflow velocity of AR outflows increases with the formation temperature which emission lines Si \textsc{vii}--Fe \textsc{xv} represent \citep{warren2011}.  The blueshift becomes larger in hotter emission line as $5\text{--}20 \, \kmpers$ for Fe \textsc{xii} (formed at $\sim 1 \times 10^6 \, \mathrm{K}$) and $10\text{--}30 \, \kmpers$ for Fe \textsc{xv} (formed at $\sim 3 \times 10^6 \, \mathrm{K}$) \citep{delzanna2008}.  The appearance of the blueshifted regions often seems to trace loop-like structures. However, it is not completely understood whether the AR outflows are related to fan loop structures \citep{warren2011,tian2011,mcintosh2012}. 

AR outflows are observed as an enhanced blue wing (EBW) component in emission line profiles of Fe \textsc{xii}--\textsc{xv}.  By fitting the line profiles by a single Gaussian, it was revealed that there is a negative correlation between blueshifts and line widths \citep{doschek2008,hara2008}, which indicates the existence of an unresolved component in the blue wing emitted from the upflowing plasma.  
This EBW does not exceed the major component by $\sim 25${\%} in terms of the intensity \citep{doschek2012}.

Previous observations have revealed properties of the outflow from the edge of active regions such as (1) location: less bright region outside the active region core, (2) magnetic topology: boundary between open magnetic fields and closed loops, and (3) velocity: reaching up to $v \sim 100 \, \kmpers$ in the coronal temperature. 
Although a number of observations have revealed those physical properties, there remains one missing quantity: the electron density of the outflow itself.  The density of an outflow region derived by using the line ratio of Fe \textsc{xii} $186.88${\AA}/$195.12${\AA} was $\simeq 7 \times 10^{8} \, \mathrm{cm}^{-3}$ \citep{doschek2008}, which is slightly lower than the typical value in active region ($n_{\mathrm{e}} \ge 10^9 \, \mathrm{cm}$).  Recently, \citet{brooks2012} carried out differential emission measure (DEM) analysis at the outflow regions.  It was revealed that the properties of DEM and also the chemical abundance are rather close to those of the active region, from which the authors concluded that the outflowing plasma originates in the active region loops.  The interchange reconnection was considered to be a candidate for accelerating the plasma into the outer atmosphere \citep{baker2009,delzanna2011}.


%% file: dns_itdn.tex

The electron density of the outflow itself should help us to better understand the nature of the outflows.  However, there have been 
few intensive attempts 
to do so until present 
\citep{patsourakos2014}.  
One point of view is that those outflows are directly linked to the coronal heating in such a way that the outflowing plasma fills the outer atmosphere and form the corona \citep{depontieu2009,mcintosh2012}.  The impulsive heating in a coronal loop induces an upflow from its footpoint, which may account for what we see as the outflow \citep{delzanna2008,hara2008}.  Outflows can be also caused by the sudden change of the pressure environment in a coronal loop \citep{bradshaw2011}.  

A theoretical estimation was recently proposed in terms of the ratio of the electron density between major 
component ($n_{\mathrm{Major}}$) and EBW component ($n_{\mathrm{EBW}}$) in coronal emission line profiles \citep{klimchuk2012}.  It was shown that if the tips of the chromospheric spicules supply the coronal plasma \citep{depontieu2011}, that ratio (here after denoted as $n_{\mathrm{EBW}}/n_{\mathrm{Major}}$) takes a value of an order of $10$--$100$, while tiny impulsive heating (\textit{i.e.}, nanoflare) creates the ratio of $0.4$--$1$ \citep{patsourakos2014}.  Thus, it was suggested that the ratio $n_{\mathrm{EBW}}/n_{\mathrm{Major}}$ can be used as a diagnostic tool which enables us to discriminate these two mechanisms in the corona.  
\citet{patsourakos2014} showed that this ratio peaks at order of unity, and suggested that type II spicules \citep{depontieu2007} cannot be the primary source of the coronal plasma.  

\begin{figure}[!t]
	\centering
	\includegraphics[width=8cm,clip]{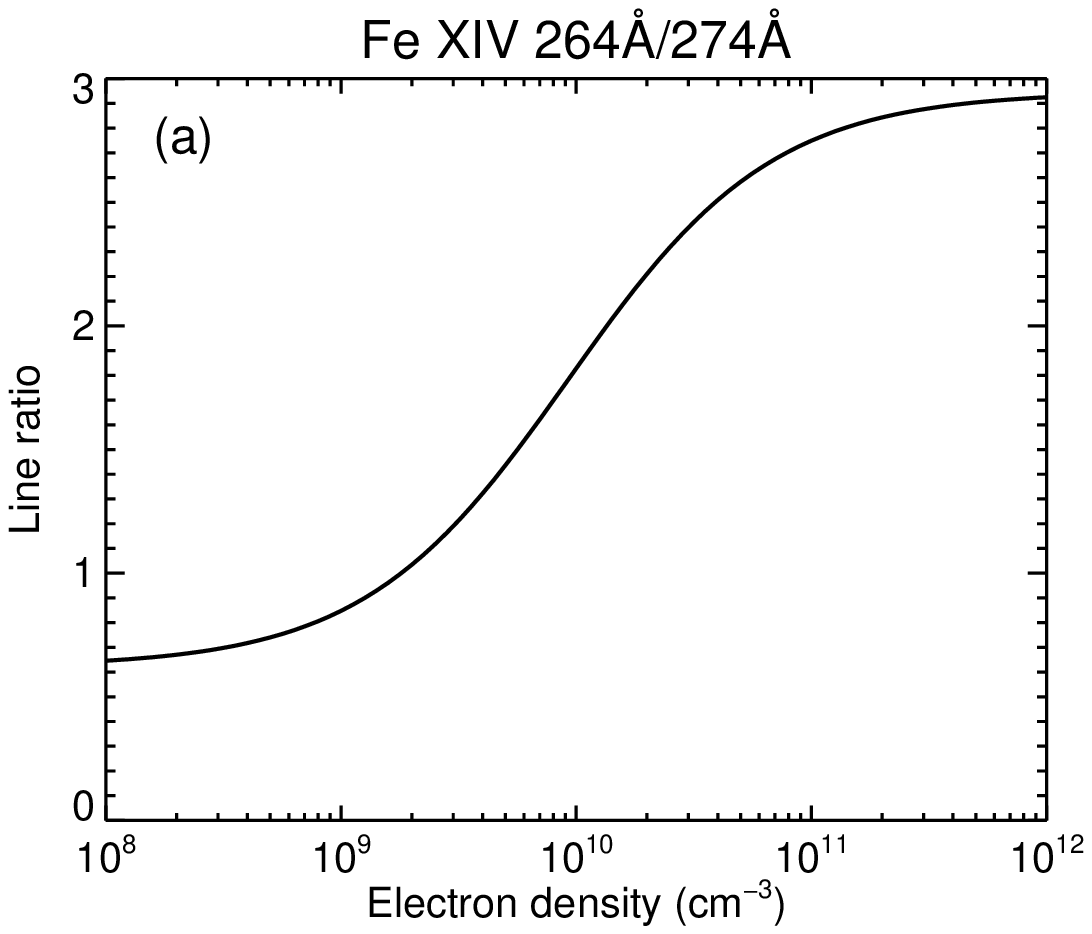}
	\includegraphics[width=8cm,clip]{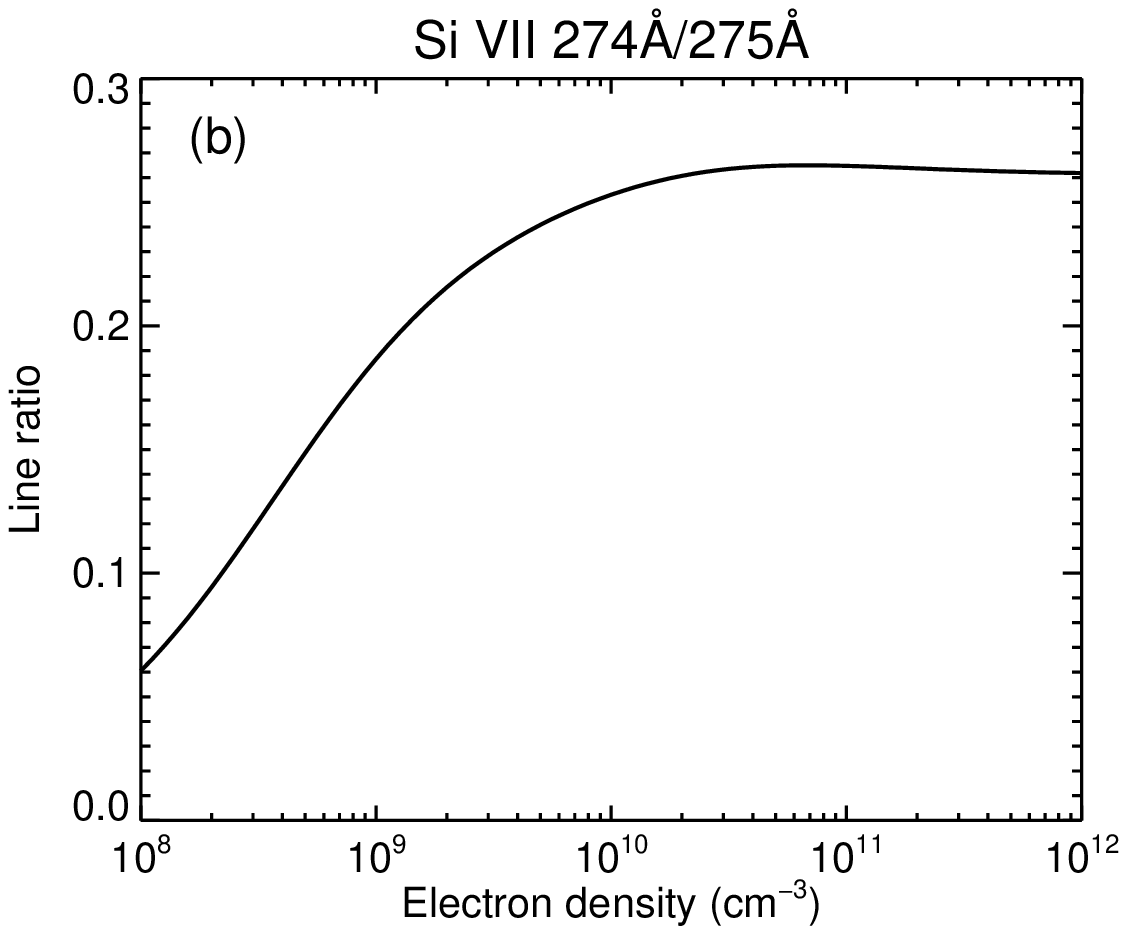}
	\caption{Theoretical line ratio calculated by CHIANTI database version 7 \citep{dere1997,landi2012}. (a) Fe \textsc{xiv} $264.78${\AA}/$274.20${\AA}. (b) Si \textsc{vii} $274.18${\AA}/$275.35${\AA}.}
  \label{fig:dns_itdn_rat_chianti}
\end{figure}

In this study, we used the spectroscopic data obtained with EIS onboard \textit{Hinode} in order to measure the electron density of the outflows.  
As a line pair suitable for our purpose, Fe \textsc{xiv} $264.78${\AA} and $274.20${\AA} were chosen because (1) those emission lines have a distinct enhanced blue wing at the outflow region which leads to better signal-to-noise ratio, (2) they consist of relatively clean emission lines and their line wings in the shorter wavelength side do not overlap with other emission lines, different from the cases for Fe \textsc{xii} $186.88${\AA}/$195.12${\AA} and Fe \textsc{xiii} $202.04${\AA}/$203.83${\AA}, and (3) the Fe \textsc{xiv} line pair is sensitive to the density range of $n_{\mathrm{e}} = 10^{8\text{--}12} \, \mathrm{cm}^{-3}$ as shown in Figure \ref{fig:dns_itdn_rat_chianti}, which is wider than other line pairs.  
The analyzed active region was as the same one as \citet{patsourakos2014}, and one of our advantages is the spatial information (\textit{i.e.}, east/west edges), which was not focused in their study.  

The following parts of this paper are structured as follows.  Section \ref{sect:obs} describes EIS observation and wavelength calibration.  Density of the outflows is derived in Section \ref{subsect:dns_diag}, and the results will be shown in Section \ref{sect:dns_results}.  We propose a new technique for line profile analysis ($\lambda$-$n_{\mathrm{e}}$) in Section \ref{chap:ndv}.  We discuss the nature of the observed outflows in Section \ref{sect:dns_dis}. Section \ref{sect:dns_sum} will provide a summary of this paper.  Two appendices describe some details in our analysis.   


%% file: dns_raster_scans.tex

In this study, we analyzed a raster scan obtained with \textit{Hinode}/EIS, which observed active region NOAA AR10978 (hereafter AR10978) at the center of the solar disk.  The scan with narrow $1''$ slit started on 2007 December 11 00:24:16UT and ended at 04:47:29UT. Field of view (FOV) was $256'' \times 256''$ and exposure time was $60 \, \mathrm{s}$.  The EIS data was processed through the standard software which detects the cosmic ray hits on the CCD pixels, subtracts the dark current bias, and corrects DN at warm pixels.  The DN is converted into the unit of intensity: $\mathrm{erg} \, \mathrm{cm}^{-2} \, \mathrm{s}^{-1} \, \mathrm{sr}^{-1} \, \text{\AA}^{-1}$.  This quantity should be called \textit{spectral intensity} in the literature. However, we use the term \textit{intensity} for the sake of simplicity.  One complicated point in the calibration is the thermal drift of the projected location on the CCD pixels due to the orbital motion of \textit{Hinode}.  We calibrated the absolute wavelength through the method developed by \citet{kamio2010}.  Since the relative position of two emission lines Fe \textsc{xiv} $264.78${\AA} and $274.20${\AA} is the most important factor in this analysis, we carried out relative wavelength calibration 
whose details are described in \citet{kitagawa2013}.  


%% file: dns_analysis_preface.tex

One of our main achievements is 
density measurement of AR outflows.  Previous observations have revealed that the density of the outflow region measured by using a line pair Fe \textsc{xii} $186.88${\AA}/$195.12${\AA} indicates $7 \times 10^{8} \, \mathrm{cm}^{-3}$ which is close to that of coronal holes rather than that of active regions \citep{doschek2008}.  However, density of the outflow itself, measured by separating its component from the major component in line profiles, has not been investigated so far. 

\begin{figure}[!t]
  \centering
  \begin{minipage}[c]{10cm}
    \includegraphics[width=10cm,clip]{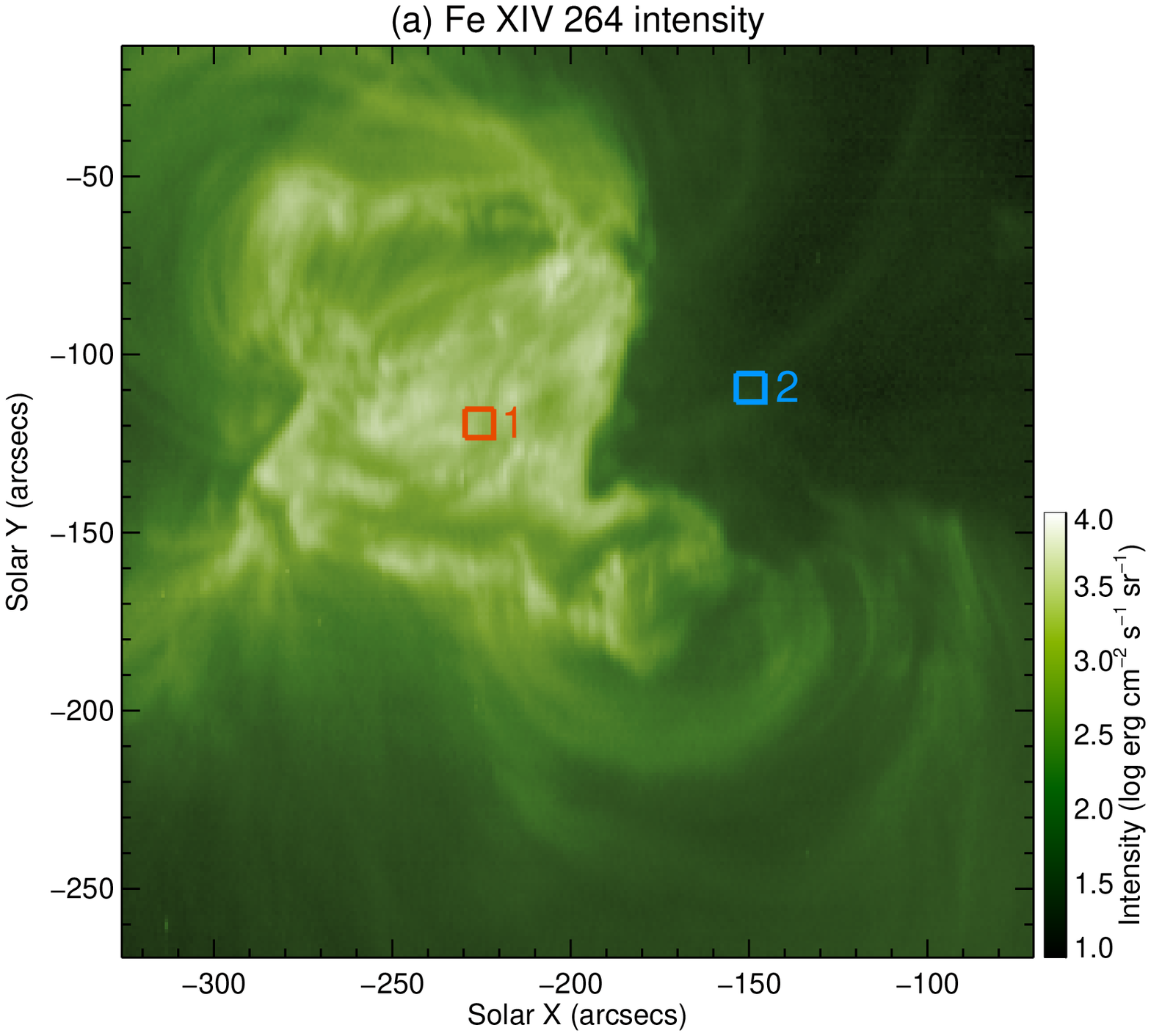}
  \end{minipage}  
  \begin{minipage}[c]{6cm}
    \includegraphics[width=6cm,clip]{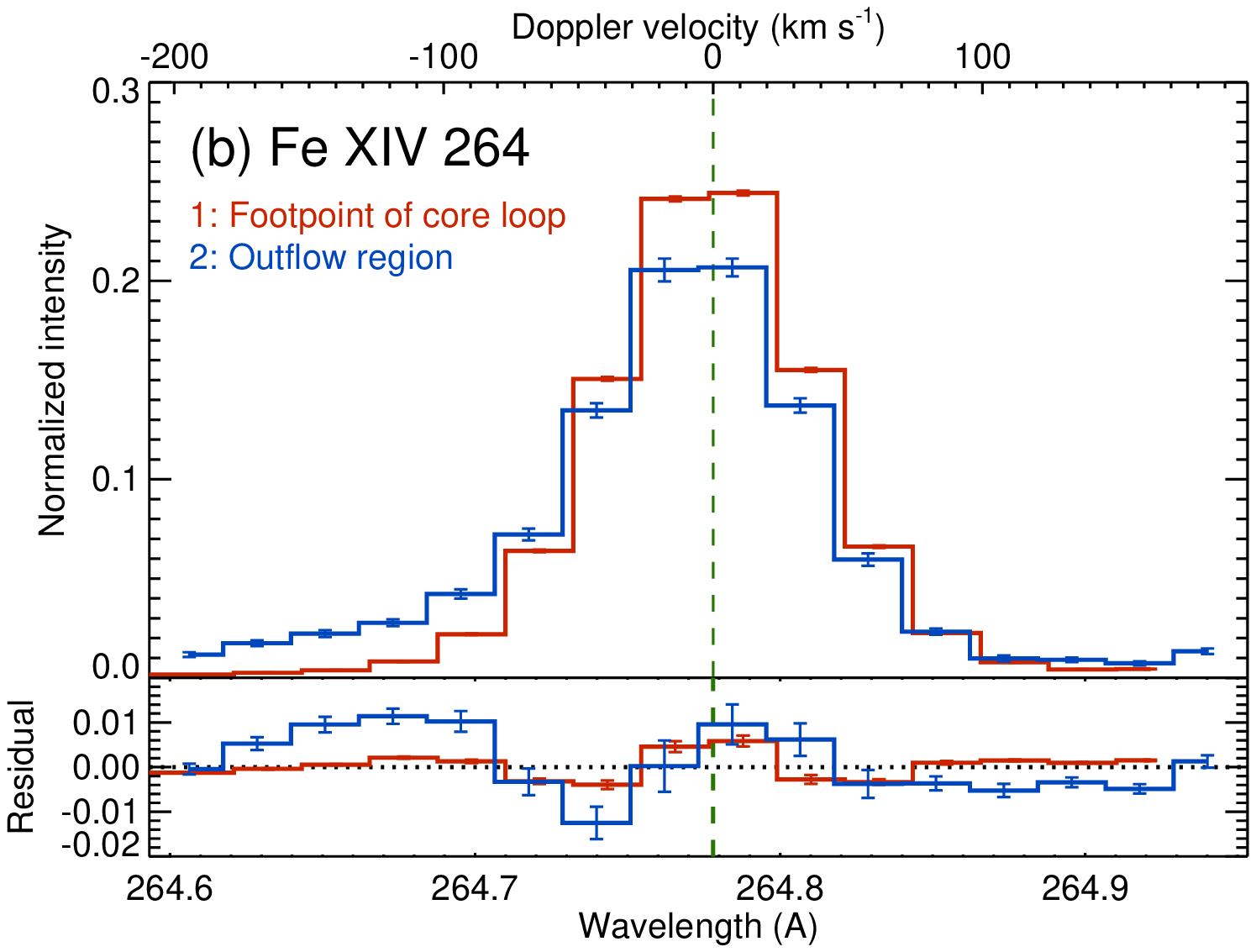}
    \includegraphics[width=6cm,clip]{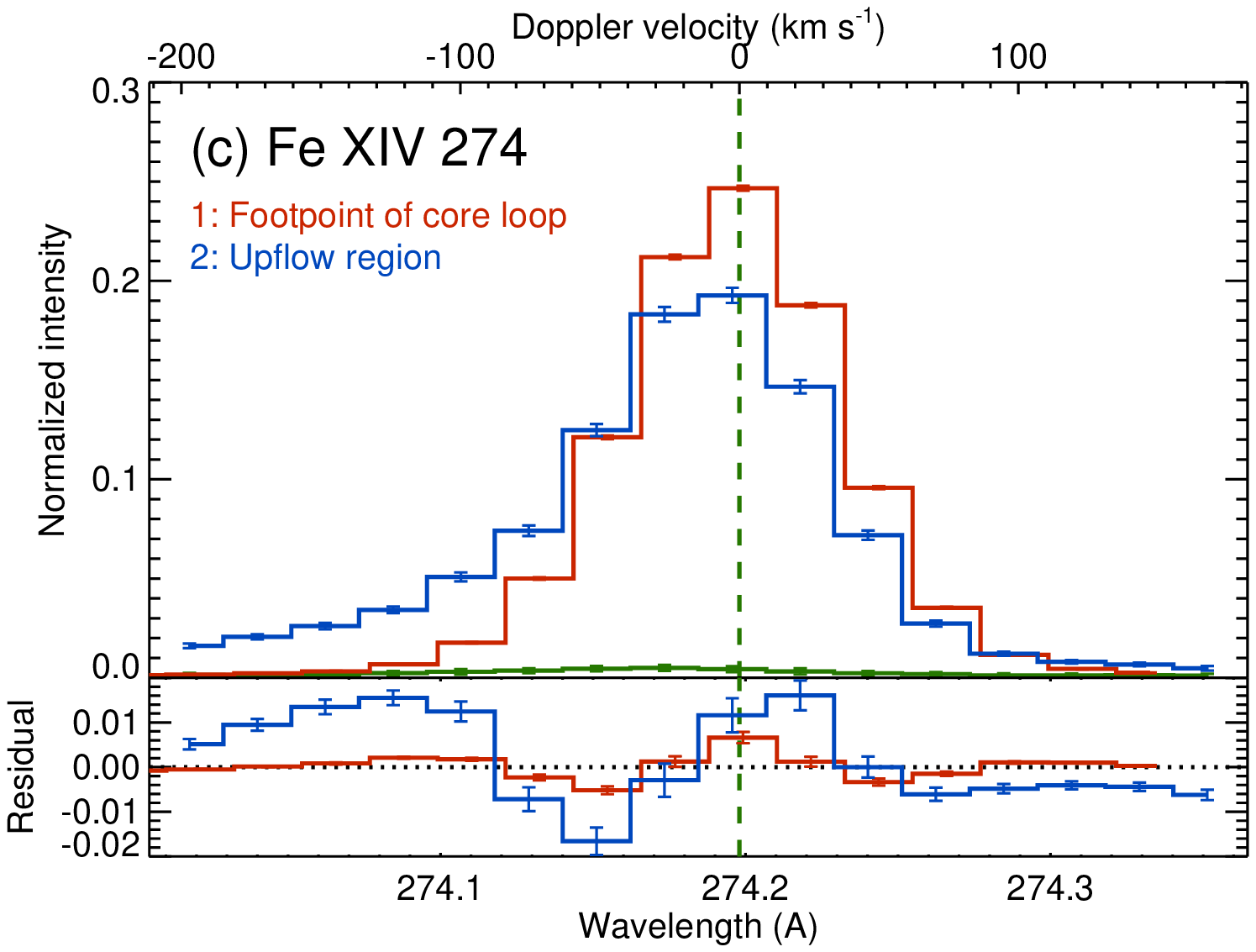}
  \end{minipage}
  \caption{Line profiles of the active region AR10978. (a) Context image of AR10978 obtained on 2007 December 11 00:24:16--04:47:29UT.  Intensity of Fe \textsc{xiv} $264.78${\AA} is shown.  Boxes numbered as 1 (\textit{red}) and 2 (\textit{blue}) respectively indicate a footpoint of core loops and the outflow region.  (b) Fe \textsc{xiv} $264.78${\AA} spectra.  (c) Fe \textsc{xiv} $274.20${\AA} spectra.  In each panel, line profiles at the footpoint of core loops (\textit{red} histogram) and at the outflow region analyzed here (\textit{blue} histogram) are shown in the \textit{upper} half.  The spectra were normalized by their integration.  Residuals from the single-Gaussian fitting of each histogram are shown in the \textit{lower} half.  The \textit{Green} histogram in panel (c) shows estimated spectrum of Si \textsc{vii} $274.18${\AA}.}
  \label{fig:dns_lp_examples}
\end{figure}

There are three reasons for the difficulties in the analysis of spectroscopic data obtained by \textit{Hinode}/EIS.  Firstly, the signals from an upflow are detected as an enhanced blue wing (hereafter, EBW) component in emission line profiles.  Examples are shown in Figure \ref{fig:dns_lp_examples}.  In each panel, line profiles at the footpoint of a core loop (\textit{red} histogram) and at the outflow region analyzed here (\textit{blue} histogram) are shown in the \textit{upper} half.  Residuals from single-Gaussian fitting of each histogram are shown in the \textit{lower} half, which is quite useful in detecting weak signals in line wing \citep{hara2008}.  There is a significant enhancement at the blue wing ($\le -100 \, \mathrm{km} \, \mathrm{s}^{-1}$) both in Fe XIV $264.78${\AA} and $274.20${\AA} as shown by \textit{blue} histograms.  \textit{Green} histogram in panel (c) shows estimated spectrum of Si \textsc{vii} $274.18${\AA} which was subtracted in the density diagnostics described later.  The EBW component is weak in most cases as seen in spectra indicated by the blue histograms shown in Figure \ref{fig:dns_lp_examples}.  In addition, EBW component is significantly dominated by the strong component 
closer to the rest wavelength,  
which makes the analysis of upflows quite uncertain. 

Secondly, the density measurement of the outflow itself needs the accurate determination of the rest wavelengths of emission lines from which we fit the two emission lines simultaneously and deduce the intensity.  
This is often laborious because we do not have the absolute measure of the wavelength corresponding to each observational spectral pixels. 

Thirdly, density measurement needs at least two emission lines from the same ion (\textit{e.g.}, Fe \textsc{xiv} as used 
in this paper).  
This means that the two emission lines should be fitted simultaneously using same parameters such as Doppler velocity and line width.  No previous studies on the outflows from the edge of active region have dealt with such fitting. 

Our 
procedure of density diagnostics 
is as follows: (1) integration of neighboring multiple pixels in order to reduce the noise, (2) determination of the wavelength position corresponding to the same Doppler velocity, (3) removal of blending Si \textsc{vii} $274.18${\AA} from Fe \textsc{xiv} $274.20${\AA} using Si \textsc{vii} $275.35${\AA} as a reference, (4) simultaneous fitting of Fe \textsc{xiv} $264.78${\AA} and $274.20${\AA}, and (5) density inversion using a theoretical curve from CHIANTI as a function of the intensity ratio.  In the following sections, each procedure will be described in detail. 


%% file: dns_analysis_integration_pixels.tex

The outflows from the edge of active regions are usually detected as an EBW in emission line profiles.  Its intensity does not exceed $\sim 25 \mathrm{\%}$ of that of the major component \citep{doschek2012}.  This makes analysis difficult since the photon noise of the major component affects the emission from EBW.  In addition, the region where the outflows can be seen is usually dark (\textit{i.e.}, small signal-to-noise ratio).  In order to improve the signal-to-noise ratio, we integrated over multiple observational pixels in space using a square box with the size of $5'' \times 5''$.  A larger integration box generally results in better signal-to-noise ratio.  However, we chose that particular size of integration box so as not to lose the information of the outflow region.  In the integration, the pixels with instrumental problems (\textit{i.e.}, hot or bad pixels) were excluded. 


%% file: dns_analysis_de-blending.tex

Fe \textsc{xiv} 274.20 potentially has a contribution from Si \textsc{vii} $274.18${\AA}, which may become significant in the vicinity of an active region because Si \textsc{vii} emission often comes from the footpoint of cool loops extending from the edge of the active region.  We need to subtract this blend from Fe \textsc{xiv} $274.20${\AA}.  In this study, the spectrum of Si \textsc{vii} $274.18${\AA} was calculated by using the observed line profile of Si \textsc{vii} $275.35${\AA} which is known to be clean (\textit{i.e.}, without any significant blend).  The intensity ratio of Si \textsc{vii} $274.18${\AA}/$275.35${\AA} is at most $0.25$ as calculated from CHIANTI version 7 \citep{dere1997,landi2012}.  The value has a dependence in the density range $10^{8} \, \mathrm{cm}^{-3} \le n_{\mathrm{e}} \le 10^{10} \, \mathrm{cm}^{-3}$, and it varies $0.06$--$0.27$ (monotonically increasing) as shown in Figure \ref{fig:dns_itdn_rat_chianti}.  First we remove the blending Si \textsc{vii} $274.18${\AA} for the case $n_{\mathrm{e}} = 10^{9} \, \mathrm{cm}^{-3}$ (Si \textsc{vii} electron density), and after that we considered three cases of the ratio corresponding to the density of $10^{8}$, $10^{9}$, and $10^{10} \, \mathrm{cm}^{-3}$.  In order to make our analysis more robust, we excluded the location where the estimated intensity of Si \textsc{vii} $274.18${\AA} exceeds $5${\%} of the Fe \textsc{xiv} intensity.  Using the theoretical ratio, the intensity of Si \textsc{vii} $275.35${\AA} was converted into that of Si \textsc{vii} $274.18${\AA}.  The spectrum of Si \textsc{vii} $275.35${\AA} was then placed at Si \textsc{vii} $274.18${\AA} taking into account the shift of Si \textsc{vii} $275.35${\AA} from the rest wavelength using the relative difference between wavelength of Si \textsc{vii} $274.18${\AA} and $275.35${\AA} (\textit{i.e.}, $1.1808${\AA}) given by CHIANTI database.  Note that since there were no locations where Si \textsc{vii} $274.18${\AA} dominates Fe \textsc{xiv} $274.20${\AA} in the data, we could not determine the relative wavelength position of the two Si \textsc{vii} lines, therefore we used the wavelength difference given by CHIANTI for the Si \textsc{vii} lines.  The data points of the estimated Si \textsc{vii} $274.18${\AA} in the wavelength direction were interpolated into the data points of Fe \textsc{xiv} $274.20${\AA} by cubic spline. Thus, we removed the blended Si \textsc{vii} $274.18${\AA} from Fe \textsc{xiv} $274.20${\AA}. 

Concerning Fe \textsc{xiv} $264.78${\AA}, there are two possible blend lines: Fe \textsc{xi} $264.77${\AA} and Fe \textsc{xvi} $265.01${\AA}.  
As for Fe \textsc{xvi} $265.01${\AA}, it is sufficiently far enough from Fe \textsc{xiv} $264.78${\AA} in non-flare situations.  Moreover, estimated peak intensity of Fe \textsc{xvi} $265.01${\AA} was around $100 \, \mathrm{erg} \, \mathrm{cm}^{-2} \, \mathrm{s}^{-1} \, \mathrm{sr}^{-1} \, \text{\AA}^{-1}$ in the observed outflow region\footnote{We estimated the intensity from Fe \textsc{xvi} $262.98${\AA} included in EIS data.  The line ratio Fe \textsc{xvi} $265.01${\AA}/$262.98${\AA} was determined in the raster scan which started from 10:25:42UT since it included the spectra of both Fe \textsc{xvi} $262.98${\AA} and $265.01${\AA}, and it resulted in the ratio of $0.083$.}, 
which is no greater than the background level of Fe \textsc{xiv} $264.78${\AA} as seen in Figure \ref{fig:fig_diff}.  
Unfortunately, our data set did not have any isolated Fe \textsc{xi} emission line, which makes difficult to remove the blending Fe \textsc{xi} from Fe \textsc{xiv} $264.78${\AA}. 
Nevertheless, our crude estimation of the intensity of Fe \textsc{xi} $264.77${\AA} from Fe \textsc{xi} $188.21$/$188.30${\AA} ($I_{\mathrm{264.77}\text{\AA}} / I_{\mathrm{188.21}\text{\AA}} \lesssim 0.03$) leads to the potential influence on Fe \textsc{xiv} $264.78${\AA} by up to $5${\%} in maximum.  
It is inferred from Appendix \ref{sect:dns_results_sivii} that the error in our results can be considered to lie within $\sim 3${\%}.  


%% file: dns_analysis_simul_fit.tex

In order to make the fitting more robust, the two emission line profiles of Fe \textsc{xiv} $264.78${\AA}/$274.20${\AA} were fitted simultaneously.  It is based on the consideration that the emission line profiles coming from the same ion species must have the same Doppler shift and the same Doppler width. 
As seen in Figure \ref{fig:dns_lp_examples}, emission line profiles of Fe \textsc{xiv} $264.78${\AA} and $274.20${\AA} from the active region core (\textit{red} histogram) are obviously symmetric, while those from the outflow region (\textit{blue} histogram) have an EBW, 
from which it is not likly considered that a strong major component hinders any signals of the upflows in the active region core.  
This EBW did not exceed the major component anywhere in the outflow region ($\le 30${\%}).  Previous observations have never shown such emission line profiles whose EBW dominates over the major component \citep{doschek2012}. 

\begin{figure}[!t]
  \centering
  \includegraphics[width=8cm,clip]{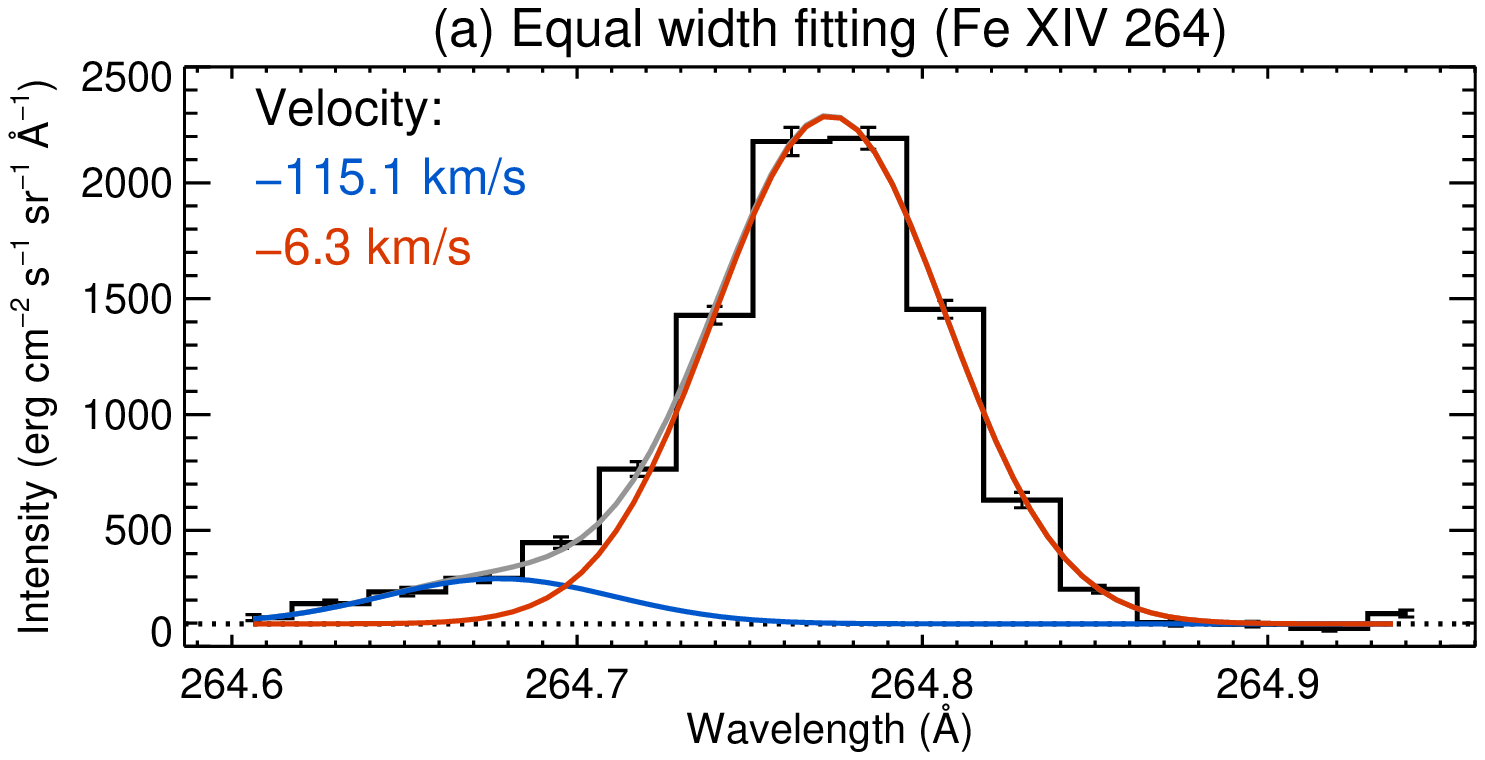}
  \includegraphics[width=8cm,clip]{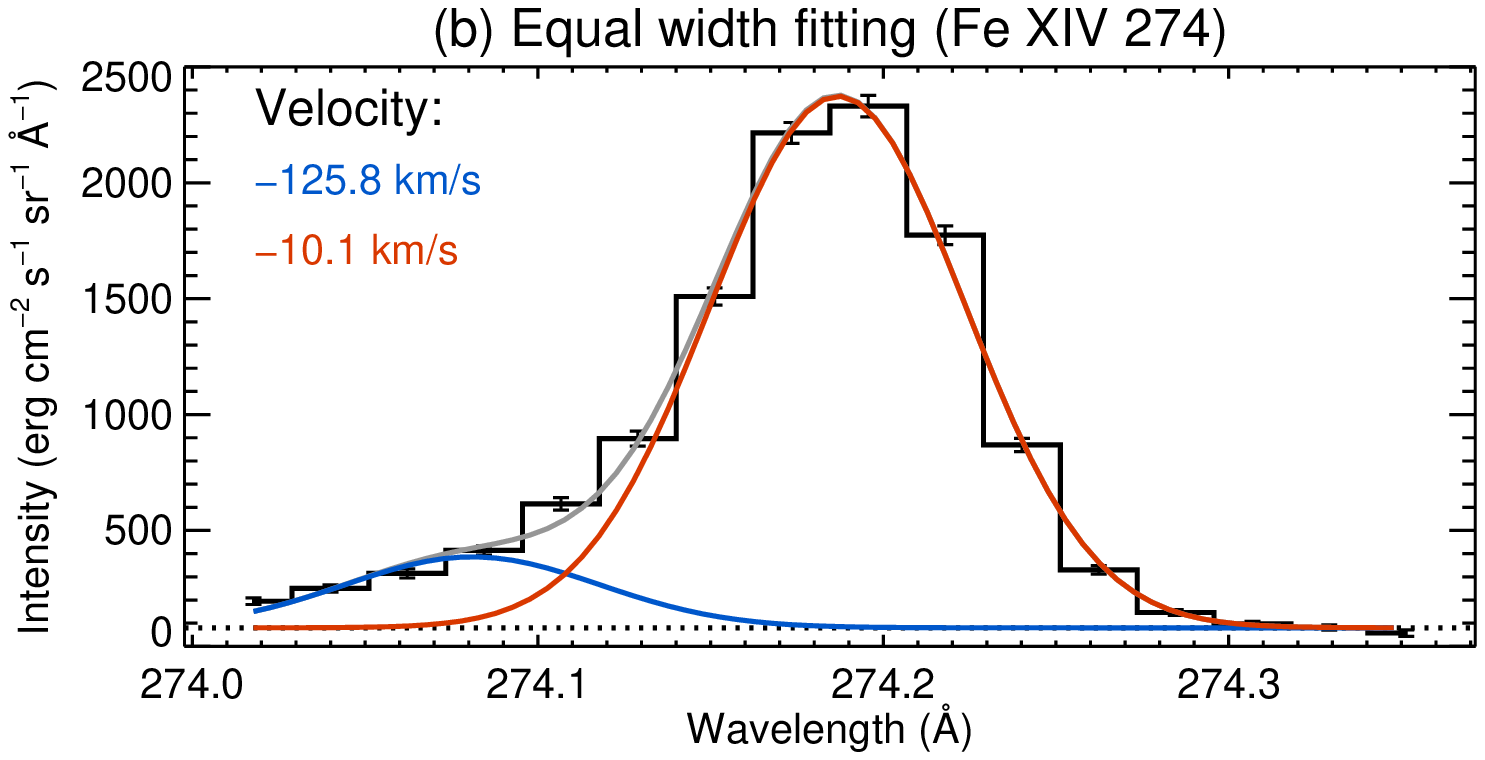}
  \includegraphics[width=8cm,clip]{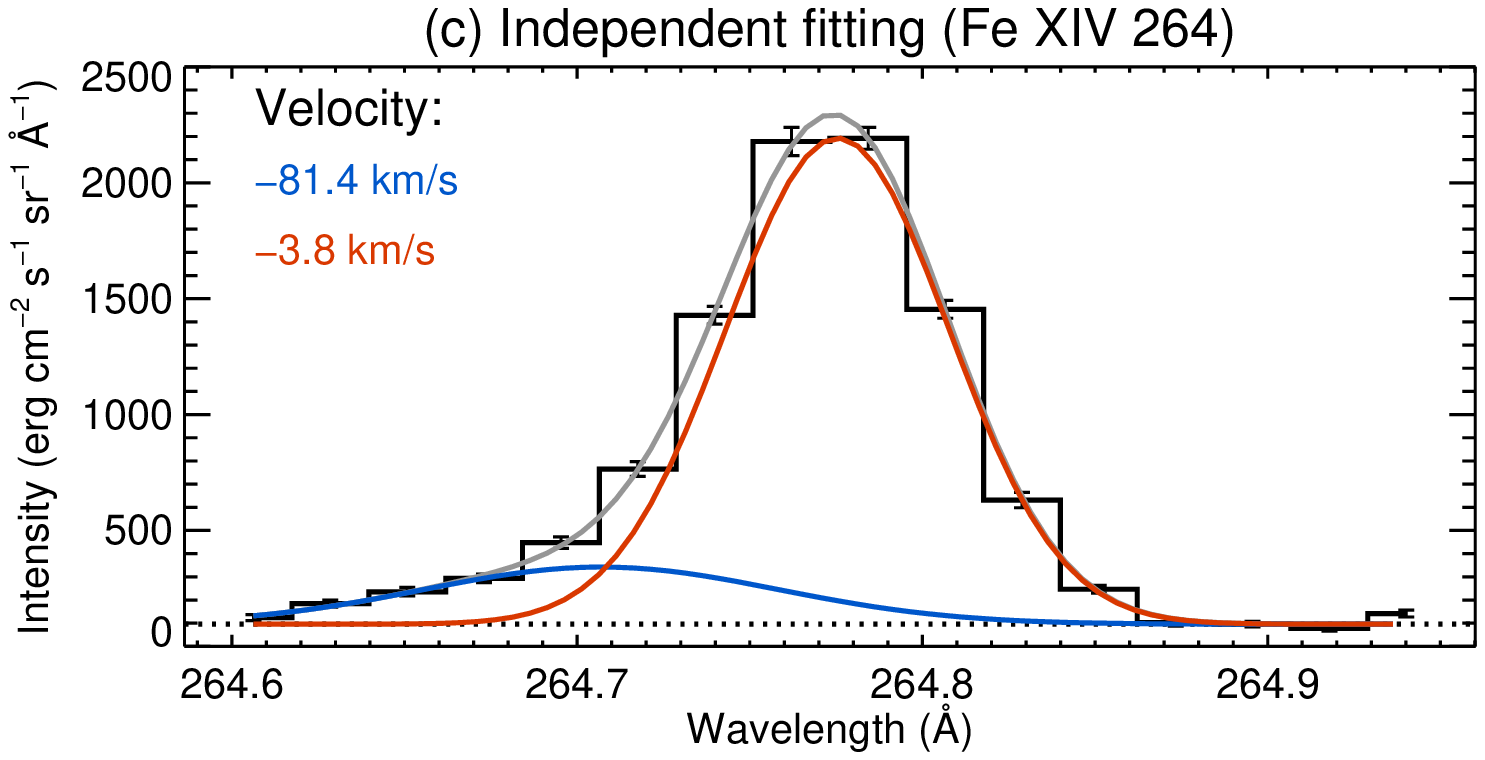}
  \includegraphics[width=8cm,clip]{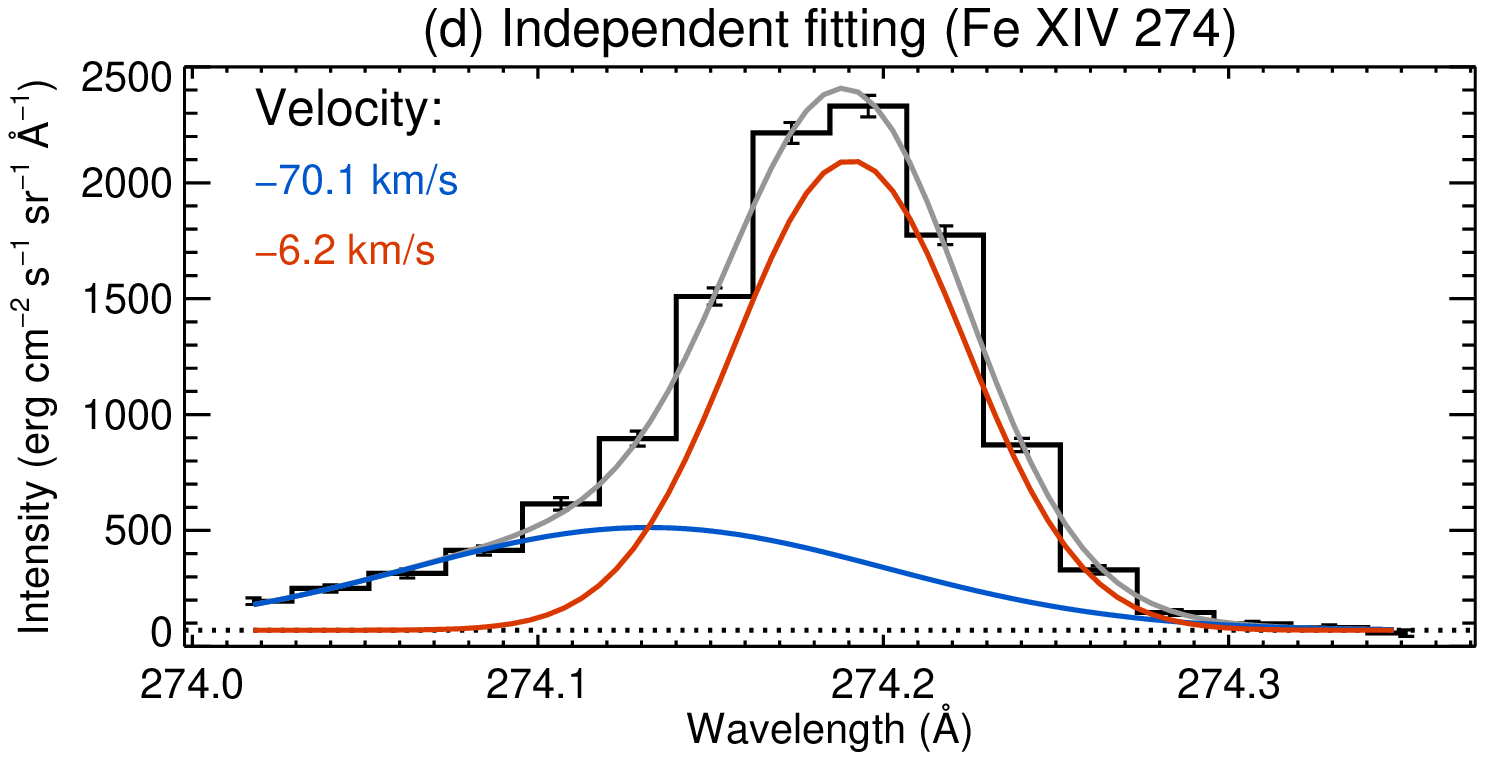}
  \includegraphics[width=8cm,clip]{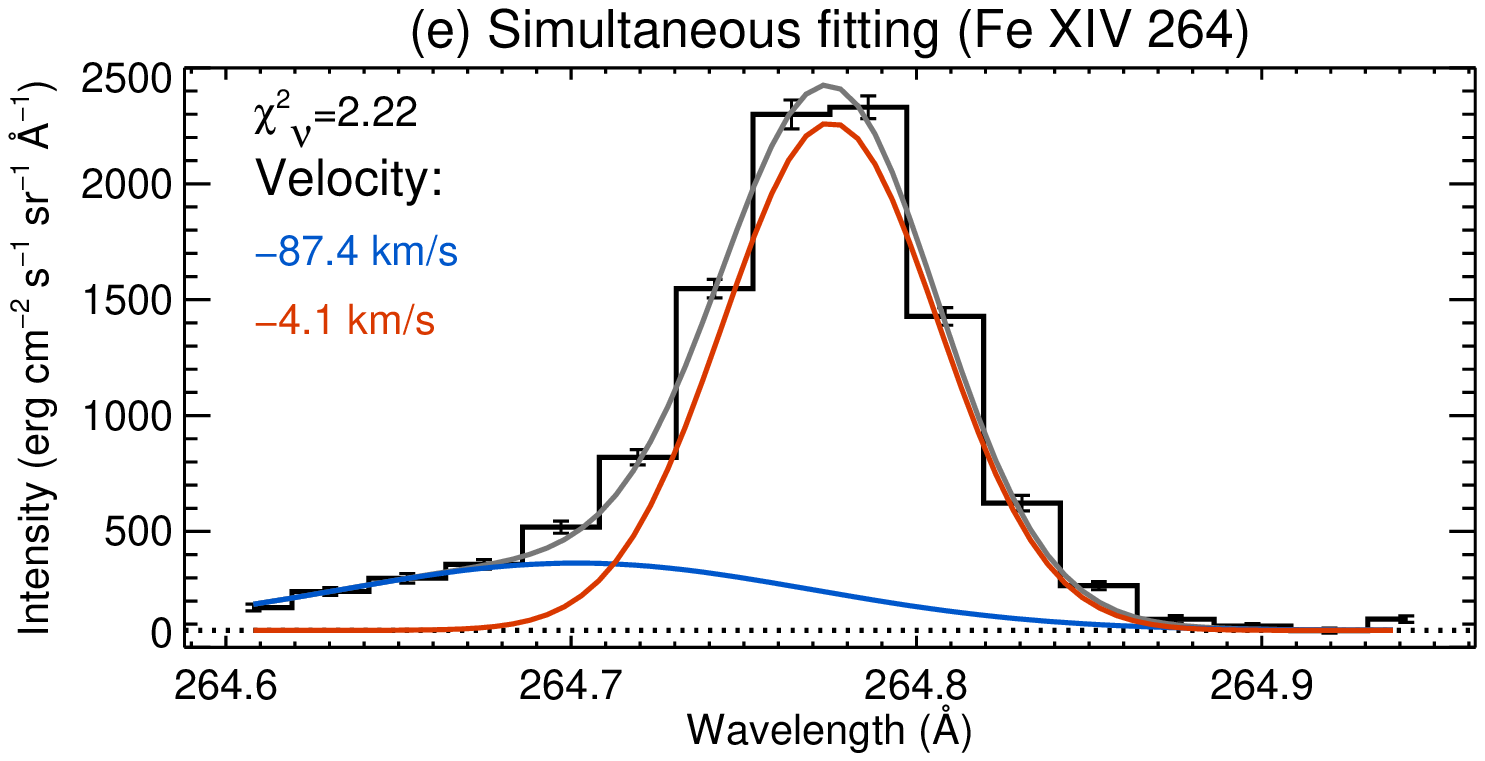}
  \includegraphics[width=8cm,clip]{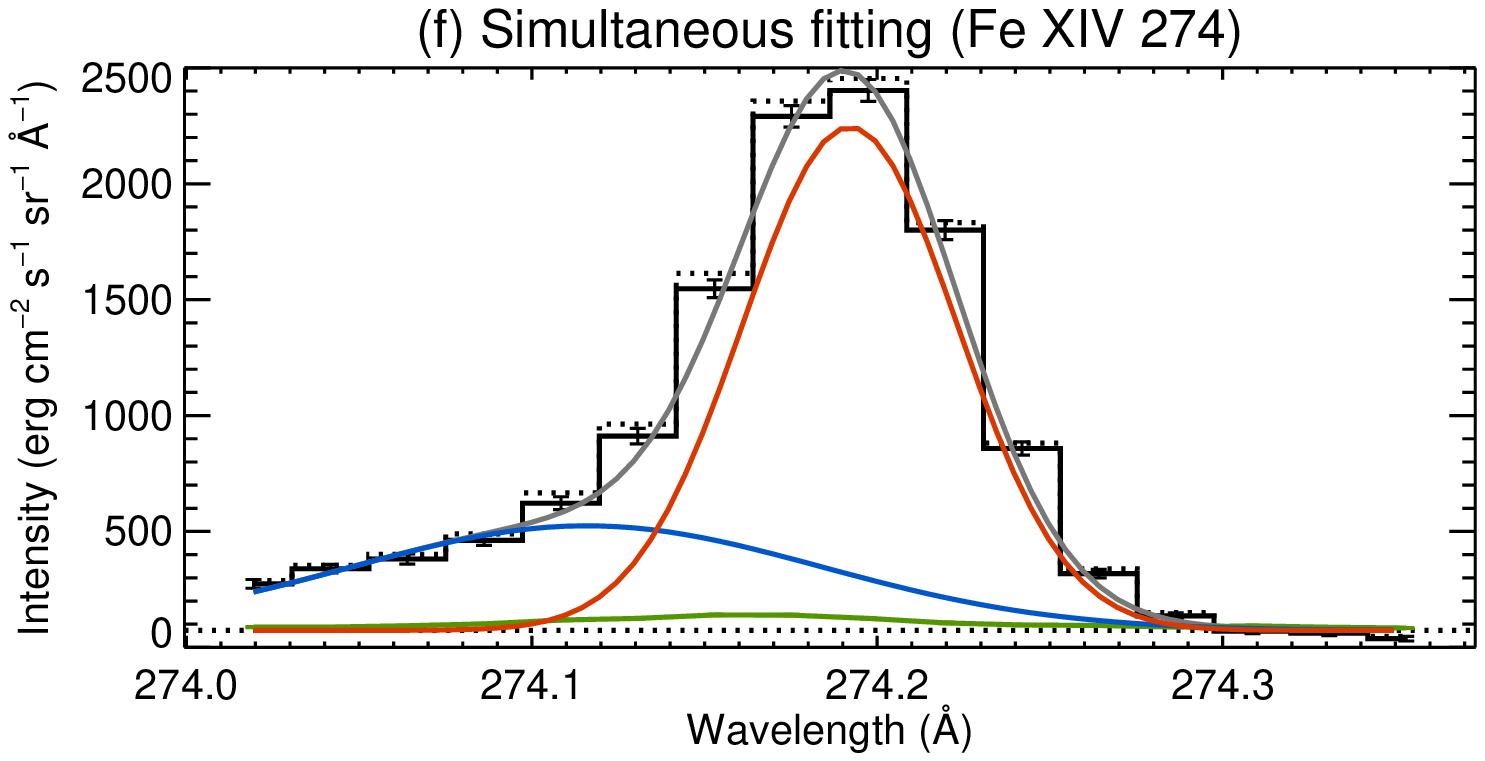}
  \caption{Fitting results for Fe \textsc{xiv} $264.78${\AA} and $274.20${\AA} obtained by three different models.  \textit{Upper row} (panels a and b): fitting with double Gaussians that have the same line width applied to each line profile independently (model 1).  \textit{Middle row} (panels c and d): fitting with double Gaussians that do not necessarily have the same line width applied to each line profile independently (model 2).  \textit{Lower row} (panels e and f): fitting with double Gaussians applied to two line profiles simultaneously without the assumption of the same line width of two components (model 3). }
  \label{fig:fig_diff}
\end{figure}

In this study, the emission line profiles of the outflow region are assumed to be composed of two Gaussian components.  
Most previous analyses on the outflows at the edge of an active region assumed that the main component and EBW have \textit{the same line width} in order to reduce avoid an unrealistic solution in the fitting parameter space, but the assumption could strongly affect the fitting \citep{bryans2010,brooks2012}.  \citet{brooks2012} mentioned that this assumption may lead to the underestimation of the intensity of EBW.  Line profile with EBW often shows rather longer tail in the line wing than could be represented by a Gaussian which has the same line width as the major component. Moreover, the assumption that the major component and EBW have the same line width is not based on the physical principles. 

In order to examine the differences in the fitting result between different constraint on the fitting parameters, we applied three fitting models to a line profile pair of Fe \textsc{xiv} $264.78${\AA}.  Line centroid and line width are respectively denoted by $\lambda$ and $W$. The suffixes below represent: ``1'' for Fe \textsc{xiv} $264.78${\AA}, ``2'' for Fe \textsc{xiv} $274.20${\AA} followed by the component either ``Major'' or ``EBW''.  
First model (model 1) assumes $W_{1, \mathrm{Major}} = W_{1, \mathrm{EBW}}$ and $W_{2, \mathrm{Major}} = W_{2, \mathrm{EBW}}$, and fits the line profiles of Fe \textsc{xiv} $264.78${\AA} and $274.20${\AA} separately with double Gaussians that have the same line width for each component.  
The second model (model 2) also fits the line profiles of the two Fe \textsc{xiv} separately, but with double Gaussians that do not necessarily have the same line width for each component.  
The third model (model 3) fits the two Fe \textsc{xiv} line profiles simultaneously by applying $\lambda_{2, \mathrm{Major}} = \alpha \lambda_{1, \mathrm{Major}}$, $\lambda_{2, \mathrm{EBW}} = \alpha \lambda_{1, \mathrm{EBW}}$ ($\alpha = 1.0355657$, \citeauthor{kitagawa2013} \citeyear{kitagawa2013}), $W_{1, \mathrm{Major}} = W_{2, \mathrm{Major}}$, and $W_{1, \mathrm{EBW}} = W_{2, \mathrm{EBW}}$.  We adopted model 3 for the electron density measurement in this study because it is physically most reasonable 
in the sense that the model calculates the parameters (line centroids and thermal widths) consistently for both emission lines and does not impose artificial restrictions on the line widths.

The results for those three models are shown in Figure \ref{fig:fig_diff}.  We obtained smaller and more blueshifted second component (EBW) with the model 1 in panels (a) and (b), which confirms the suggestion in \citet{brooks2012}.  In contrast, larger and less blueshifted EBWs were obtained with models 2 and 3 as clearly seen in panels (c)--(f).  In addition to this, the line widths of EBW component were much broader for models 2 and 3 than for model 1.  
It is not clear at present whether the increased widths may indicate superposition of multiple upflow components, which will be another point to be revealed in the future.
The comparison between those three models shows that the results in previous analyses probably underestimate the intensity of EBW with an artificial assumption that two components in line profile have the same line width.  Moreover, independent fitting applied to two emission lines causes a discrepancy as seen in panel (c) and (d).  The Doppler velocity of EBW component was $-81.4 \, \kmpers$ for Fe \textsc{xiv} $264.78${\AA} while it was $-70.1 \, \kmpers$ for Fe \textsc{xiv} $274.20${\AA}.  Note that the rest wavelengths were determined from a limb observation on 2007 December 6, so these Doppler velocities have an uncertainty of $10 \, \kmpers$ at most.


%% file: dns_analysis_dens_inv.tex

Now the densities of EBW and the major component can be obtained by referring to the theoretical intensity ratio of Fe \textsc{xiv} $264.78${\AA}/$274.20${\AA} as a function of electron density shown in Figure \ref{fig:dns_itdn_rat_chianti}.  The intensity ratio monotonically increases within the density range of $10^{8} \, \mathrm{cm}^{-3} \le n_{\mathrm{e}} \le 10^{12} \, \mathrm{cm}^{-3}$.  The electron density in the solar corona generally falls between $10^{8} \, \mathrm{cm}^{-3}$ (for coronal holes) and $10^{11} \, \mathrm{cm}^{-3}$ (for flare loops), so the intensity ratio of Fe \textsc{xiv} $264.78${\AA}/$274.20${\AA} is quite useful.  The error in the density was calculated by using the 1-$\sigma$ error in the intensity ratio.  The electron density is obtained from
\begin{equation}
  n_{\mathrm{e}} = F^{-1} \left( \frac{I_{264}}{I_{274}} \right) \, \mathrm{,}
\end{equation}
where $F^{-1}$ is the inverse function of the theoretical intensity ratio, and $I_{264}$ and $I_{274}$ are respectively the observed intensity of Fe \textsc{xiv} $264.78${\AA} and $274.20${\AA}.  Using $\sigma_{I_{264} / I_{274}}$ as the error of observed intensity ratio, we estimate the error of the density $\sigma_{n_{\mathrm{e}}}$ as
\begin{equation}
  n_{\mathrm{e}} \pm \sigma_{n_{\mathrm{e}}} 
  = F^{-1} \left( \frac{I_{264}}{I_{274}} \pm \sigma_{I_{264} / I_{274}} \right) \, \mathrm{.} 
\end{equation}
The error $\sigma_{n_{\mathrm{e}}}$ was not dealt symmetrically in this definition, which comes from the fact the function $F$ has a curvature which can not be negligible compared to $\sigma_{I_{264} / I_{274}}$. 


%% file: dns_results_1G.tex

\begin{figure}[!t]
  \centering
  \includegraphics[width=8cm,clip]{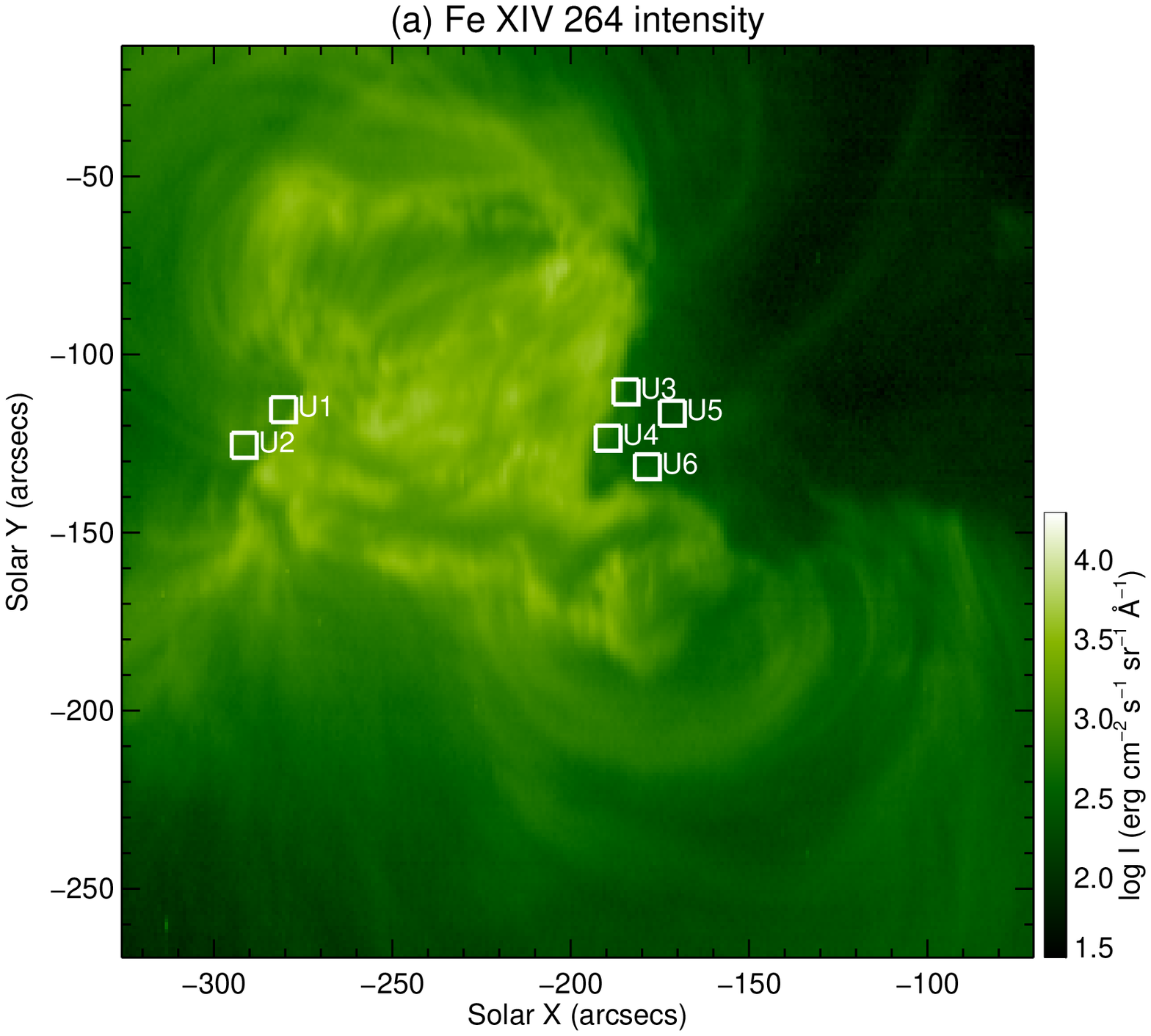}
  \includegraphics[width=8cm,clip]{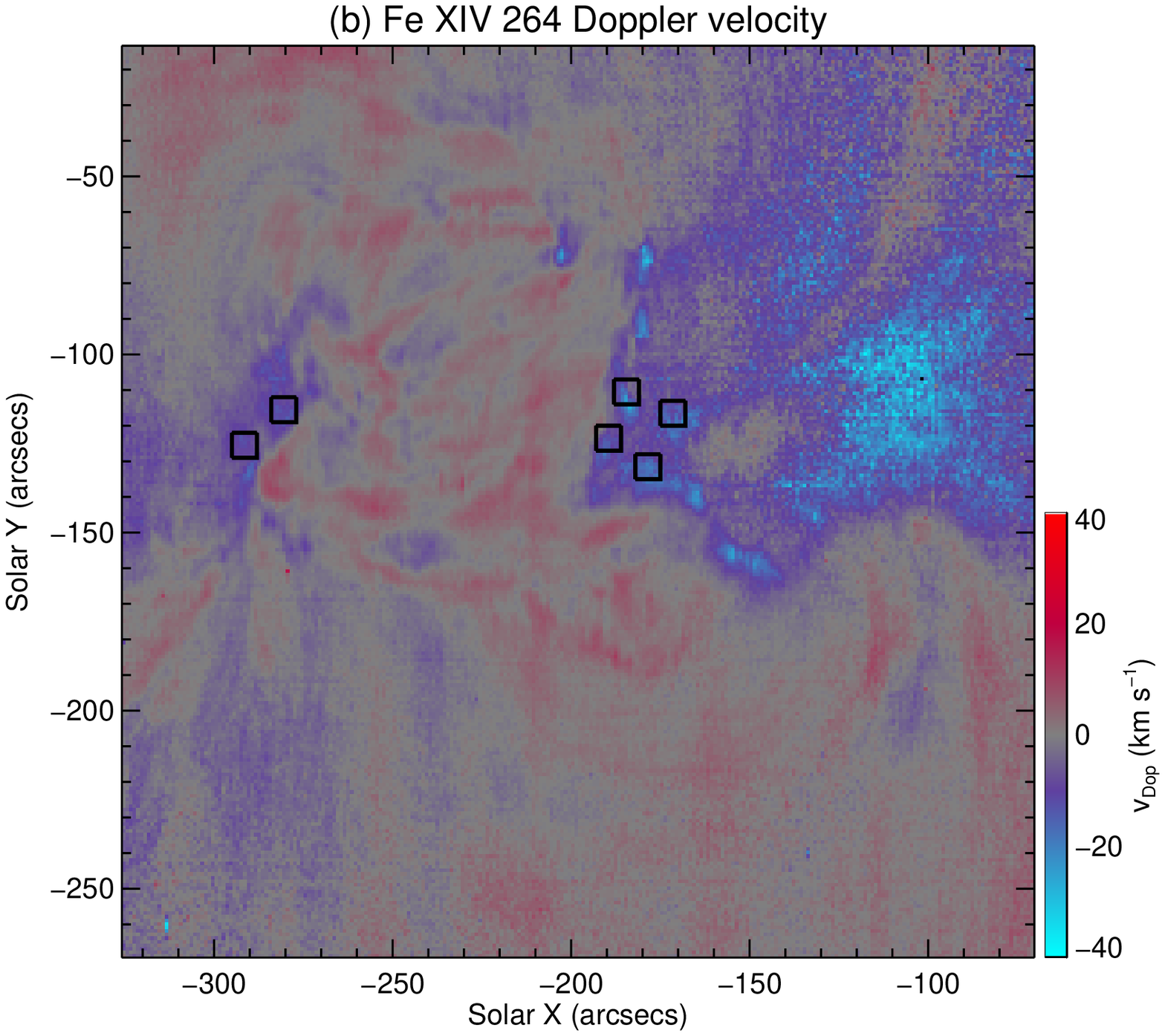}
  \includegraphics[width=8cm,clip]{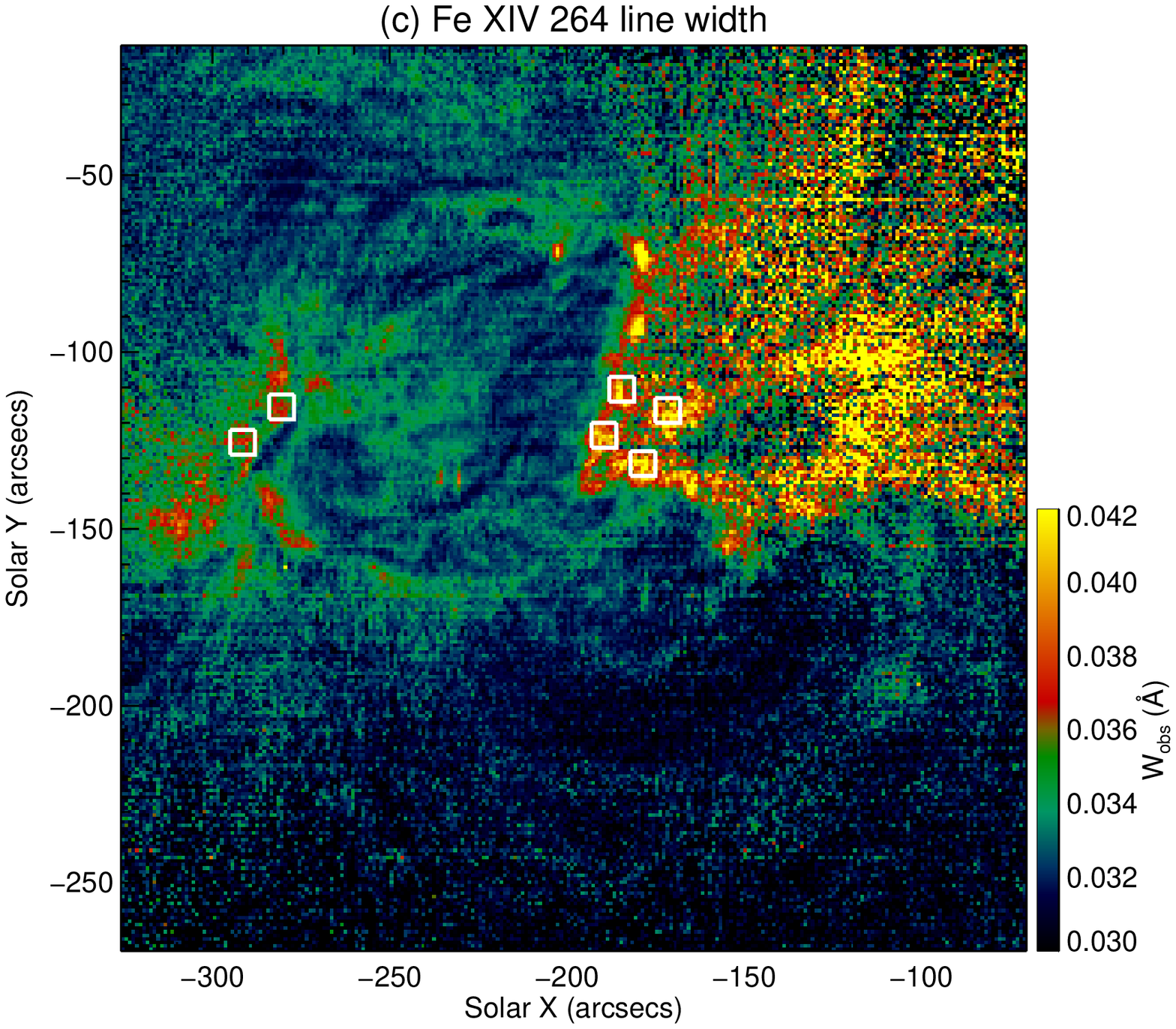}
  \includegraphics[width=8cm,clip]{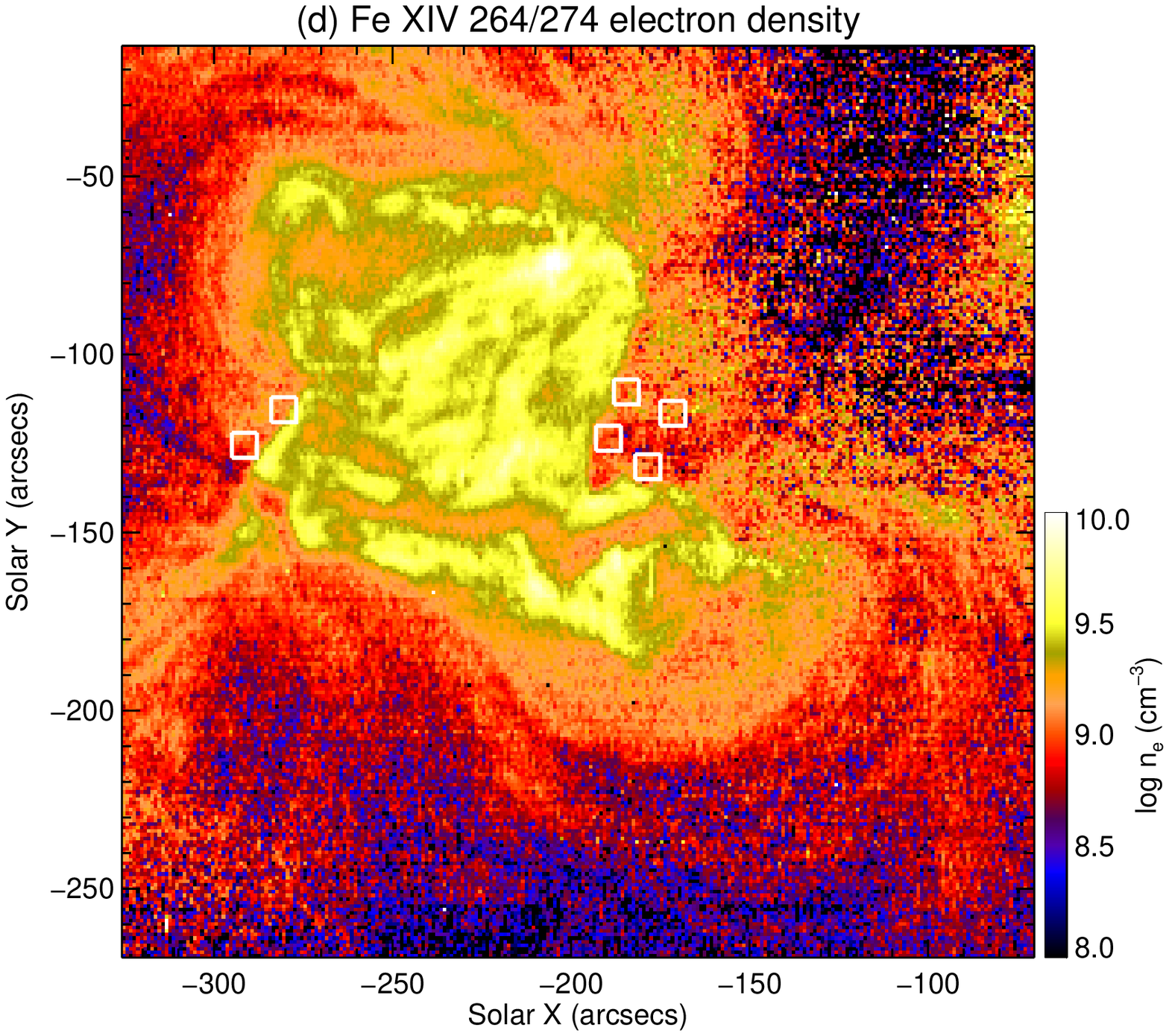}
  \caption{Physical quantities deduced from single Gaussian fitting for Fe \textsc{xiv} $264.78${\AA} obtained on 2007 December 11 00:24:16--04:47:29UT. (a) Intensity of Fe \textsc{xiv} $264.78${\AA}. (b) Doppler velocity of Fe \textsc{xiv} $264.78${\AA}. (c) Line width of Fe \textsc{xiv} $264.78${\AA}. (d) Electron density derived from the line ratio Fe \textsc{xiv} $264.78${\AA}/$274.20${\AA}.}
  \label{fig:fexiv_1G_map}
\end{figure}

First we describe the results deduced from single-Gaussian fitting.  As described above, line profiles at the outflow regions are known to have a distorted shape which cannot be well represented by single Gaussian. Nonetheless, the results deduced from single-Gaussian fitting may be useful because the fitting is much more robust in terms of the freedom of variables (\textit{e.g}, 4 parameters for single Gaussian with constant background and 7 parameters for double Gaussians).  Figure \ref{fig:fexiv_1G_map} shows the map of intensity, Doppler velocity, line width of Fe \textsc{xiv} $264.78${\AA}, and electron density derived from the line ratio Fe \textsc{xiv} $264.78${\AA}/$274.20${\AA}.  The blending Si \textsc{vii} $274.18${\AA} was taken into account and subtracted by referring to Si \textsc{vii} $275.35${\AA}.  It is clear from panel (b) that the outflow regions are present (\textit{i.e.}, blueshift) at the east/west edge of the active region core around 
$(x, y) = (-280'', -120'')$
 and $(-175'', -125'')$.  Panel (c) shows that the line width at those outflow regions is larger than other locations by $\Delta W = 0.020\text{--}0.027$\text{\AA} (square root of the difference of squared line width) equivalent to $\delta v = 20 \text{--} 30 \, \kmpers$, which is similar to a result reported previously \citep{doschek2008,hara2008}.  The electron density at the outflow regions is $n_{\mathrm{e}} = 10^{8.5\text{--}9.5} \, \mathrm{cm}^{-3}$, which is lower than that at the core ($n_{\mathrm{e}} \geq 10^{9.5} \, \mathrm{cm}^{-3}$).  

We defined the outflow regions as the locations (1) where the line width of Fe \textsc{xiv} $264.78${\AA} is enhanced, and (2) which can be separated from fan loops seen in Si \textsc{vii} intensity map (not shown here).  The selected six regions are indicated by \textit{white} boxes in each map (numbered by U1--U6 as written in panel (a), whose size is $8'' \times 8''$.  Those regions are located beside the bright core as seen in the intensity map (panel a).  We hereafter refer to U1--U2 as the eastern outflow region and U3--U6 as the western outflow region. 


%% file: dns_results_n.tex

\begin{figure}[!t]
  \centering
  \includegraphics[width=8cm,clip]{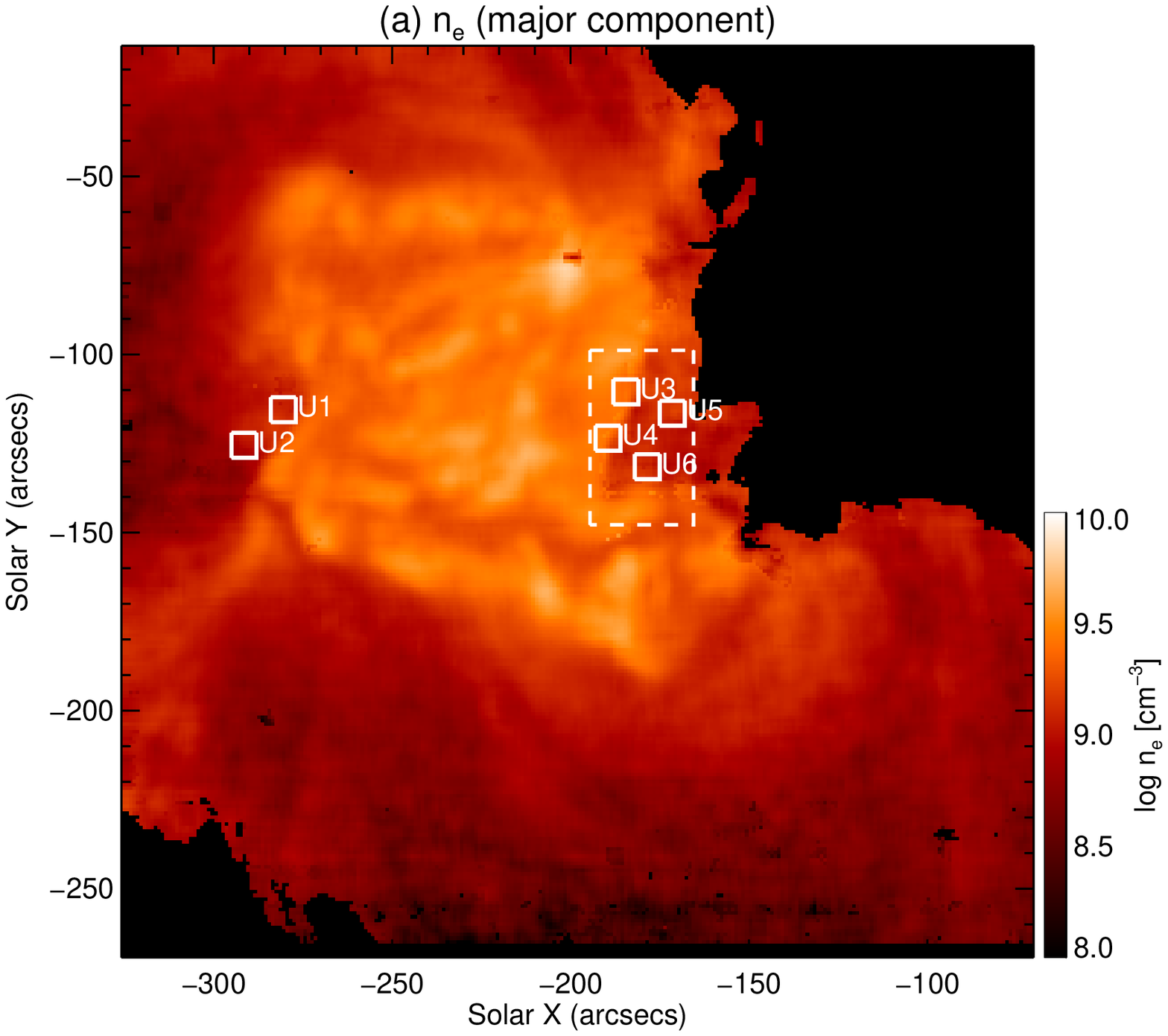}
  \includegraphics[width=8cm,clip]{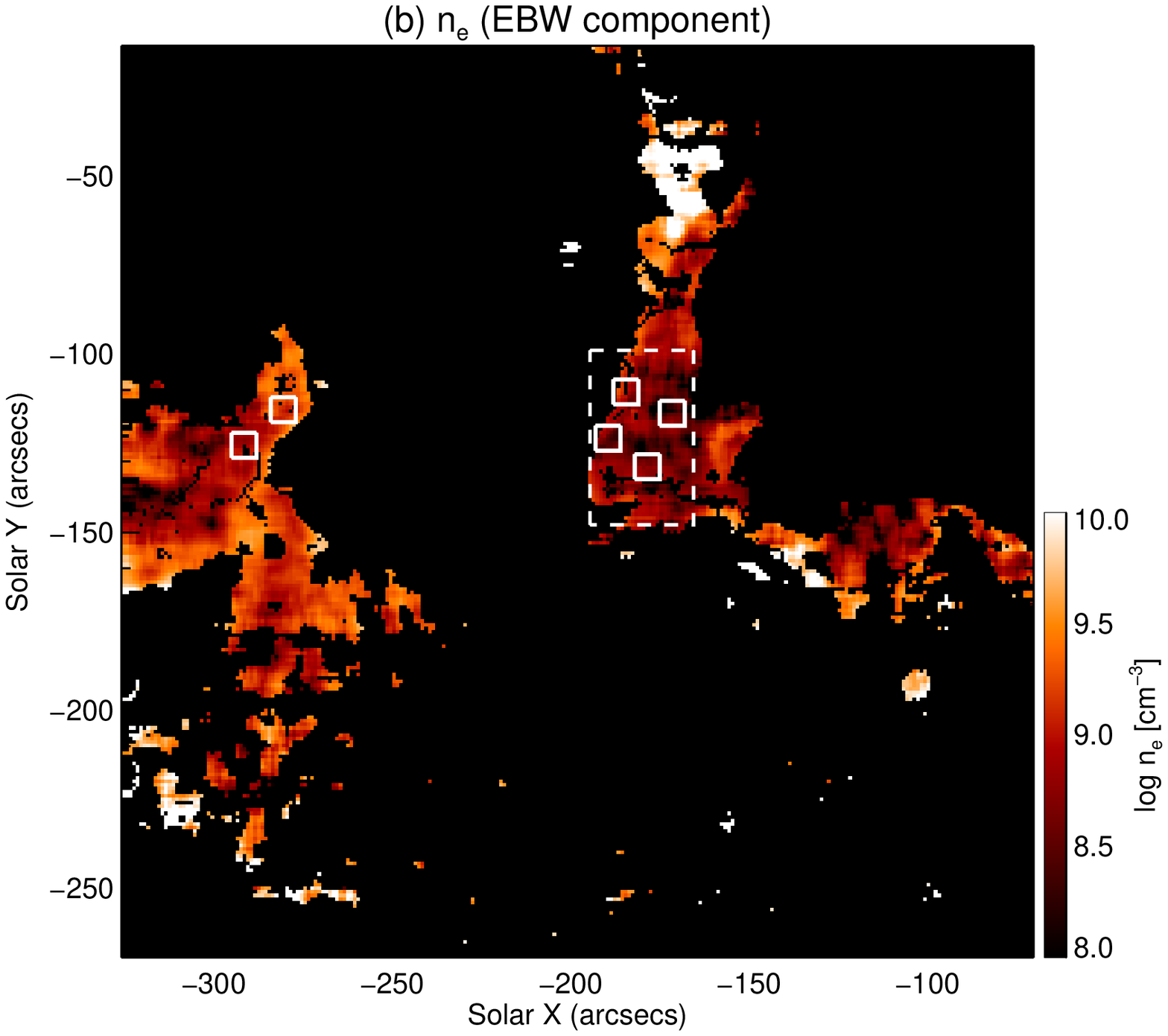}
  \caption{Electron density map deduced from two Gaussian fitting of an emission line pair Fe \textsc{xiv} $264.78${\AA}/$274.20${\AA} obtained by the raster scan on 2007 December 11 00:24:16--04:47:29UT.  (a) Electron density of the major component.  (b) Electron density of EBW component.  Same color contour are used in the two panels.  Pixels where the peak intensity of the major component ($I_{\mathrm{Major}}$) did not exceed $2.0 \times 10^{3} \, \mathrm{erg} \, \mathrm{cm}^{-2} \, \mathrm{s}^{-1} \, \mathrm{sr}^{-1} \, \text{\AA}^{-1}$ were masked by \textit{black}.  \textit{White} boxes numbered U1--U6 are the same as those in Figure \ref{fig:fexiv_1G_map}.  The \textit{white dashed} box indicate the entire western outflow region.}
  \label{fig:dens_map}
\end{figure}

The electron density of EBW component was measured through the analysis described in Section \ref{subsect:dns_diag}.  Figure \ref{fig:dens_map} shows the distributions of electron density for the major component ($n_{\mathrm{Major}}$) in panel (a) and EBW component ($n_{\mathrm{EBW}}$) in panel (b).  Pixels where the peak intensity of the major component ($I_{\mathrm{Major}}$) did not exceed $2.0 \times 10^{3} \, \mathrm{erg} \, \mathrm{cm}^{-2} \, \mathrm{s}^{-1} \, \mathrm{sr}^{-1} \, \text{\AA}^{-1}$ were masked by \textit{black}.  This threshold was determined by using the scatter plot of intensity and electron density of the major component shown in Appendix \ref{sect:dns_app_in}.  Pixels falling into the next three conditions were displayed, and others were masked by \textit{black}.  (1) $I_{\mathrm{Major}} \ge 2.0 \times 10^3 \, \mathrm{erg} \, \mathrm{cm}^{-2} \, \mathrm{s}^{-1} \, \mathrm{sr}^{-1} \, \text{\AA}^{-1}$.  (2) The intensity of EBW component ($I_{\mathrm{EBW}}$) exceeds $3${\%} of that of the major component ($I_{\mathrm{EBW}}/I_{\mathrm{Major}} \ge 0.03$).  (3) The difference between the Doppler velocity of EBW component ($v_{\mathrm{EBW}}$) and that of the major component ($v_{\mathrm{Major}}$) satisfies $v_{\mathrm{EBW}} - v_{\mathrm{Major}} < -30 \, \mathrm{km} \, \mathrm{s}^{-1}$ (\textit{i.e.}, the two components are well separated).

\begin{figure}[!t]
  \centering
  \begin{minipage}[c]{8cm}
    \includegraphics[width=8cm,clip]{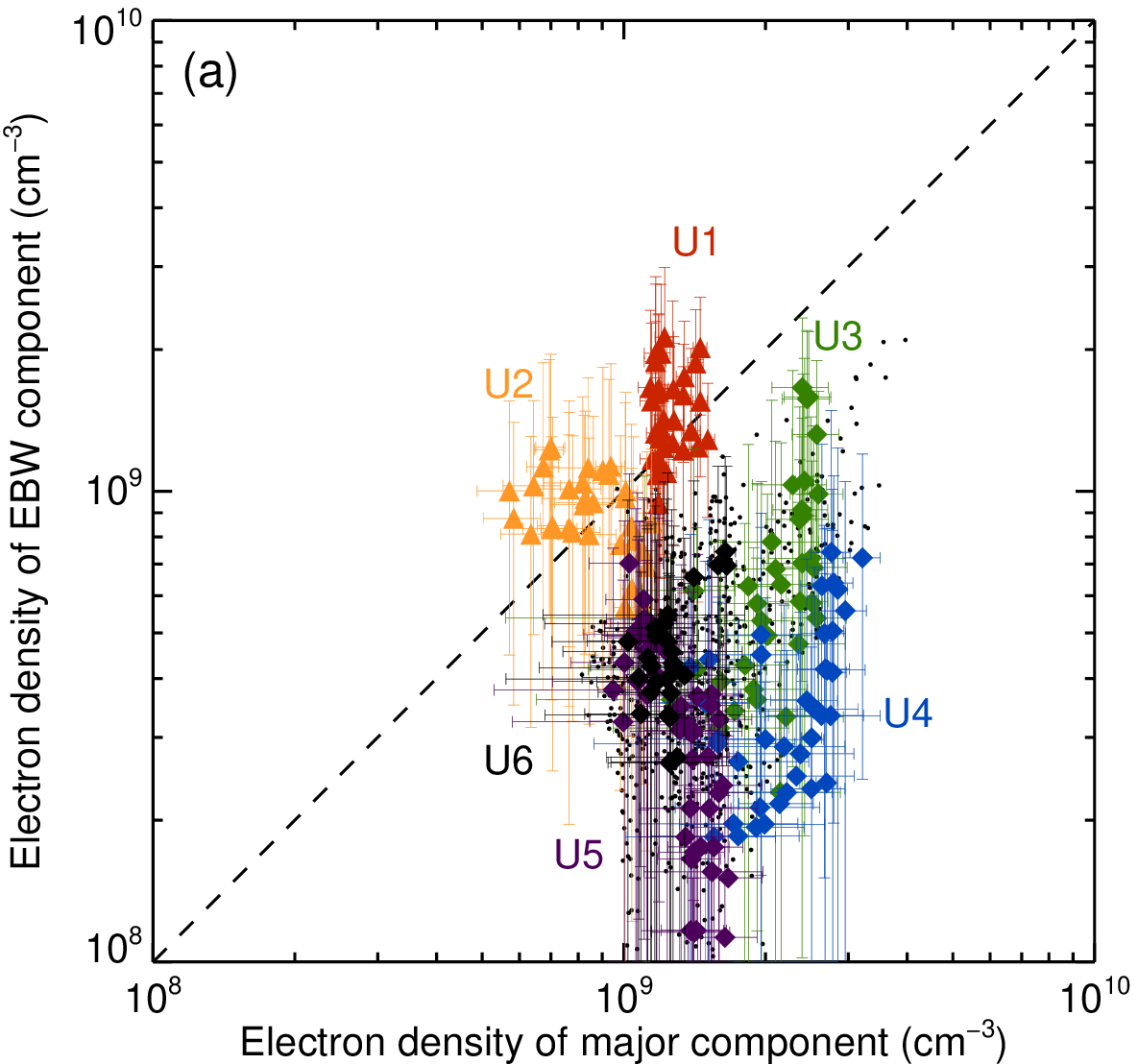}
  \end{minipage}
  \begin{minipage}[c]{8cm}
    \includegraphics[width=8cm,clip]{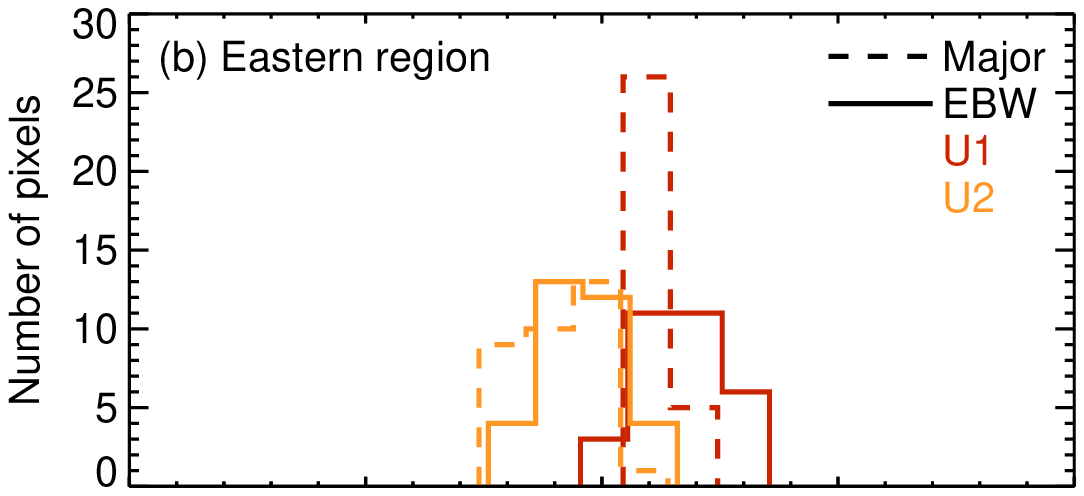}
    \includegraphics[width=8cm,clip]{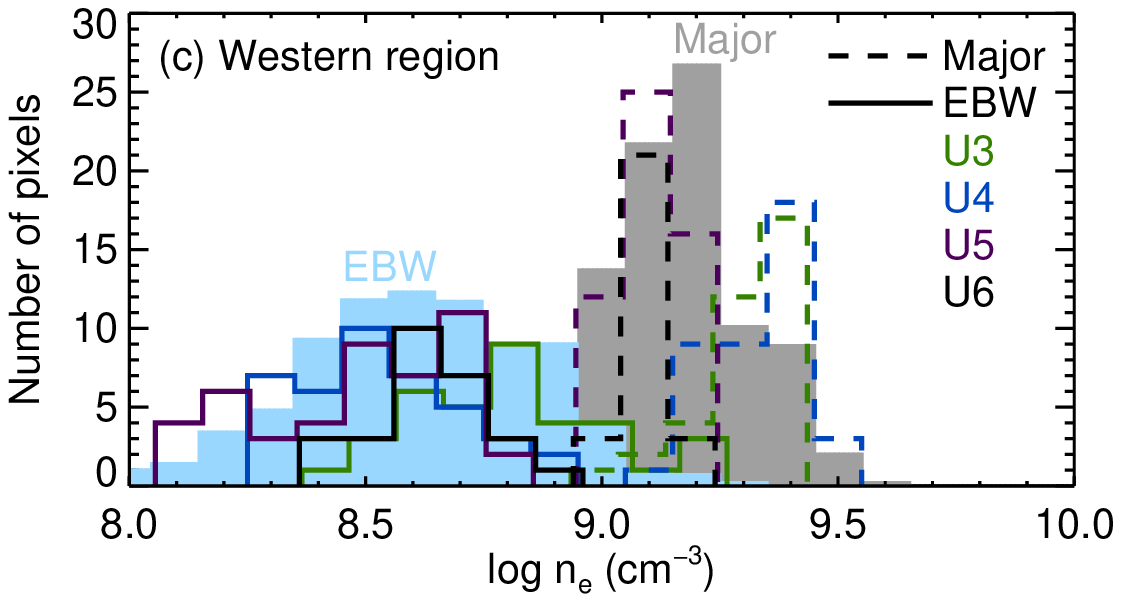}
  \end{minipage}
  \caption{(a) Scatter plot for Fe \textsc{xiv} electron density of the major component vs.\ that of EBW component.  \textit{Colors} show the selected region indicated by \textit{white} boxes in Figure \ref{fig:fexiv_1G_map}.  \textit{Triangles} (\textit{Diamonds}) represent the data points in the eastern (western) outflow regions.  Numbers beside data points correspond to the name of the \textit{white} boxes.  \textit{Black dots} show the electron density for the western outflow region indicated by a \textit{white dashed} box in Figure \ref{fig:dens_map}.  The dashed line indicates the point where two densities equal each other.  (b) Histograms for the electron density of the major component (\textit{dotted}) and EBW component (\textit{solid}) in the eastern outflow region.  (c) Histograms for the electron density of the major component (\textit{dotted}) and EBW component (\textit{solid}) in the western outflow region.  The \textit{gray} (the major component) and \textit{turquoise} (EBW component) histograms in the background are made for the entire western outflow region.  Those two histograms are multiplied by $0.1$.}
  \label{fig:dns_sct}
\end{figure}

The relationship of electron density between the major component and EBW component are shown in Figure \ref{fig:dns_sct}.   Scatter plot in panel (a) shows the electron density for the outflow regions U1--U6 (\textit{colored symbols}) and for the entire western outflow region indicated by the \textit{white dashed} box in Figure \ref{fig:dens_map} (\textit{black dots}).  The eastern outflow regions (U1--U2) and west ones (U3--U6) exhibit different characteristics.  The scatter plots for U1--U2 indicate $n_{\mathrm{Major}} \leq n_{\mathrm{EBW}}$, while those for U3--U6 indicate $n_{\mathrm{Major}} \geq n_{\mathrm{EBW}}$.  Panels (b) and (c) show the same data but in histograms for which colors again indicate the selected outflow regions.  The \textit{gray} (the major component) and \textit{turquoise} (EBW component) histograms in the background of panel (c) are made for the entire western outflow region.  Those two histograms clearly indicate that $n_{\mathrm{EBW}}$ ($10^{8.61 \pm 0.24} \, \mathrm{cm}^{-3}$) is smaller than $n_{\mathrm{Major}}$ ($10^{9.18 \pm 0.13} \, \mathrm{cm}^{-3}$) at the entire western outflow region, which confirms that our selection of the studied regions was not arbitrary.  
Note that our results for the western outflow region (U3--U6) roughly consistent with those of \citet{patsourakos2014} (\textit{i.e.}, $n_{\mathrm{EBW}} / n_{\mathrm{Major}} \lesssim 1$).  

%% file: dns_results_h.tex

\begin{figure}[!t]
  \centering
  \begin{minipage}[c]{8cm}
    \includegraphics[width=8cm,clip]{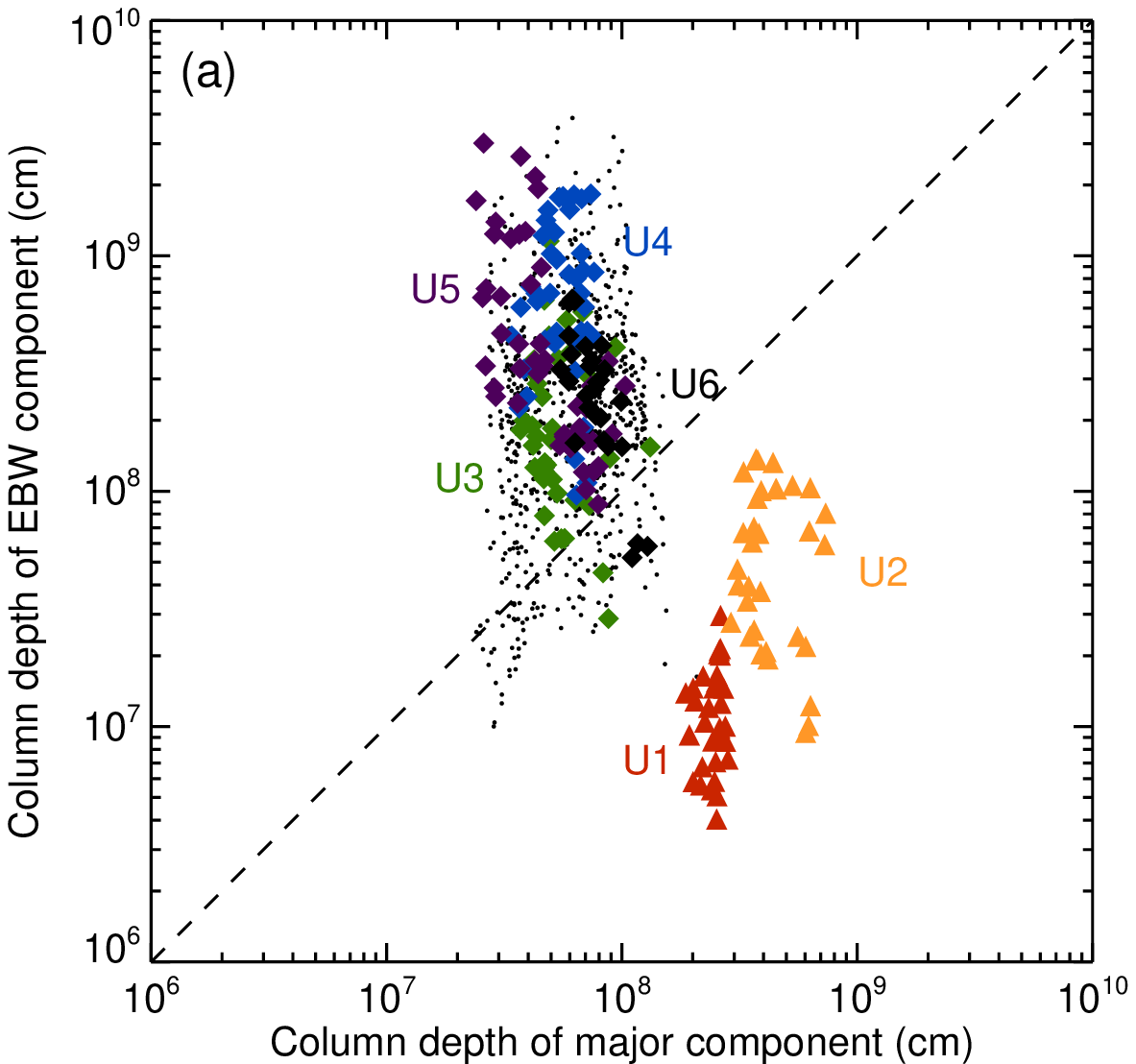}
  \end{minipage}
  \begin{minipage}[c]{8cm}
    \includegraphics[width=8cm,clip]{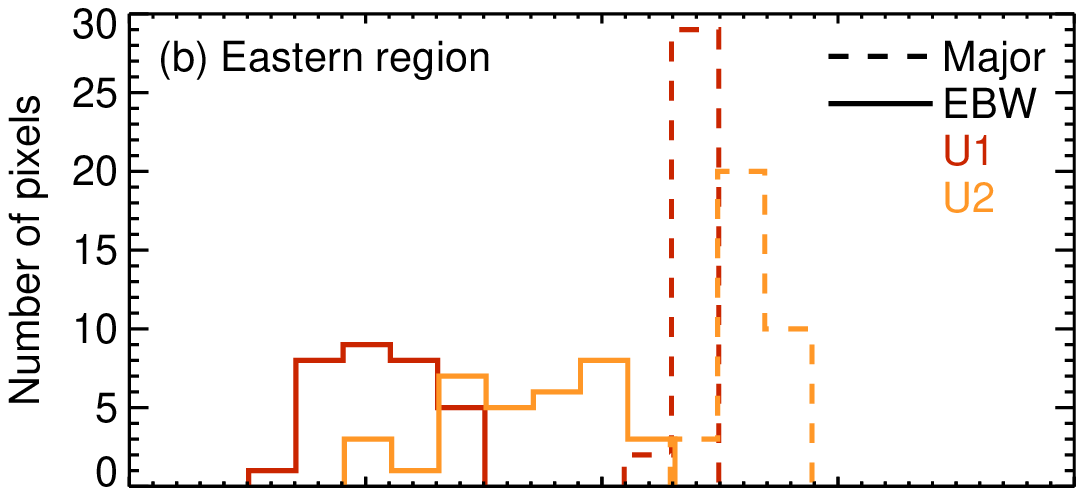}
    \includegraphics[width=8cm,clip]{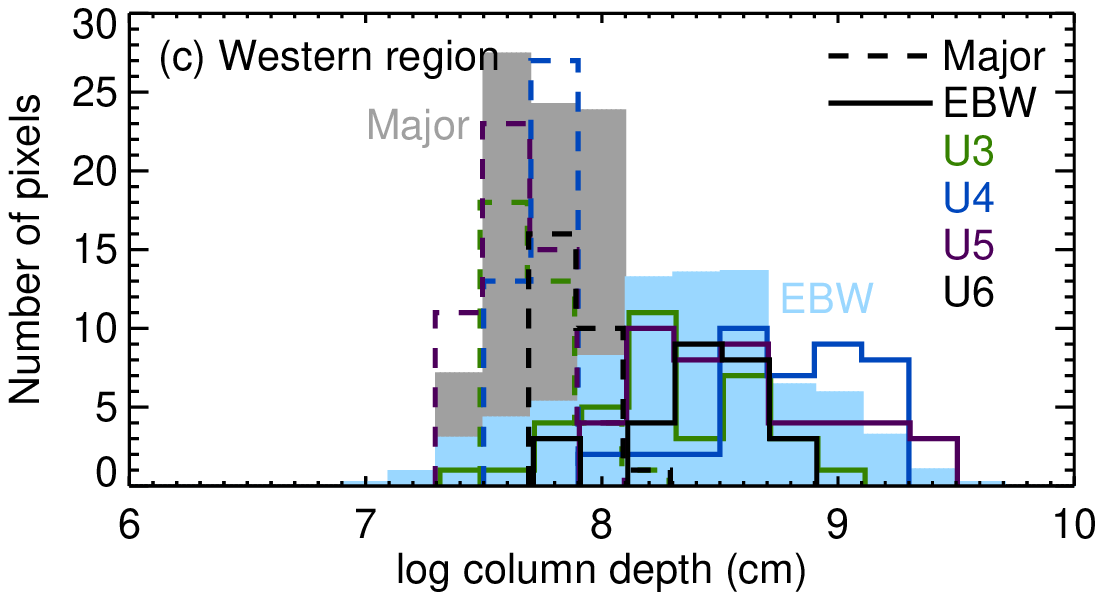}
  \end{minipage}
  \caption{%
    (a) Scatter plot for column depth of the major component vs.\ that of EBW component.  \textit{Colors} show the selected region indicated by \textit{white} boxes in Figure \ref{fig:fexiv_1G_map}.  \textit{Triangles} (\textit{Diamonds}) represent the data points in the eastern (western) outflow regions.  Numbers beside data points correspond to the names of the \textit{white} boxes.  \textit{Black dots} show the column depth for the entire western region indicated by a \textit{white dashed} box in Figure \ref{fig:dens_map}.  The \textit{dashed} line indicates the point where two quantities equal each other.  (b) Histograms for column depth of the major component (\textit{dotted}) and EBW component (\textit{solid}) in the eastern outflow region.  (c) Histograms for column depth of the major component (\textit{dotted}) and EBW component (\textit{solid}) in the western outflow region.  The \textit{gray} (the major component) and \textit{turquoise} (EBW component) histograms in the background are made for the entire western outflow region.  Those two histograms are multiplied by $0.1$.
  }
  \label{fig:column_depth}
\end{figure}

Using the obtained electron density for each component in Fe \textsc{xiv} line profiles, the column depth of each component can be calculated.  We use the equation for the column depth including the filling factor, 
\begin{equation}
  h^{\ast} = hf = \dfrac{I}{n_{\mathrm{e}}^2 G (n_{\mathrm{e}}, \, T)} \, \text{,}
  \label{eq:column_depth}
\end{equation}
where $f$ is the filling factor, $I$ is the intensity of an emission line, $n_{\mathrm{e}}$ is the electron density, and $G (n_{\mathrm{e}}, \, T)$ is the contribution function of an emission line.  The quantity $h^{\ast}$ physically represents the plasma volume per unit area along the line of sight.  Here the temperature substituted to Equation (\ref{eq:column_depth}) was simply assumed to take a single value $T_{\mathrm{f}}$ at which the contribution function $G (n_{\mathrm{e}}, \, T)$ becomes maximum ($\log T_{\mathrm{f}} \, [\mathrm{K}]=6.30$ for the Fe \textsc{xiv} lines used here).  Panel (a) in Figure \ref{fig:column_depth} shows a scatter plot for the column depth of the major component ($h_{\mathrm{Major}}$) and that of EBW component ($h_{\mathrm{EBW}}$).  Colored symbols respectively indicate the studied regions (U1--U2 for the eastern outflow region, and U3--U6 for the western outflow region).  Similar to the result for the electron density, the eastern and western outflow regions exhibit different characteristics: $h_{\mathrm{Major}} \ge h_{\mathrm{EBW}}$ in the eastern region, and $h_{\mathrm{Major}} \le h_{\mathrm{EBW}}$ in the western region.  Panels (b) and (c) display the same data in the form of histograms for the eastern and western outflow region respectively.  The \textit{gray} and \textit{turquoise} histograms in the background of panel (c) show the results for the entire western outflow region indicated by a white dashed box in Figure \ref{fig:dens_map}.  Table \ref{tab:dns_cmpl} shows the column depths averaged in each studied region.  

The result $h_{\mathrm{Major}} \le h_{\mathrm{EBW}}$ in the western outflow regions (U3--U6) means that the upflow dominates over the 
major 
component in terms of the volume, opposite to the composition ratio of emission line profile itself.  The value of $h_{\mathrm{EBW}} \simeq 10^{8.0\text{--}9.0} \, \mathrm{cm}$ can be understood by considering that the inclination of the magnetic field lines in the western outflow region was $30^{\circ} \text{--} 50^{\circ}$ 
(given the potential field calculation) 
and the horizontal spatial scale of the region was the order of $10'' (\sim 10^{9} \, \mathrm{cm})$, which leads to the vertical height of nearly the same amount.  On the other hand, it is clearly indicated that $h_{\mathrm{EBW}}$ is smaller than $h_{\mathrm{Major}}$ by up to one order of magnitude in the eastern outflow region (U1--U2).  This means that the upflows possess only a small fraction compared to the 
plasma 
characterized by the major component in line profiles.  
The Doppler velocities, derived electron densities, and the column depths for the studied outflow regions are listed in Table \ref{tab:dns_cmpl}.

Note that in the line profile analysis, we assumed that the electron density corresponding to the temperature of Si \textsc{vii} (\textit{i.e.}, the transition region; hereafter $n_{\mathrm{Si \textsc{vii}}}$) was $10^{9} \, \mathrm{cm}^{-3}$.  We discuss this assumption and its influence on our results in Appendix \ref{sect:dns_results_sivii}.  

\input{tab_dns_cmpl.tex}


%% file: tab_dns_cmpl.tex
\begin{table}[!t]
  \centering
  \caption{Doppler velocities, electron densities, and column depths of EBW component and the major component derived through the double-Gaussian fitting applied to Fe \textsc{xiv} $264${\AA}/$274${\AA}.  Note that the Doppler velocities listed in the table are calculated by using limb spectra observed independently on 2007 December 6 as a reference of zero velocity, which leads to errors up to $10 \, \kmpers$ at most mainly originating in the absolute wavelength calibration.}
  \begin{tabular}{lrrrrrr}
    \toprule
    & \multicolumn{3}{c}{EBW component} 
    & \multicolumn{3}{c}{The major component} 
    \\
    \cmidrule(lr){2-4}
    \cmidrule(lr){5-7}
    \rule[-2pt]{0pt}{13pt}
    & \multicolumn{1}{c}{$v_{\mathrm{Dop}} \, (\kmpers)$} 
    & \multicolumn{1}{c}{$\log n_{\mathrm{e}} \, [\mathrm{cm}^{-3}]$}
    & \multicolumn{1}{c}{$\log h \, [\mathrm{cm}]$}
    & \multicolumn{1}{c}{$v_{\mathrm{Dop}} \, (\kmpers)$} 
    & \multicolumn{1}{c}{$\log n_{\mathrm{e}} \, [\mathrm{cm}^{-3}]$}
    & \multicolumn{1}{c}{$\log h \, [\mathrm{cm}]$}
    \\
    \midrule
    \multicolumn{2}{l}{Eastern outflow region} & \rule[-2pt]{0pt}{15pt} & & & & \\
    U1     & $-92.4 \pm \mspace{9mu} 2.4$ & $9.17 \pm 0.09$ & $7.03 \pm 0.22$ & $-4.7 \pm 0.9$ & $9.10 \pm 0.04$ & $8.38 \pm 0.05$ \\
    U2     & $-84.8 \pm 21.4$ & $8.95 \pm 0.09$ & $7.67 \pm 0.34$ & $-3.6 \pm 1.7$ & $8.93 \pm 0.09$ & $8.64 \pm 0.12$ \\
    \midrule
    Ave.{} & $-88.8 \pm 15.2$ & $9.06 \pm 0.14$ & $7.36 \pm 0.43$ & $-4.2 \pm 1.4$ & $9.01 \pm 0.11$ & $8.51 \pm 0.16$ \\
    \multicolumn{2}{l}{Western outflow region} & \rule[-2pt]{0pt}{18pt} & & & & \\
    U3     & $-61.4 \pm 15.7$ & $8.79 \pm 0.21$ & $8.25 \pm 0.35$ & $-0.6 \pm 2.6$ & $9.31 \pm 0.09$ & $7.74 \pm 0.12$ \\
    U4     & $-56.3 \pm 15.2$ & $8.53 \pm 0.17$ & $8.80 \pm 0.34$ & $ 3.4 \pm 2.5$ & $9.34 \pm 0.10$ & $7.74 \pm 0.10$ \\
    U5     & $-73.2 \pm 12.2$ & $8.48 \pm 0.21$ & $8.59 \pm 0.41$ & $-1.3 \pm 1.0$ & $9.11 \pm 0.07$ & $7.67 \pm 0.18$ \\
    U6     & $-54.3 \pm 13.3$ & $8.64 \pm 0.12$ & $8.40 \pm 0.29$ & $-0.8 \pm 1.1$ & $9.10 \pm 0.05$ & $7.89 \pm 0.10$ \\
    \midrule
    Ave.{} & $-62.0 \pm 16.0$ & $8.60 \pm 0.22$ & $8.53 \pm 0.41$ & $ 0.1 \pm 2.7$ & $9.22 \pm 0.14$ & $7.74 \pm 0.15$ \\
    \bottomrule
  \end{tabular}
  \label{tab:dns_cmpl}
\end{table}


%% file: ndv_itdn.tex

We modeled the spectra by the composition of two Gaussians in the above analysis.  However, it is difficult to prove whether or not this assumption is suitable for the outflow regions.  There are two alternative approaches to dealing with such a spectrum consisting of more than two Gaussians.  One way is to adopt multiple-Gaussian functions (more than two components) and resolve multiple flows existing in a emission line.  The more free parameters we use, the spectra would be fitted with less $\chi^2$.  But this does not mean that we extracted a great deal of useful physical information from the spectra.  The number of local minima increase with the complexity of the fitting model, and the fitting process becomes an ill-posed problem.  

The other way is our new type of plot without assuming any fitting model.  Each spectral bin in a spectrum pair is used to derive electron density at each bin, which we refer to as ``$\lambda$-$n_{\mathrm{e}}$ diagram'' hereafter.  With this method, we measured the electron density of the plasma which have the speed of $v_{\mathrm{Dop}}=c \, (\lambda - \lambda_0)/\lambda_0$ ($\lambda_0$: rest wavelength), which is a function of wavelength.  Consider a density-sensitive pair of spectra $\phi_1 (\lambda)$ and $\phi_2 (\lambda)$ emitted from the same degree of an ion.  These emission lines must have the same Doppler velocity because they came from the same degree of the ion, so after converting the variable $\lambda$ into Doppler velocity $v_{\mathrm{Dop}}$ as denoted by $\phi_i^* (v_{\mathrm{Dop}}) = \phi_i (\lambda)$ ($i=1,2$), we can calculate the electron density as a function of the Doppler velocity
\begin{equation}
  n_{\mathrm{e}}^{*} (v_{\mathrm{Dop}}) = 
    R^{-1}
    \left[
      \frac{\phi_2^* (v_{\mathrm{Dop}})}{\phi_1^* (v_{\mathrm{Dop}})} 
    \right] \, \text{.}
    \label{eq:ndv_rprsnt}
\end{equation}
The derived $n_{\mathrm{e}}^{*} (v_{\mathrm{Dop}})$ can be converted into a function of wavelength in either spectrum, $n_{\mathrm{e}} (\lambda)$, by the Doppler effect equation.  Function $R (n_{\mathrm{e}})$ is the ratio of intensities from two emission lines which is a function of electron density, so when we know the intensities of two emission lines which are represented as
\begin{align}
  I_1 = \int \phi_1 (\lambda) d \lambda \, \text{,} \\
  I_2 = \int \phi_2 (\lambda) d \lambda \, \text{,} 
\end{align}
electron densities can be usually derived by
\begin{equation}
  n_{\mathrm{e}} = 
    R^{-1}
    \left( 
      \frac{I_2}{I_1} 
    \right) \, \text{.}
\end{equation}
Note that we used the same curve as shown in panel (a) in Fig.~\ref{fig:dns_itdn_rat_chianti} for the function $R (n_{\mathrm{e}})$.  This assumes that $R (n_{\mathrm{e}})$ is the same for all wavelengths in the range of interest, which we have not investigated in detail. 

As shown in the above equations, $\lambda$-$n_{\mathrm{e}}$ diagram represents that of the particles which move with that speed, in other words, we do not obtain the electron density of the whole plasma as an ensemble of Maxwellian distribution.  We emphasize that the advantage of our method using Equation (\ref{eq:ndv_rprsnt}) is that even if we do not know the precise functional form of spectra, it gives us the electron density as a function of Doppler velocity without any modeling.  


%% file: ndv_method.tex

Making $\lambda$-$n_\mathrm{e}$ diagram contains the following processes: (1) subtraction of blending emission line, (2) adjusting wavelength scale of Fe \textsc{XIV} $264.78${\AA} to $274.20${\AA} by interpolation, and (3) density inversion at each spectral pixel. Since the blend of an emission line Si \textsc{vii} $274.18${\AA} into Fe \textsc{xiv} $274.20${\AA} was already described in Section \ref{sec:de-blend}, here we explain only processes (2) and (3). 


%% file: ndv_method_2.tex

Since the EIS instrument does not have absolute wavelength scale, the corresponding wavelength location of the same velocity in Fe \textsc{xiv} $264.78${\AA} and $274.20${\AA} must be determined from the data itself as described in 
\citet{kitagawa2013}.  
Using obtained relation $\lambda_{\mathrm{obs,}274}/\lambda_{\mathrm{obs,}264} = 1.0355657$ $(\pm 0.0000044)$, each wavelength value imposed on the spectral window of Fe \textsc{xiv} $264.78${\AA} was projected onto the values on the spectral window of Fe \textsc{xiv} $274.20${\AA} by the scaling 
\begin{equation}
  \tilde{\lambda_{i}} = \alpha \lambda_{264,i} \, \, (\alpha = 1.0355657) \, \text{,}
  \label{eq:ndv_scaling}
\end{equation}
where a number $i$ indicates the $i$th spectral pixel in a spectrum of $264.78${\AA}.

Since the wavelength values of each bin of projected Fe \textsc{xiv} $264.78${\AA} do not generally coincide with those of Fe \textsc{xiv} $274.20${\AA}, the projected spectrum was interpolated by a cubic spline in order to align two Fe \textsc{xiv} spectra in identical wavelength bins.  


%% file: ndv_method_3.tex

We can calculate the ratio of spectral intensity Fe \textsc{xiv} $264.78${\AA}/$274.20${\AA} at each spectral bin.  Now we are able to derive the electron density in the same way described in Section \ref{sec:dns_inv}.  Because intensity at each spectral bin has larger errors compared to the integrated intensity (\textit{e.g.}, double-Gaussian fitting), the estimated errors for the electron density in the $\lambda$-$n_{\mathrm{e}}$ diagram become large especially for the line wing.


%% file: ndv_test.tex

In order to test the validity of $\lambda$-$n_{\mathrm{e}}$ method, we synthesized spectra of Fe \textsc{xiv} $264.78${\AA} and $274.20${\AA} taking into account the spectral resolution of EIS and instrumental broadening.  The spectra were composed of two components which represent plasma at the rest and an upflow.  While the physical parameters for the major rest component (peak, Doppler velocity, and width) were fixed, those for a minor blueshifted component (\textit{i.e.}, upflow) were taken as variables.  We made $\lambda$-$n_{\mathrm{e}}$ diagrams for the minor component with
\vspace{-5pt}
\begin{itemize}
\item electron density of $8.50$, $8.75$, $9.00$, $9.25$, and $9.50$ 
  in the unit of $\log \, \mathrm{cm}^{-3}$, 
  \vspace{-5pt}
\item intensity of $1$, $5$, $10$, $15$, and $20 \, \mathrm{\%}$ 
  (ratio to the major component in Fe \textsc{xiv} $274.20${\AA}), 
  \vspace{-5pt}
\item Doppler velocity of $0$, $-50$, $-100$, $-150$, 
  and $-200\,\mathrm{km}\,\mathrm{s}^{-1}$, 
  \vspace{-5pt}
\item thermal width of $2.0$, $2.5$, $3.0$, $3.5$, and $4.0\,\mathrm{MK}$.
\end{itemize}
\vspace{-5pt}
The nonthermal width was not considered in this test because essentially it does not produce any differences.  
In this paper, the tests only for electron density and Doppler velocity will be given below, since the dependence on them are significant.  The other two variables (\textit{i.e.}, intensity and thermal width) do not have strong effects and are described in the author's PhD thesis \citep{kitagawa2013}.  


%% file: ndv_test_density.tex

The most important point on $\lambda$-$n_{\mathrm{e}}$ diagram is whether it reflects the electron density of the components which compose spectrum properly or not.  In order to test that, we synthesized the spectra which are composed of the major component at the rest which has the fixed electron density of $\log n_{\mathrm{e}} [\mathrm{cm}^{-3}]=9.0$ and the minor component which has a variable electron density. Five cases ($\log n_{\mathrm{e}} \, [\mathrm{cm}^{-3}] = 8.50$, $8.75$, $9.00$, $9.25$, and $9.50$) were analyzed, where the peak ratio of the minor/major component was $15\mathrm{\%}$ with fixed upflow speed $v=-100 \, \mathrm{km} \, \mathrm{s}^{-1}$.  Panels (a) and (b) of Figure \ref{fig:test_density} show the spectra of Fe \textsc{xiv} $264.78${\AA} and $274.20${\AA} respectively.  Colors (\textit{blue}, \textit{turquoise}, \textit{yellow}, \textit{green}, and \textit{red}) indicate the five cases calculated here.  After converting the wavelength scale of $264.78${\AA} to $274.20${\AA}, $\lambda$-$n_{\mathrm{e}}$ were obtained as shown in panel (c) of Figure \ref{fig:test_density}.  The triangles in panel (c) indicate centroid and electron density of the given minor component.  It is clear that those $\lambda$-$n_{\mathrm{e}}$ diagrams clearly reflect the changes of the electron density from $\log n_{\mathrm{e}} \, [\mathrm{cm}^{-3}] = 8.50\text{--}9.50$.  Despite the spectra being composed of only two components, $\lambda$-$n_{\mathrm{e}}$ diagrams do not become a step function but a smooth function.  This is natural because the two Gaussians in the spectra contribute to each other by their overlapping wings.  We claim that the method proposed here ($\lambda$-$n_{\mathrm{e}}$) is a good indicator of the electron density of components in the spectrum. 


%% file: ndv_test_velocity.tex

\begin{figure}[!t]
  \centering
  \begin{minipage}[c]{5.0cm}
    \includegraphics[width=5.0cm,clip]{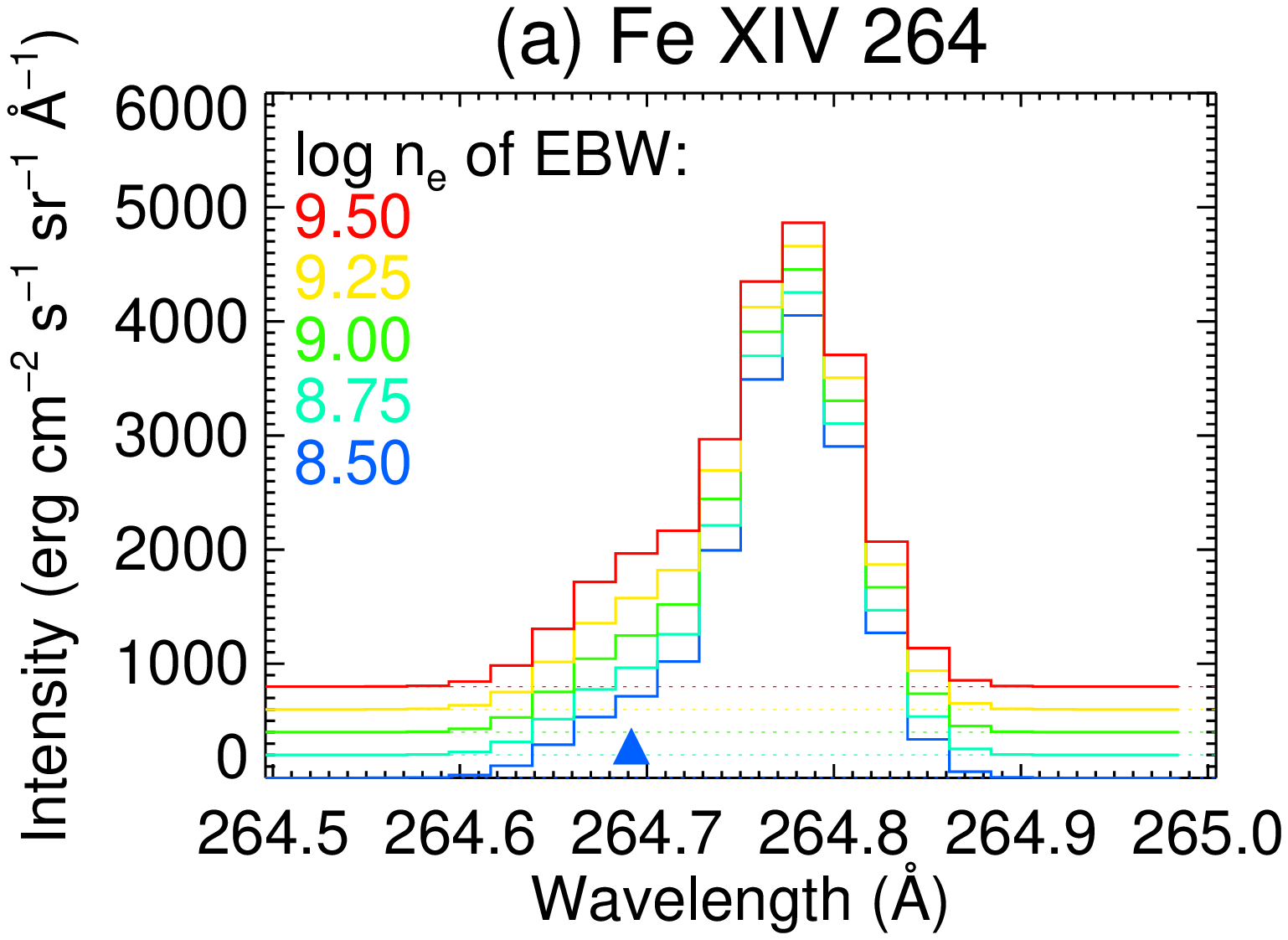}
    \includegraphics[width=5.0cm,clip]{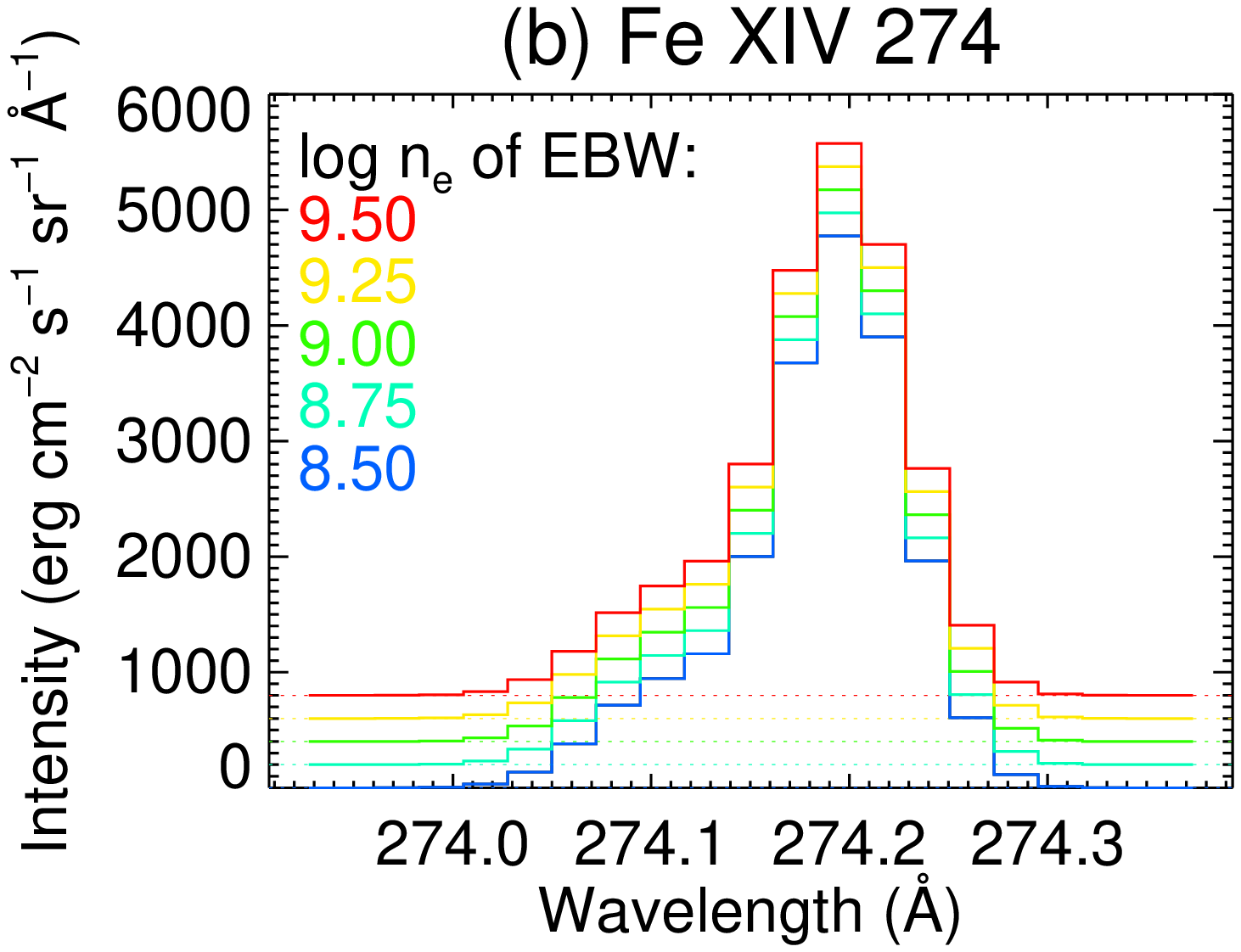}
  \end{minipage}
  \begin{minipage}[c]{8.5cm}
    \includegraphics[width=8.5cm,clip]{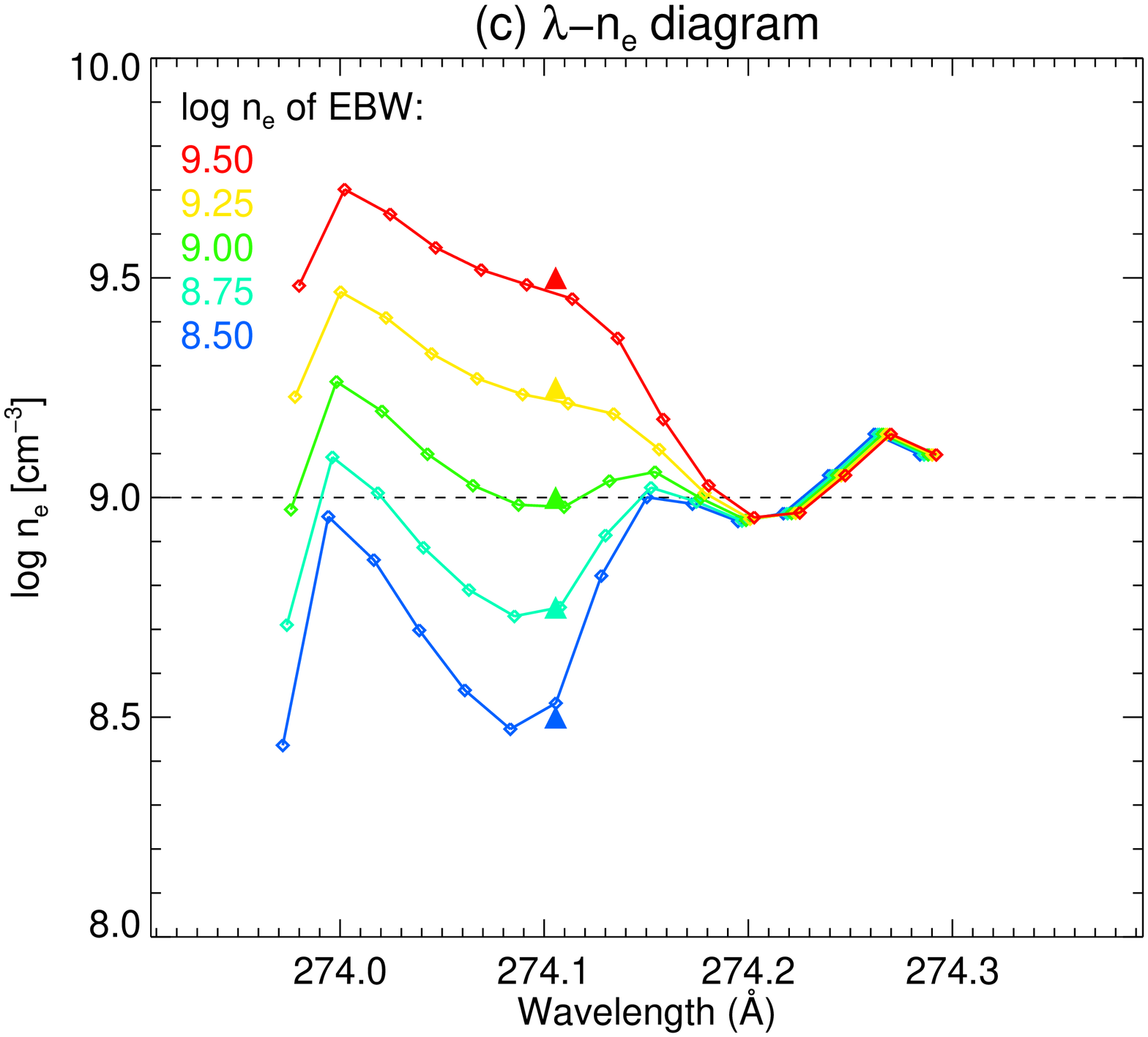}
  \end{minipage}
  \begin{minipage}[c]{5.0cm}
    \includegraphics[width=5.0cm,clip]{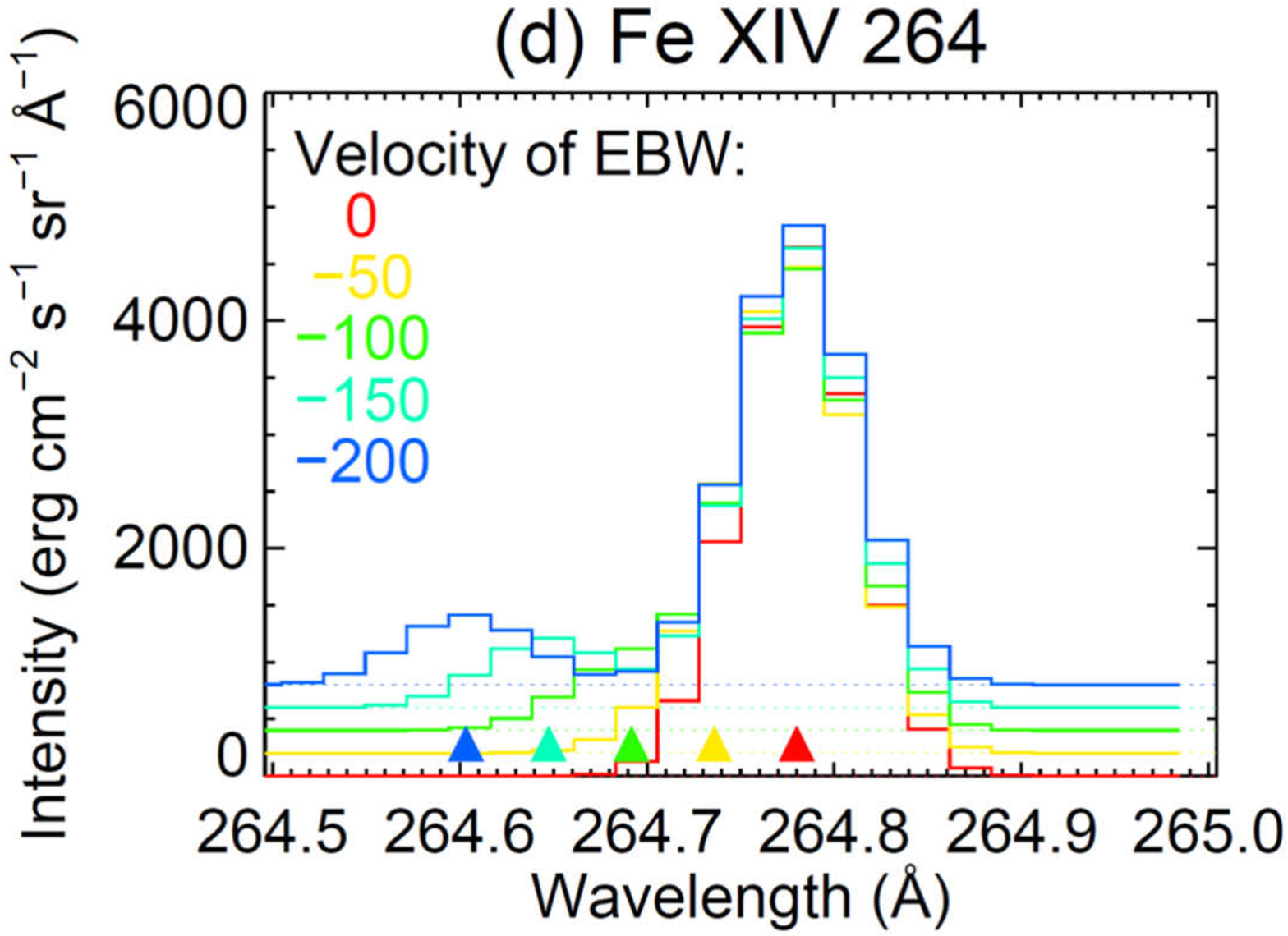}
    \includegraphics[width=5.0cm,clip]{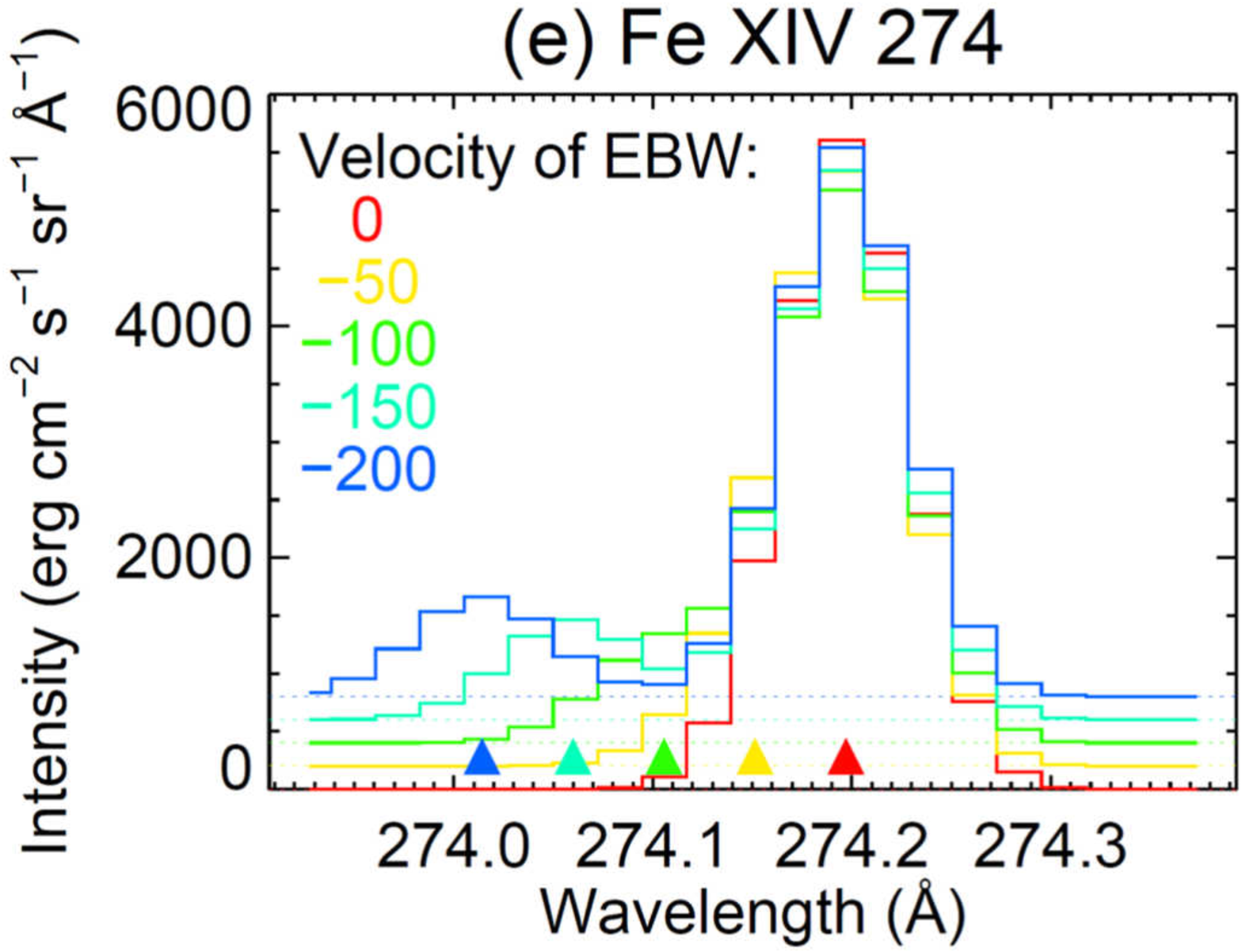}
  \end{minipage}
  \begin{minipage}[c]{8.5cm}
    \includegraphics[width=8.5cm,clip]{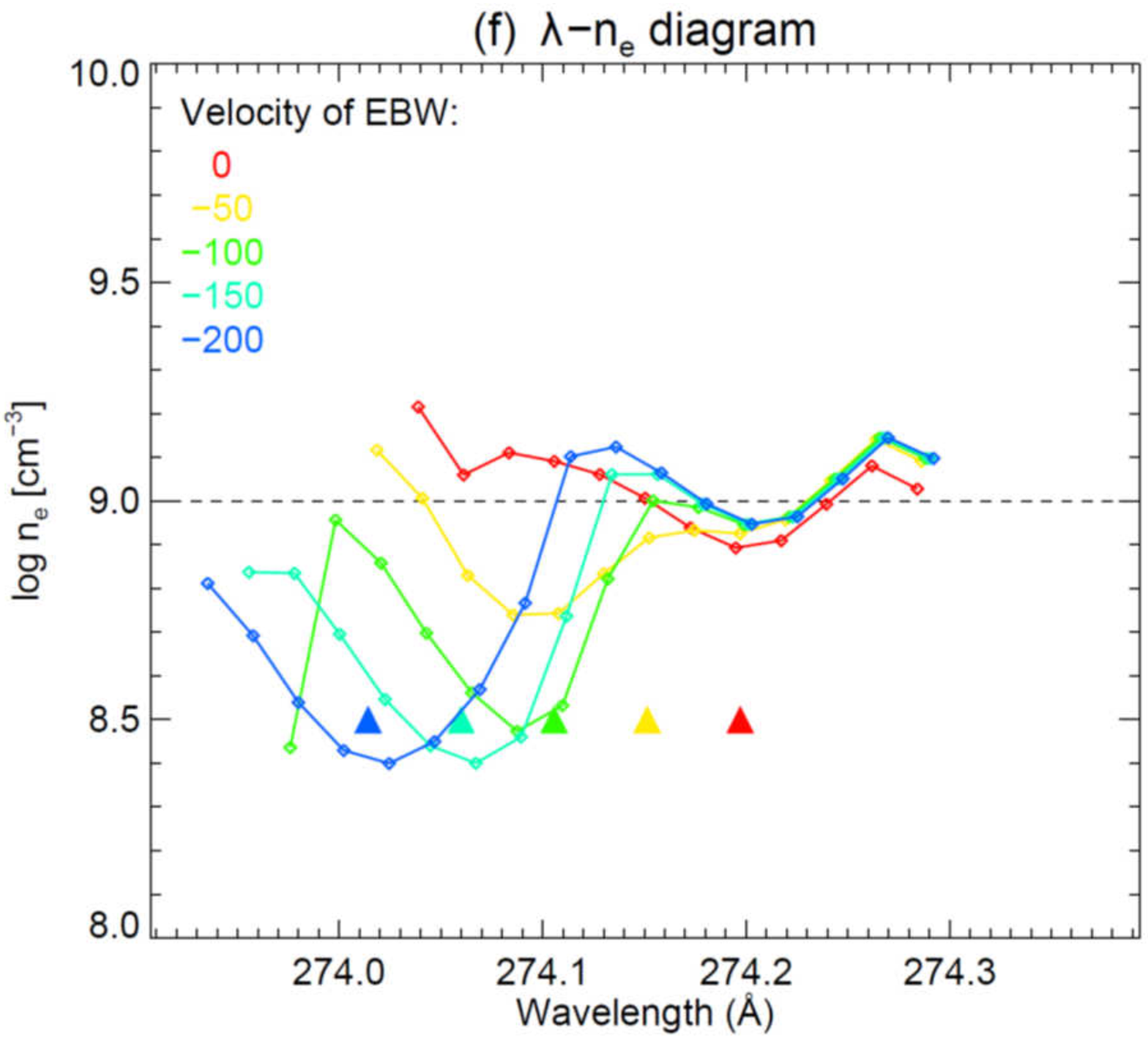}
  \end{minipage}
  \caption{
    \textit{Upper panels:}  
    (a) synthetic line profiles of Fe \textsc{xiv} $264.78${\AA},
    (b) those of Fe \textsc{xiv} $274.20${\AA}, and 
    (c) $\lambda$-$n_\mathrm{e}$ diagrams.  
    Each color indicates different electron density of the minor blueshifted component (\textit{blue}: $8.50$, \textit{turquoise}: $8.75$, \textit{green}: $9.00$, \textit{yellow}: $9.25$, and \textit{red}: $9.25$ in the unit of $\log \, \mathrm{cm}^{-3}$). Electron density of the major component was fixed to $\log n_{\mathrm{e}} \, [\mathrm{cm}^{-3}] = 9.00$. The triangles in panel (c) indicate centroid and electron density of the given minor component. 
    \textit{Lower panels:}  
    (d) synthetic line profiles of Fe \textsc{xiv} $264.78${\AA},
    (e) those of Fe \textsc{xiv} $274.20${\AA}, and 
    (f) $\lambda$-$n_\mathrm{e}$ diagrams.
    Each color indicates different velocity of the minor blueshifted component (\textit{blue}: $0\,\kmpers$, \textit{turquoise}: $-50\,\kmpers$, \textit{green}: $-100\,\kmpers$, \textit{yellow}: $-150\,\kmpers$, and \textit{red}: $-200\,\kmpers$).  The major rest component was at rest ($0\,\kmpers$).
  }
  \label{fig:test_density}
\end{figure}

The dependence of $\lambda$-$n_{\mathrm{e}}$ diagram on the Doppler velocity of the minor component is obvious.  The spectra of Fe \textsc{xiv} $264.78${\AA} and $274.20${\AA}, and $\lambda$-$n_{\mathrm{e}}$ diagrams are shown in panels (d), (e), and (f) respectively of Figure \ref{fig:test_density}.  Colors indicate the five cases for variable Doppler velocity calculated (\textit{blue}: $0\,\kmpers$, \textit{turquoise}: $-50\,\kmpers$, \textit{green}: $-100\,\kmpers$, \textit{yellow}: $-150\,\kmpers$, and \textit{red}: $-200\,\kmpers$).  Major rest component was at rest ($0\,\kmpers$) with the electron density of $\log n_{\mathrm{e}} \, [\mathrm{cm}^{-3}] = 9.0$.  The triangles in panel (f) indicate centroid and electron density of the given minor component. The relative intensity of the minor component is $15 \, \mathrm{\%}$ of that of the major component and the electron density of the minor component was set to $\log n_{\mathrm{e}} \, [\mathrm{cm}^{-3}] = 8.5$ in all five cases here.  The location of dips in $\lambda$-$n_{\mathrm{e}}$ diagram well represent the centroid position of the input minor component when two components are separated so that the spectrum is dominated by themselves near their centroids.  This is not the case for $v=-50 \, \kmpers$ (\textit{i.e.}, \textit{yellow}), where those two components are not separated so clearly.  In this case, $\lambda$-$n_{\mathrm{e}}$ diagram gradually decreases from longer to shorter wavelength.  One advantage of the method described here is that we are able to know the tendency of electron density of upflow/downflow without any fitting to the spectrum which might produce spurious results occasionally.

The tests for the four variables (\textit{i.e.}, density, intensity, velocity, and thermal width) indicate that the method proposed here ($\lambda$-$n_{\mathrm{e}}$ diagram) is a powerful diagnostic tool for coronal plasma which may be constituted of several component along the line of sight and form non-single-Gaussian line profile. In the next section, we exploit this $\lambda$-$n_{\mathrm{e}}$ diagram so that the result obtained by double-Gaussian fitting would be confirmed (\textit{i.e.}, upflows are more tenuous than the rest component). 

\clearpage


%% file: ndv_sv.tex

\begin{figure}[!t]
  \centering
  \includegraphics[width=12cm,clip]{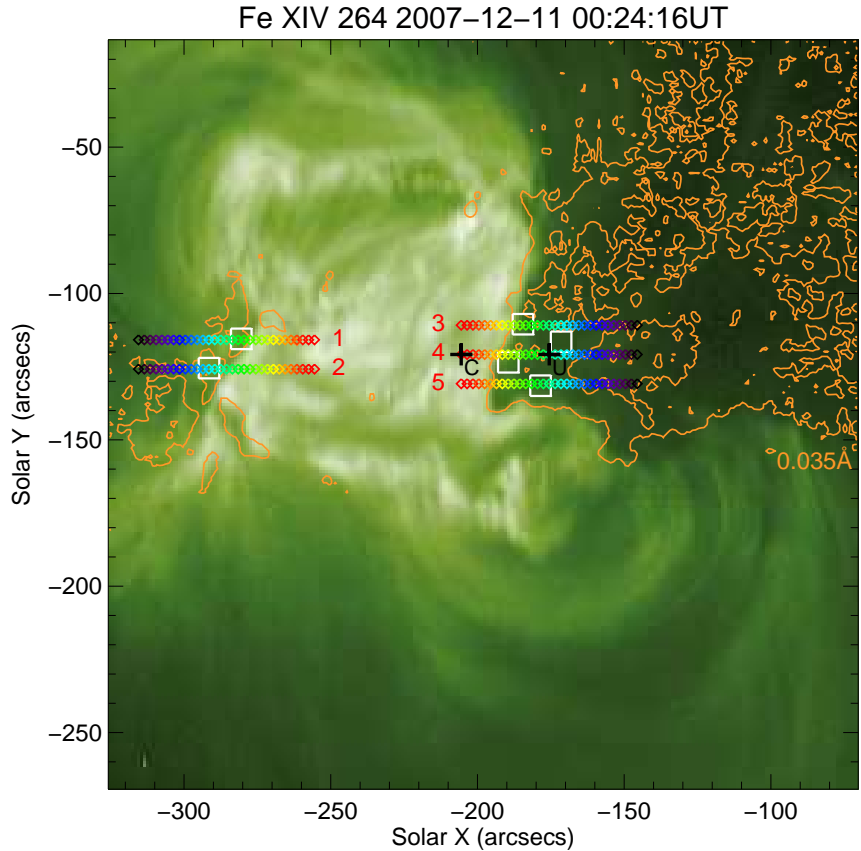}
  \caption{Intensity map of Fe \textsc{xiv} $264.78${\AA} obtained with EIS.  Five arrays of colored diamonds (\textit{red}--\textit{violet}) are the locations where $\lambda$-$n_\mathrm{e}$ diagrams were made.  The locations cut across the active region core and the outflow region.  \textit{Orange} contour indicates the line width of $0.035${\AA}.}
  \label{fig:eis_map_lp_sv}
\end{figure}

The electron density of the outflow region in AR 10978 is investigated through $\lambda$-$n_{\mathrm{e}}$ diagram here.  Figure \ref{fig:eis_map_lp_sv} shows intensity map of Fe \textsc{xiv} $264.78${\AA} obtained with EIS.  \textit{Orange} contours indicate the line width of $0.035${\AA}, which becomes an indication of the outflows.  Five horizontal arrays of colored diamonds (\textit{red}--\textit{violet}) which cut across the active region core and the outflow region are the locations where we made $\lambda$-$n_{\mathrm{e}}$ diagrams.  First, we look at the location indicated by \textit{black plus} signs named \textsf{C} (core) and \textsf{U} (outflow).  

\begin{figure}[!t]
  \centering
  \includegraphics[width=8cm,clip]{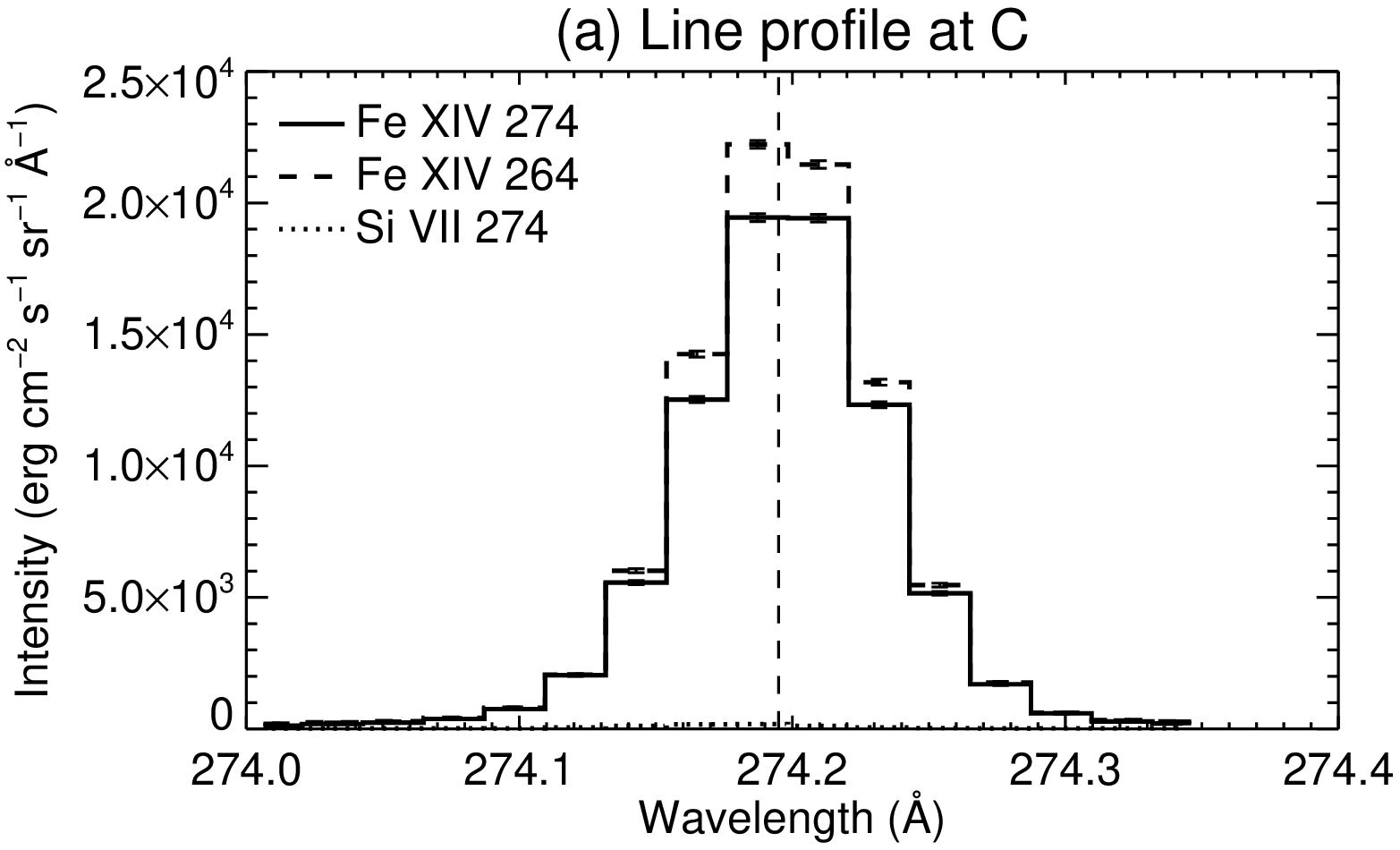}
  \includegraphics[width=8cm,clip]{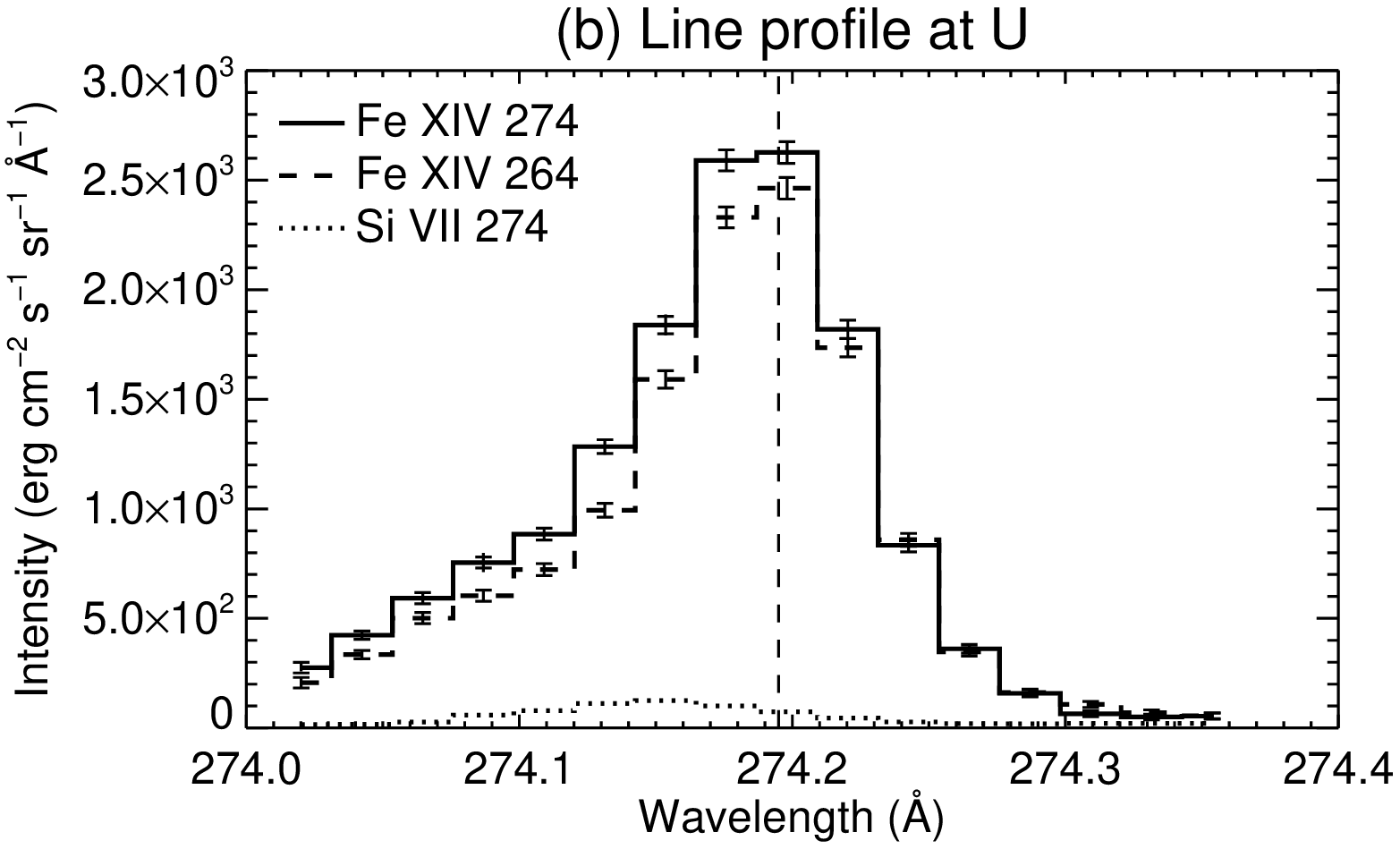}
  \includegraphics[width=8cm,clip]{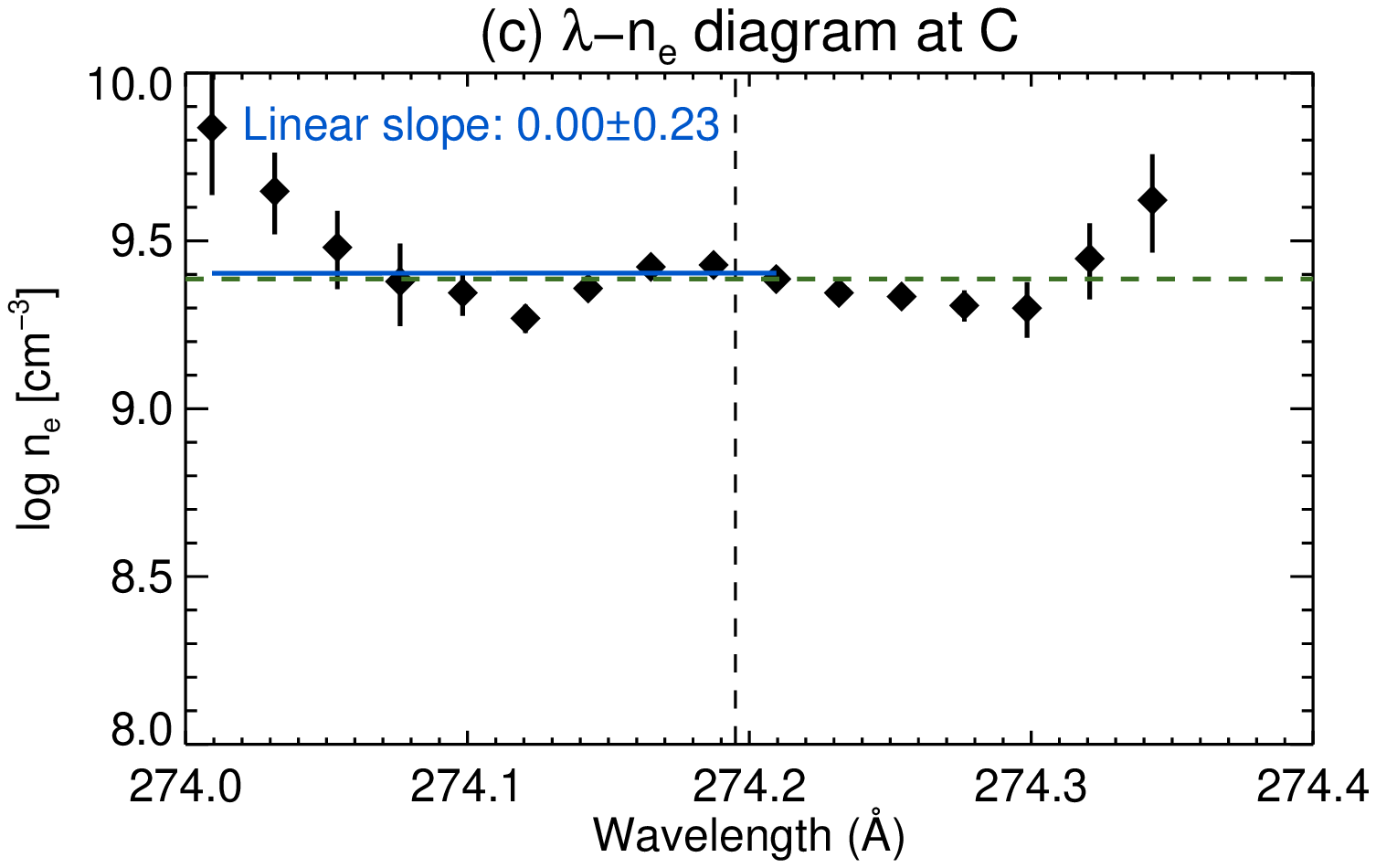}
  \includegraphics[width=8cm,clip]{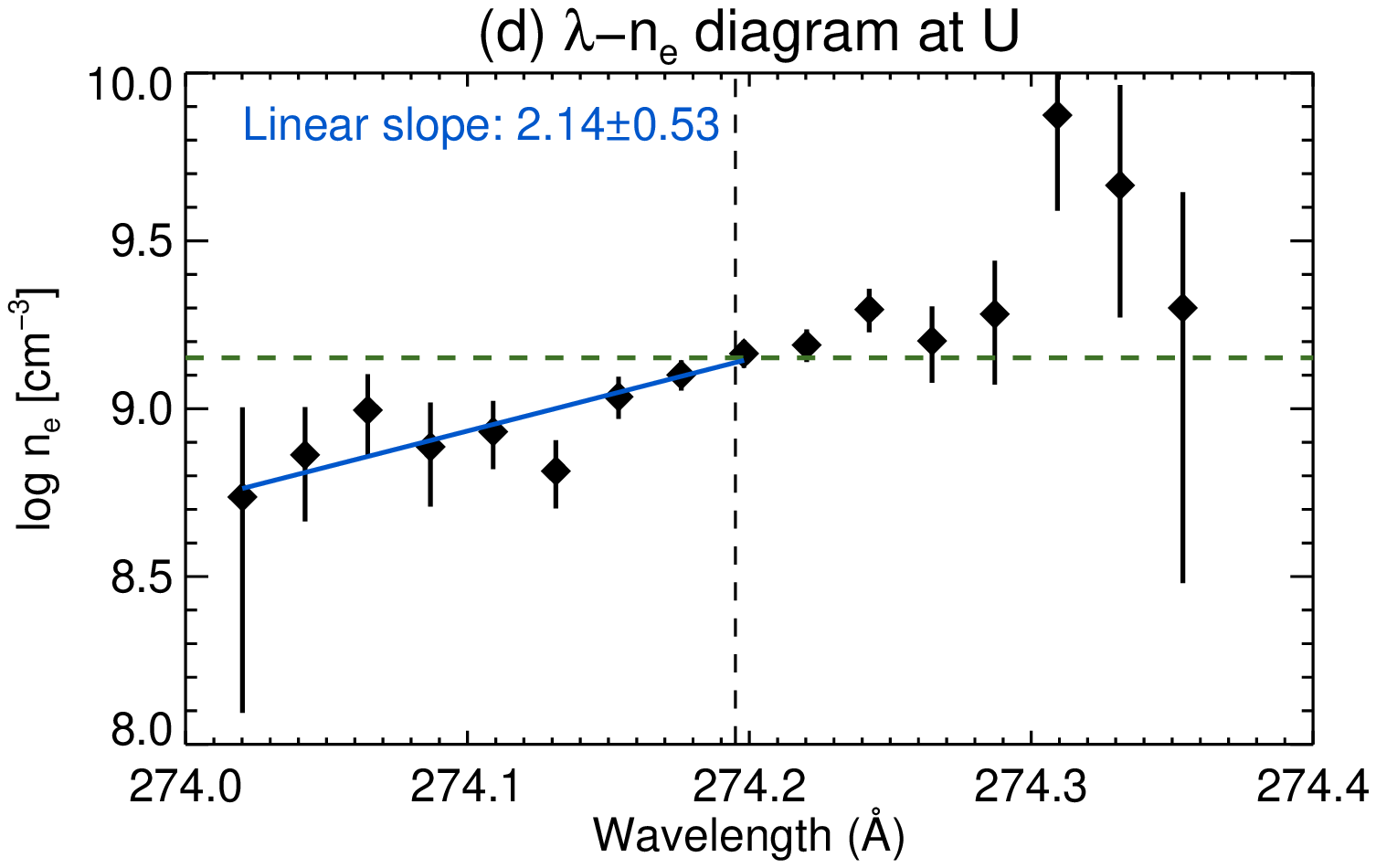}
  \caption{(a) Line profiles of Fe \textsc{xiv} $274.20${\AA} (\textit{solid})/$264.78${\AA} (\textit{dashed}; adjusted to the wavelength scale of $274.20${\AA}) and Si \textsc{vii} $274.18${\AA} (\textit{dotted}) estimated from $275.35${\AA} at the active region core.  (b) Line profiles at the outflow region.  (c) $\lambda$-$n_{\mathrm{e}}$ diagram at the active region core.  (d) $\lambda$-$n_{\mathrm{e}}$ diagram at the outflow region.}
  \label{fig:lp_and_ndv_274scale_example}
\end{figure}

In Figure \ref{fig:lp_and_ndv_274scale_example}, the line profiles of Fe \textsc{xiv} $274.20${\AA}, interpolated $264.78${\AA} and estimated Si \textsc{vii} $274.18${\AA} (see Section \ref{sec:de-blend}) are shown by \textit{solid}, \textit{dashed}, and \textit{dotted} spectrum respectively in panel (a) for the active region core and (b) for the outflow region.  We can see an enhanced blue wing in line profiles of Fe \textsc{xiv} in the outflow region.  The \textit{vertical dashed} lines indicate rough reference of the rest wavelength position $\lambda=274.195${\AA} which was the average line centroid above the limb in the 2007 December 18 data (possible error up to $0.01${\AA}). 

Panels (c) and (d) in Figure \ref{fig:lp_and_ndv_274scale_example} show the $\lambda$-$n_{\mathrm{e}}$ diagram for the active region core and the western outflow region respectively.  The \textit{horizontal green dotted} line in each plot indicates the electron density averaged in the neighboring three spectral bins which are nearest to $\lambda=274.20${\AA} (\textit{i.e.}, rest wavelength).  Those $\lambda$-$n_{\mathrm{e}}$ diagrams in the two locations exhibit a different behavior at the shorter wavelength side around $\lambda=274.00\text{--}274.20${\AA}: the diagram in the active region core is roughly constant while that in the western outflow region slightly decreases at the shorter wavelength.  The number written in the upper left corner of each plot indicates the linear slope fitted within the wavelength range $\lambda \leq 274.20${\AA}.  This implies that the electron density of the outflows (\textit{i.e.}, shorter wavelength side) is smaller than that of the major 
component
closer to the rest wavelength.  

\begin{figure}[!t]
  \includegraphics[width=8cm,clip]{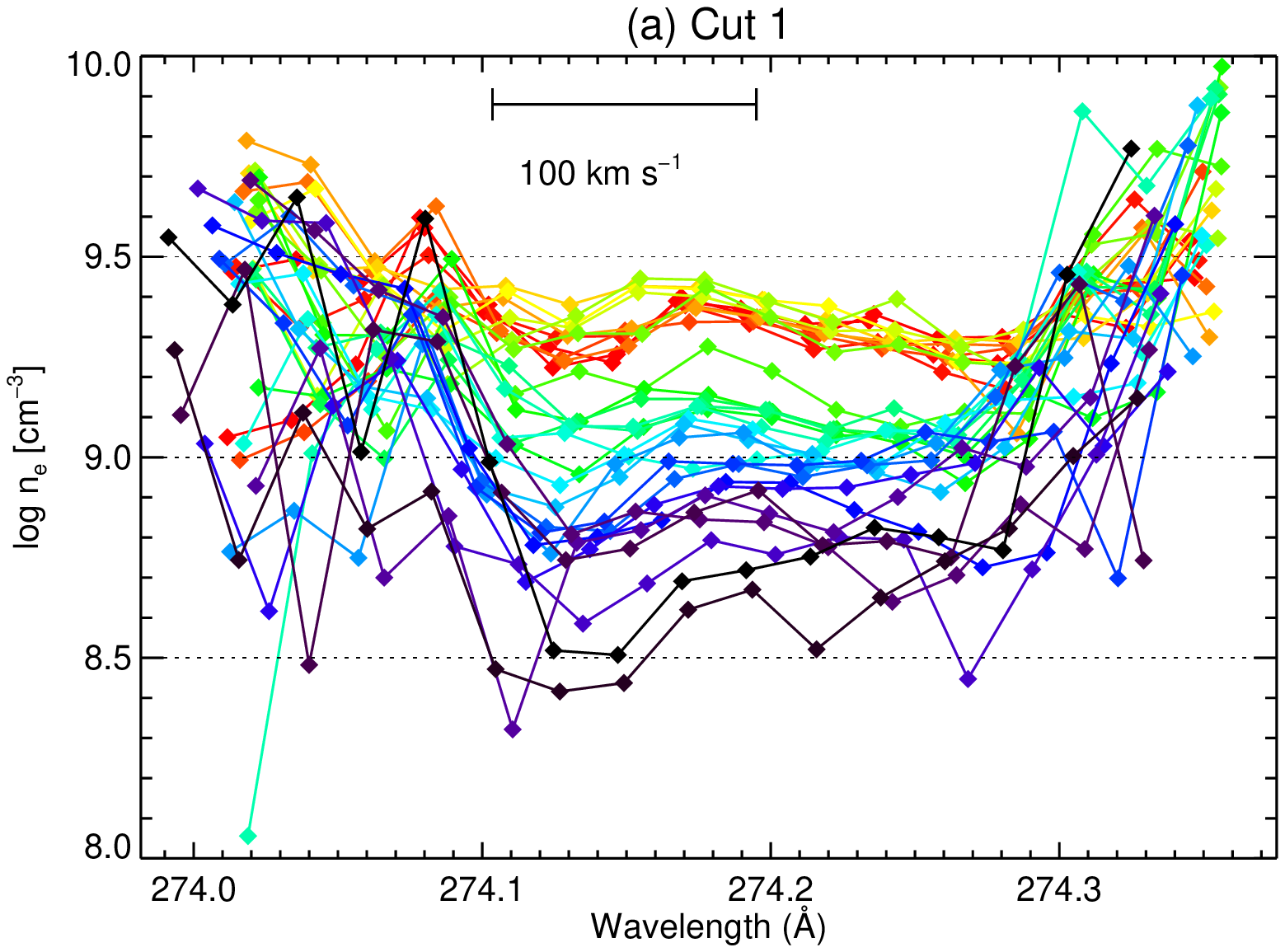}
  \includegraphics[width=8cm,clip]{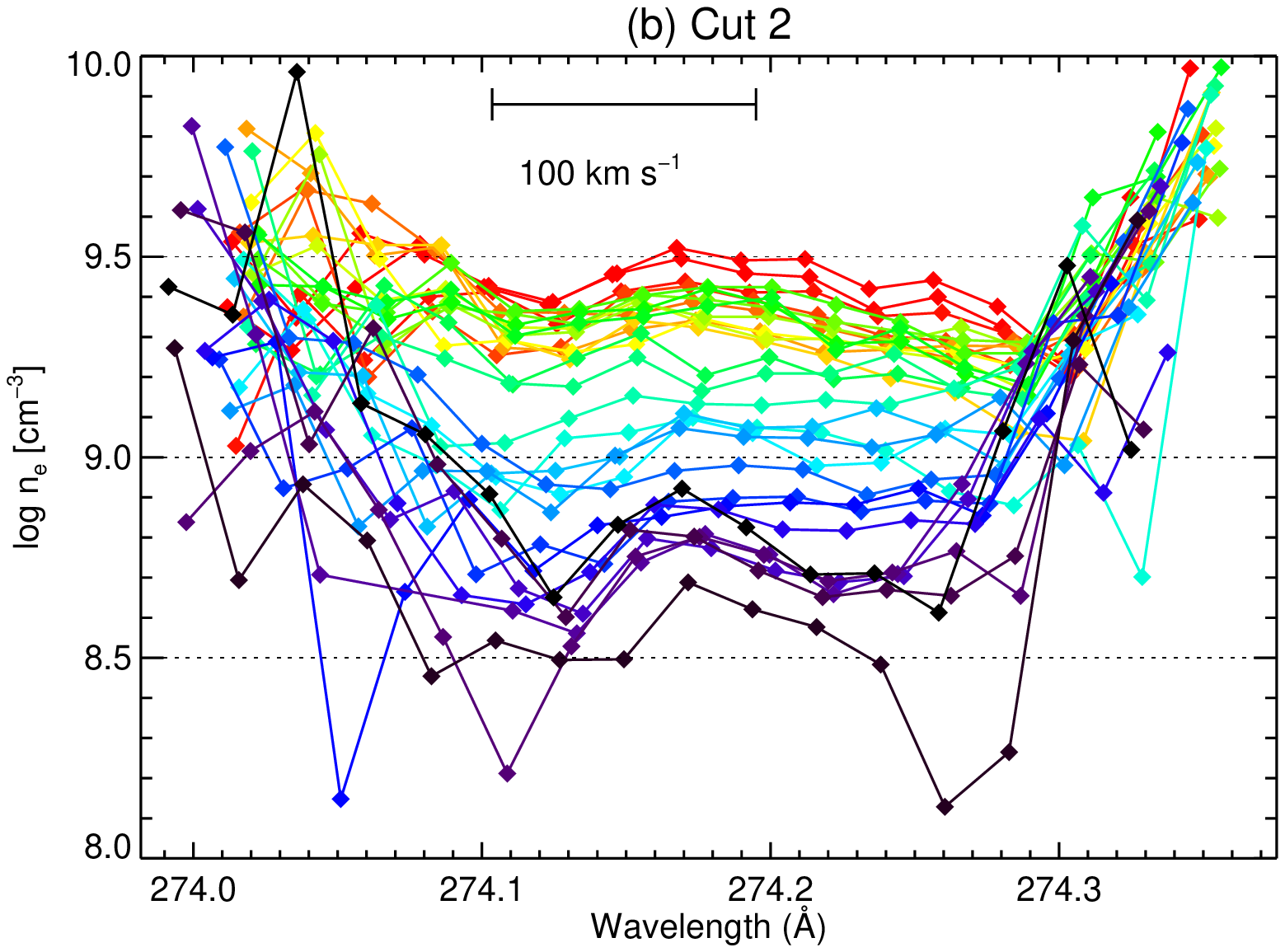}
  \includegraphics[width=8cm,clip]{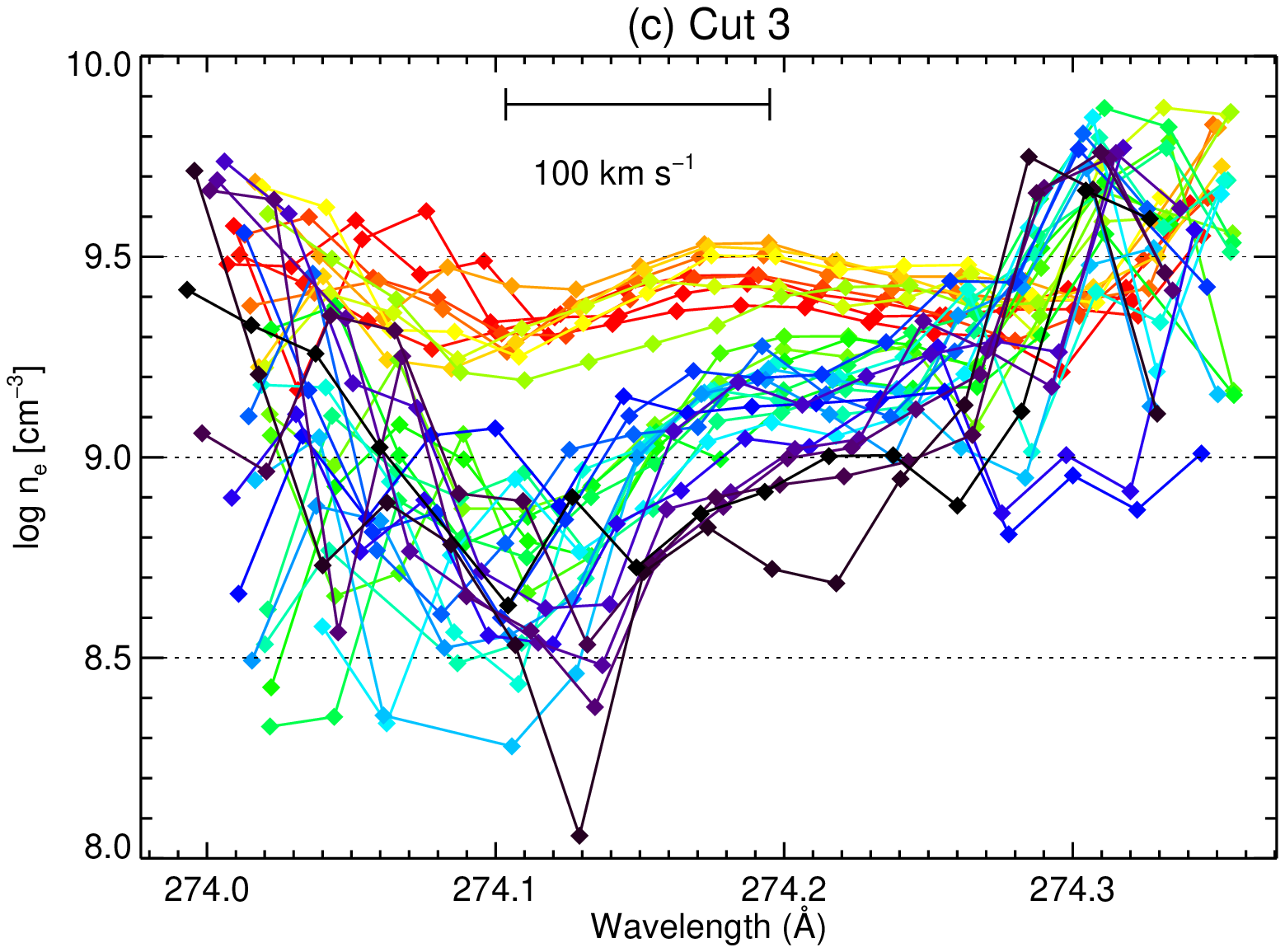}
  \includegraphics[width=8cm,clip]{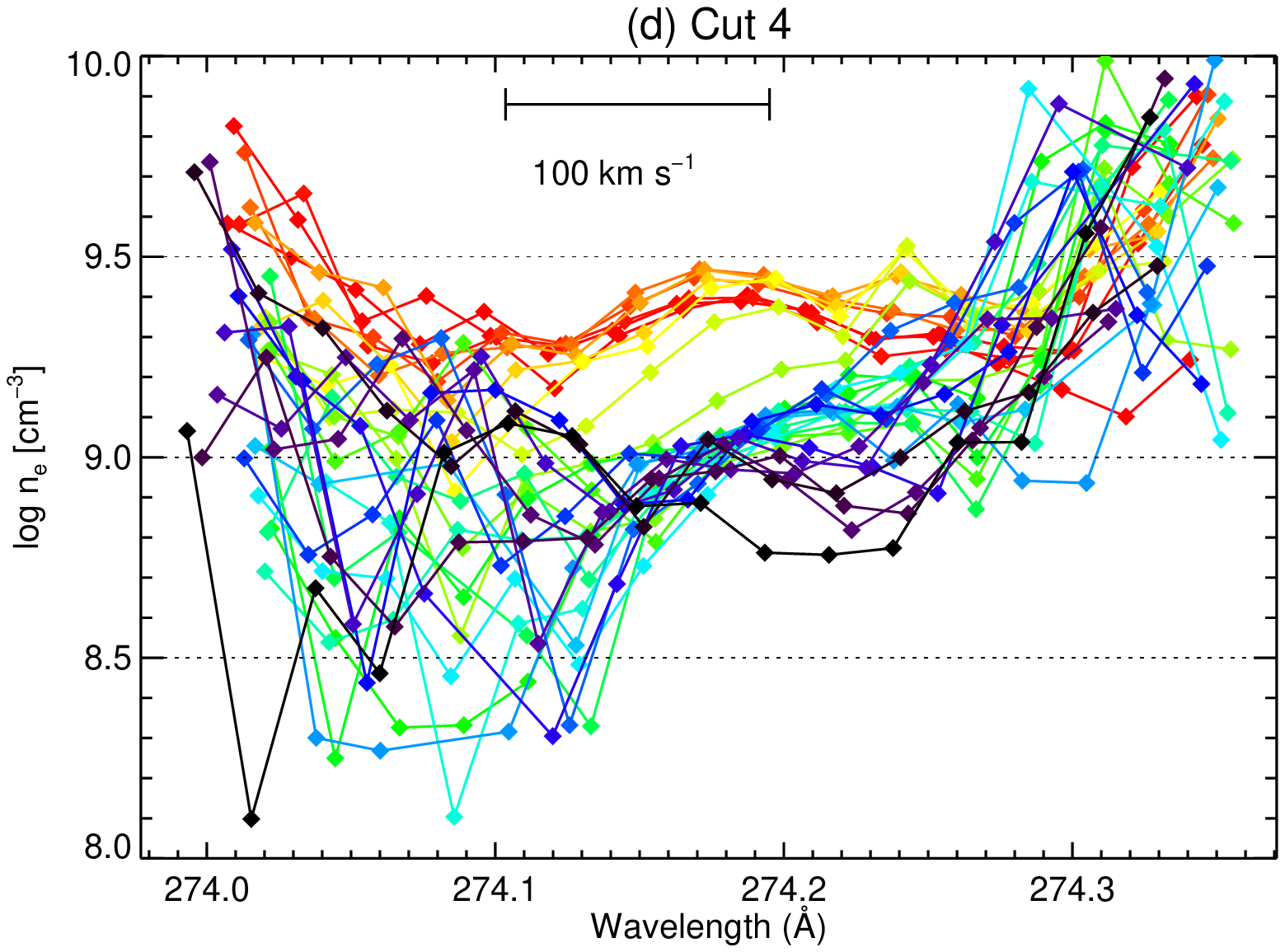}
  \includegraphics[width=8cm,clip]{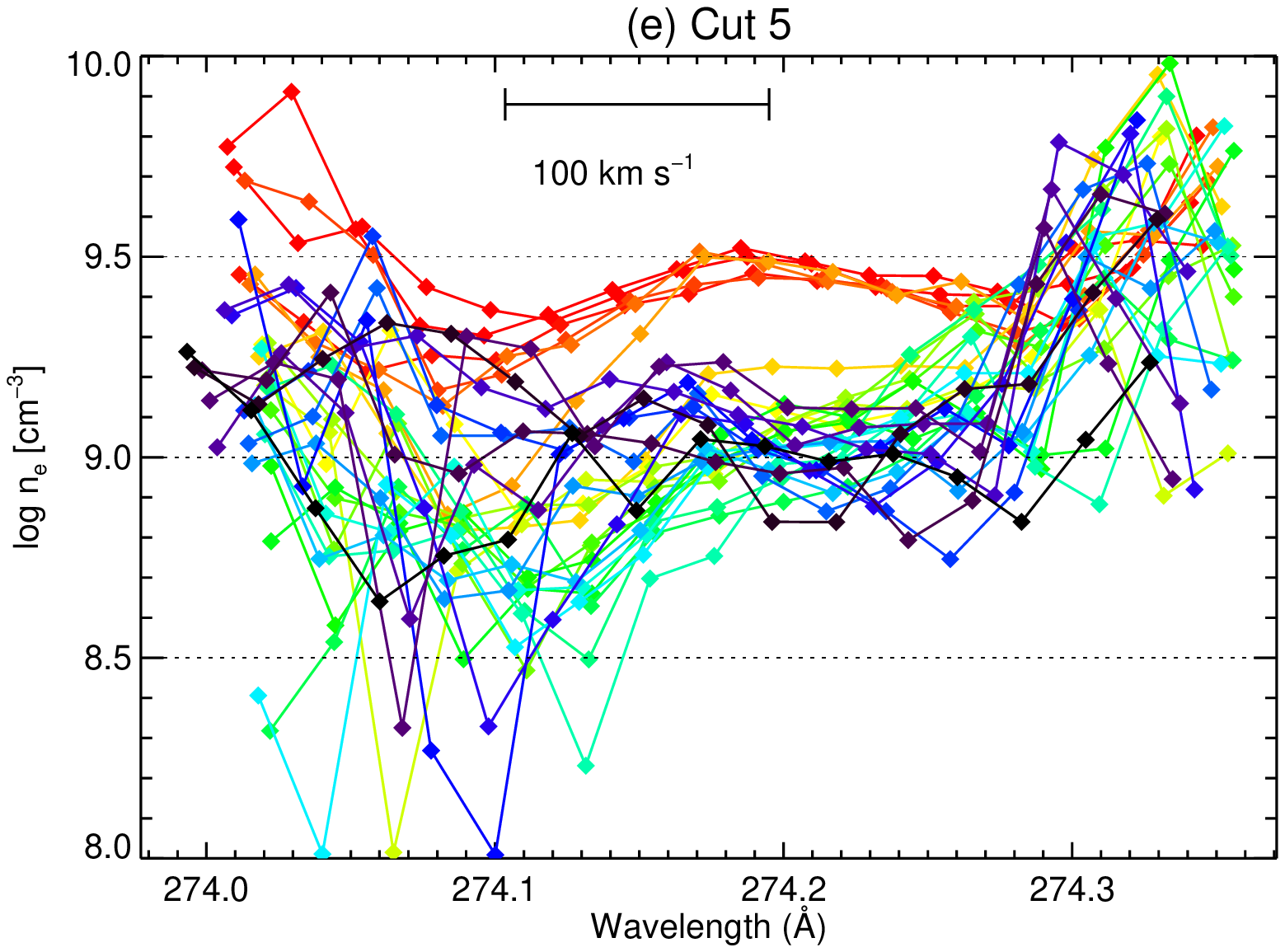}
  \caption{$\lambda$-$n_{\mathrm{e}}$ diagrams at the locations indicated by colored diamonds in Figure \ref{fig:eis_map_lp_sv} (Cuts 1 and 2; including the eastern outflow region, Cuts 3--5; including the western outflow region).}
  \label{fig:eis_ndv_sv}
\end{figure}

In order to confirm the above implication more robustly, we see the variation of $\lambda$-$n_{\mathrm{e}}$ diagram along $x$ direction from the active region core to the outflow regions.  The selected region spans from the active region core (\textit{red} diamond) to the outflow region (\textit{violet} diamond) as seen in Figure \ref{fig:eis_map_lp_sv}.  The boundary of the active region core corresponds to the color between \textit{yellow} and \textit{light green}.  The $\lambda$-$n_{\mathrm{e}}$ diagrams at each cut (1--5) are plotted in Figure \ref{fig:eis_ndv_sv}.  We can see clear changes of the $\lambda$-$n_{\mathrm{e}}$ diagrams with colors.  The $\lambda$-$n_{\mathrm{e}}$ diagrams for cut 1 show a small hump around $274.00$--$274.10${\AA} representing that EBW component has larger electron density than the major component, though the hump at almost all locations (\textit{red}--\textit{black}) might mean that it was caused by an anomalous pixel (\textit{e.g.}, warm pixel).  Both for cut 1 and 2, the diagrams show flat or slightly decreasing behavior as a function of wavelength at all locations.  These behaviors are consistent with the result obtained in Section \ref{sect:dns_results} (region U1 and U2) which indicated that the electron density of the outflows in the eastern edge is almost the same or slightly larger.  On the other hand, in the western outflow region (cut 3--5), those for the outflow region show a dip around $274.10${\AA}.  This wavelength corresponds to $v = - 110 \, \kmpers$ for the emission line Fe \textsc{xiv} $274.20${\AA}, from which it is implied that the outflows in the western edge are composed of less dense plasma compared to the 
plasma 
characterized by 
around $274.20${\AA} 
existing along the line of sight.  Note that this velocity does not mean that of the upflows because no fitting was applied in $\lambda$-$n_{\mathrm{e}}$ diagram.

The electron density of EBW component evaluated from $\lambda$-$n_{\mathrm{e}}$ diagrams around $\lambda=274.10${\AA} was $\log n_{\mathrm{e}} \, [\mathrm{cm}^{-3}]=9.0\text{--}9.2$ in the eastern outflow region, and $\log n_{\mathrm{e}} \, [\mathrm{cm}^{-3}] = 8.5\text{--}9.0$ in the western outflow region, which also coincides with the result obtained through the double-Gaussian fitting.  By exploiting $\lambda$-$n_{\mathrm{e}}$ diagram as a new diagnostic tool, we can now support the results obtained in Section \ref{sect:dns_results}.  

%% file: dns_discuss_n.tex

\citet{depontieu2011} proposed that the tip of the spicule is heated up to the coronal temperature (though the heating mechanism has not been revealed), and is injected to the higher atmosphere where the heated plasma form the corona.  The electron density of upflows from the tips of the spicules is estimated by Equation (10) in \citet{klimchuk2012} which considers the mass conservation,
\begin{equation}
  n_{\mathrm{UP,}\,s} \delta h_s = n_c h_c A,
  \label{eq:nup_spicule_1}
\end{equation}
where $n_{\mathrm{UP,}\,s}$ is the electron density of an upflow (a suffix $s$ denotes spicule), $\delta$ is the fraction of the spicule that is heated to coronal temperatures, $h_s$ is the height of the spicule, $n_c$ is the coronal density after the tip of the spicule expands into the corona, $h_c$ is the length of coronal loops, and $A$ is the expansion factor of the cross section of coronal loops from the chromosphere to the corona. Using typical coronal values: $n_c \simeq 10^9 \, \mathrm{cm}^{-3}$, $h_c \simeq 5 \times 10^9 \, \mathrm{cm}$, $\delta \simeq 10 \, \mathrm{\%}$ \citep{depontieu2011}, $h_s \simeq 10^9 \, \mathrm{cm}$ in the maximum height, and $A \sim 10$ (this factor has not been determined precisely yet, but is larger than unity), the electron density of upflows is estimated as
\begin{equation}
  n_{\mathrm{UP,}\,s} 
  \simeq 
  5 \times 10^{11} 
  \left( \frac{n_c}{10^9 \, \mathrm{cm}^{-3}} \right)
  \left( \frac{h_c}{5 \times 10^9 \, \mathrm{cm}} \right)
  \left( \frac{A}{10} \right)
  \left( \frac{\delta}{0.1} \right)^{-1}
  \left( \frac{h_s}{10^9 \, \mathrm{cm}} \right)^{-1}
    \, \mathrm{cm}^{-3} \, \text{.}
  \label{eq:nup_spicule_2}
\end{equation}

For impulsive heating, giving the typical energy content of nanoflares (\textit{i.e.}, $10^{24} \, \mathrm{erg}$) and considering the enthalpy flux as a response of the transition region below the corona leads to
\begin{equation}
  \frac{5}{2} p v_{\mathrm{UP,}\, \mathrm{i}} 
  = 
  \frac{E_{\mathrm{i}}}{\pi r^2_{\mathrm{st}} \tau_{\mathrm{i}}} \, \text{,}
  \label{eq:nup_impulsive_1}
\end{equation}
where $p$ is the gas pressure of the upflow, $v_{\mathrm{UP,} \, \mathrm{i}}$ is the speed of the upflow, $E_{\mathrm{i}}$ is the released energy by the impulsive heating, 
$r_{\mathrm{st}}$
 is the radius of the coronal strand (\textit{i.e.}, thin coronal loop as an elemental structure), and $\tau_{\mathrm{i}}$ is the duration of the impulsive heating.  Kinetic energy flux can be neglected because the upflow speed is around half the speed of sound ($\simeq 200 \, \kmpers$ at $\logt = 6.3$), which means the ratio of the kinetic energy flux to the enthalpy flux is the order of $0.1$.  Typical parameters $E_{\mathrm{i}} \sim 10^{24} \, \mathrm{erg}$, $v_{\mathrm{UP, i}} \simeq 100 \, \kmpers$, 
$r_{\mathrm{st}} \sim 10^{7\text{--}8} \, \mathrm{cm}$
and $\tau_{\mathrm{i}} \sim 10\text{--}100 \, \mathrm{s}$ (this value contains a large degree of uncertainty because of a lack of knowledge at present) imply 
\begin{align}
  n_{\mathrm{UP,}\, \mathrm{i}} 
  \simeq
  5 \times 10^{8\text{--}10} 
  \left( \frac{E_{\mathrm{i}}}{10^{24} \, \mathrm{erg}} \right)
  \left( \frac{r_{\mathrm{st}}}{10^{7\text{--}8} \, \mathrm{cm}} \right)^{-2}
  \left( \frac{\tau_{\mathrm{i}}}{10 \, \mathrm{s}} \right)^{-1}
  \left( \frac{T_{\mathrm{i}}}{10^6 \, \mathrm{K}} \right)^{-1}
  \left( \frac{v_{\mathrm{UP,}\,\mathrm{i}}}{10^7 \, \mathrm{cm} \, \mathrm{s}^{-1}} \right)^{-1}
  \, \mathrm{cm}^{-3} \, \text{,}
  \label{eq:nup_impulsive_2}
\end{align}
for which we used $p = 2 n_{\mathrm{UP,}\,\mathrm{i}} k_{\mathrm{B}} T_{\mathrm{i}}$ where $n_{\mathrm{UP,}\,\mathrm{i}}$ is the electron density of the upflow and $T_{\mathrm{i}}$ is its temperature.  
Recent observation by Hi-C \citep{kobayashi2014} indicated that the width of coronal strands is around $450 \, \mathrm{km}$ \citep{brooks2013}.  
Note that $E_{\mathrm{i}}$ and $\tau_{\mathrm{i}}$ have not been observationally constrained well so far, though we used values which are considered to be reasonable at present.  
It is clear that the predicted electron density estimated by adopting the typical coronal values from the spicule and impulsive heating significantly exceed the derived upflow density ($n_\mathrm{EBW} \leq 10^9 \, \mathrm{cm}^{-3}$ in our analysis).  
Equation (\ref{eq:nup_impulsive_2}) can be used to estimate the parameter range where the predicted upflow density becomes similar to the observed value since there is much uncertainty in the parameter $\tau_{\mathrm{i}}$.  
For example, 
If the heating continues for $\tau_{\mathrm{i}} = 500 \, \mathrm{s}$, Equation (\ref{eq:nup_impulsive_2}) leads to $n_{\mathrm{UP,}\,\mathrm{i}} \simeq 10^{7\text{--}9} \, \mathrm{cm}^{-3}$ (\textit{i.e.}, near the obtained upflow density) for the case 
other parameters 
keep their
 typical value.  


%% file: dis_mass.tex

We estimate the mass flux of the outflowing plasma $F_{\mathrm{out}}$ in the western outflow region by using Doppler velocity and electron density of EBW component.  The electron density was $n_{\mathrm{e}} \simeq 10^{8.7} \, \mathrm{cm}^{-3}$, and the Doppler velocity was ${-60} \, \kmpers$.  The total area ($S$) of the entire western outflow region was roughly $30'' \times 40''$ ($S \simeq 6 \times 10^{18} \, \mathrm{cm}^{2}$).  Considering the inclination angle of the magnetic field of $30^{\circ} \text{--} 50^{\circ}$ as calculated by potential field extrapolation of an MDI magnetogram (mentioned in Section \ref{sect:dis_ew}), the speed of the outflow is roughly thought to be $v \simeq 70\text{--}90 \, \kmpers$.  Thus, $F_{\mathrm{out}}$ can be estimated as $F_{\mathrm{out}}= 2 n_{\mathrm{e}} \mu v S = (4\text{--}5) \times 10^{10} \, \mathrm{g} \, \mathrm{s}^{-1}$ where $\mu$ is a mean mass of ions which was set to $1 \times 10^{-24} \, \mathrm{g}$.  For a comparison, we also evaluate the total mass contained in the active region.  Using volume of $V = (100'')^{3} = 4 \times 10^{29} \, \mathrm{cm}^{3}$ and typical density $n_{\mathrm{e}} = 10^{9\text{--}10} \, \mathrm{cm}^{-3}$, the total mass $M_{\mathrm{AR}}$ is evaluated as $M_{\mathrm{AR}} = 2 n_{\mathrm{e}} \mu V = 8 \times 10^{14\text{--}15} \, \mathrm{g}$. 

This implies that if the mass in the active region is actually lost by the outflow \citep{brooks2012}, the time scale of the mass drain becomes $\tau_{\mathrm{out}}=M_{\mathrm{AR}} / F_{\mathrm{out}} = 2 \times 10^{4\text{--}5} \, \mathrm{s}$ (\textit{i.e.}, several hours to a couple of days).  Since the lifetime of active regions is much longer than this time scale, up to several weeks, the active region needs a certain mechanism to provide the plasma continuously.  We note that the outflow region is localized at the edge of the active region, which means that a limited part of the active region is involved in the outflow.  In contrast to this mass drain scenario, the extrapolated magnetic field lines rooted in the outflow region were connected to near the opposite edge of the active region according to the potential field calculation.  The opposite side of the outflow region exhibit almost zero velocity, which indicates that the mass would accumulate from the outflow region.  This leads to the picture that the outflow actually provides the active region with the plasma.  However, the Doppler velocity map show a blueshifted pattern extending to the north west from the western outflow region, which may indicate that it is connected to far higher atmosphere.  We must take into account the temporal evolution of the magnetic field in order to confirm the validity of these scenarios which will be studied in the near future. 


%% file: dis_ew.tex

\begin{figure}[!t]
  \centering
  \includegraphics[width=14cm,clip]{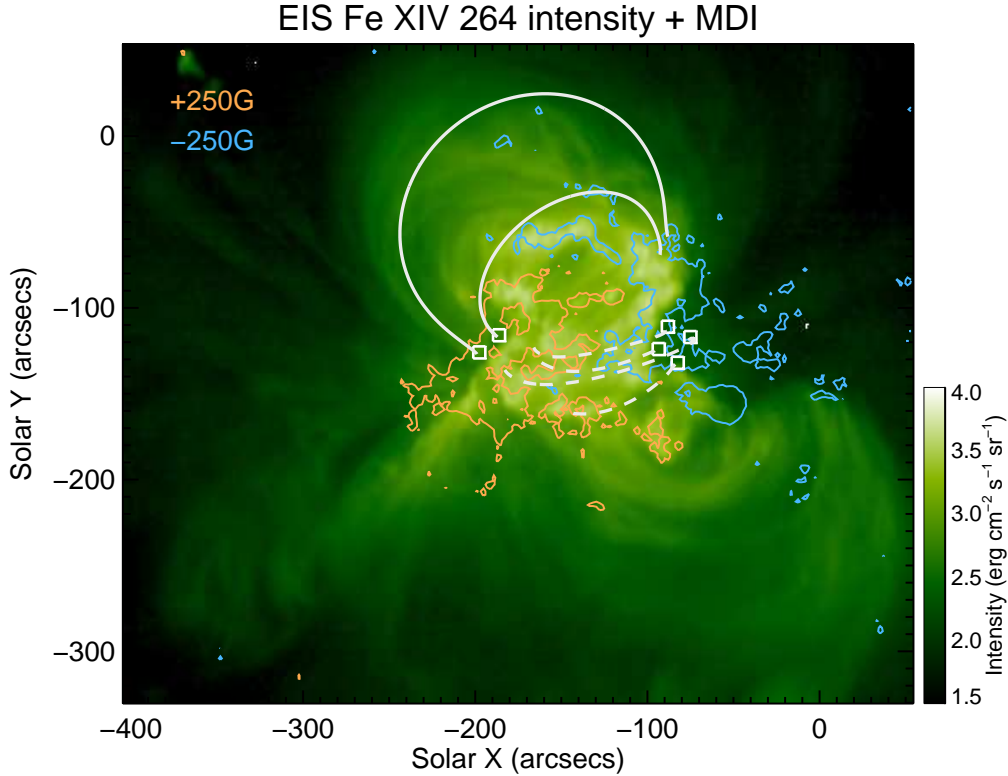}
  \caption{EIS Fe \textsc{xiv} $264.78${\AA} intensity map.  \textit{Orange} (\textit{Turquoise}) contours indicate a magnetic field strength of $+250 \, ({-250}) \, \mathrm{G}$ in an MDI magnetogram taken during the EIS scan.  Six \textit{white} boxes are located at the position corresponding to the studied locations.  \textit{White} lines rooted at those boxes indicate the magnetic field lines extrapolated from the magnetogram.}
  \label{fig:dis_ew_map}
\end{figure} 

Here we discuss some implications for the coronal formation (\textit{i.e.}, heating) from the viewpoint of the outflows.  The differences of derived quantities in those two outflow regions are listed in Table \ref{tab:dns_cmpl}.  The topology of magnetic field lines can be inferred from the extrapolated field lines and the Doppler velocity map.  We calculated the potential magnetic field from an MDI magnetogram taken during the EIS scan which started from 10:25:42UT, since its FOV is larger than that of the EIS scan used for the density diagnostics, and is large enough to include the entire active region.  In order to confirm the connectivity of the magnetic field lines rooted at the studied locations, we drew projected field lines onto the intensity map of Fe \textsc{xiv} $264.78${\AA} as shown in Figure \ref{fig:dis_ew_map}.  The outflow regions U1--U6 are indicated by \textit{white} boxes.  Note that since the intensity map was derived from the EIS scan which started from 10:25:42UT, we took into account the solar rotation to identify the locations of those boxes.  The contours with \textit{orange} (\textit{turquoise}) indicate a magnetic field strength of $+250 \, ({-250}) \, \mathrm{G}$ in the MDI magnetogram.  

Two \textit{solid white} lines trace coronal loops, therefore we regarded the topology of the eastern outflow region as closed, 
which can be also seen as a coherent pattern tracing the coronal loops in the Doppler velocity maps.  Four \textit{dashed white} lines rooted at the western outflow region are connected to the opposite polarity around $(x,y)=(-160'', -150'')$, but the Doppler velocity maps clearly show that the blueshifted feature extends into the far west from which we suspected the topology of the western outflow region would be open.  The closed loops rooted at the eastern outflow region are brighter than the open structures extending from the western outflow region by one order of magnitude.  This might reflect the length of each structure in the sense that the upflow easily fills a closed loop while it flows without obstacles in an open structure, which produces denser plasma in the closed loop.  
Note that \citet{culhane2014} suggested that the eastern outflow region is actually connected to the heliosphere through a two-step reconnection process.  We may need further observations (\textit{e.g.}, statistical) in the future mission to clarify this point.  

As a consequence, it leads to the implication that the upflow from the bottom of the corona becomes dense in the closed loop because of the pressure balance between the corona and the transition region, which is consistent with our result that the electron density of EBW component was larger in the eastern region than in the western region (see Table \ref{tab:dns_cmpl}).  Although the difference in the electron density of the major component would not be insignificant, the relationship of the column depth (\textit{i.e.}, larger $h_{\mathrm{Major}}$ in the eastern outflow region than in the western outflow region) may represent that the eastern outflow region consists of more coronal loops than the western outflow region.  

\begin{figure}[!t]
  \centering
  \includegraphics[width=16cm,clip]{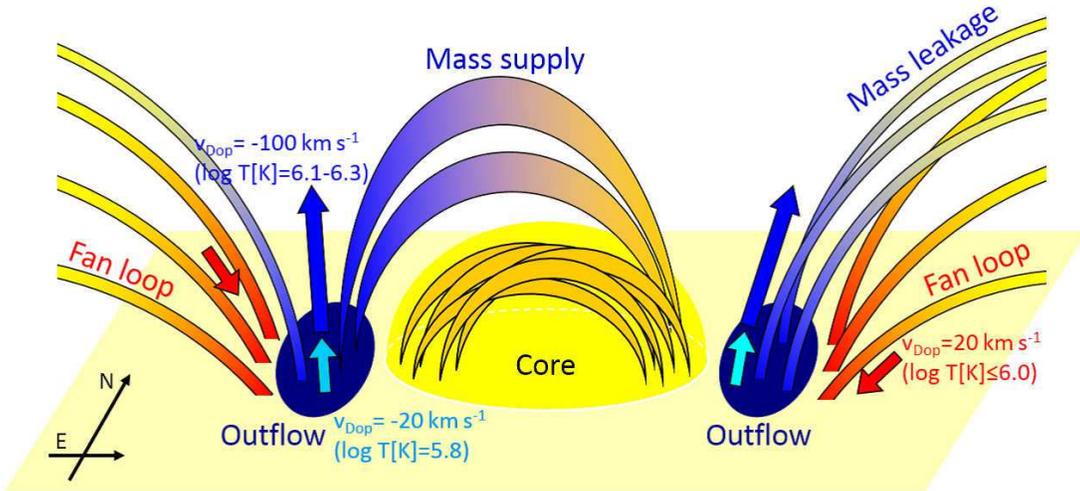}
  \caption{Schematic picture of active region outflows.}
  \label{fig:dis_mass_ew}
\end{figure}

We have evaluated mass leakage from the western outflow region in Section \ref{sect:dis_mass}.  
A study on the first ionization potential (FIP) bias in the active region by \citet{brooks2011} also suggested that the western outflow region connects to the slow solar wind.  
In contrast, 
 the closed topology of the eastern outflow region may actually imply mass supply to the active region.  If this is the case for a portion of the outflow region, it means that the outflow plays a crucial role in the coronal heating by supplying hot plasma into coronal loops.  We suggest a possible picture in Figure \ref{fig:dis_mass_ew} as a summary of this discussion. 

%% file: dns_sum.tex

The electron density of the outflow from the edges of NOAA AR10978 was measured by using an emission line pair Fe \textsc{xiv} $264.78${\AA}/$274.20${\AA}.  The upflow component was extracted from an enhanced blue wing (EBW) in Fe \textsc{xiv} line profiles through double-Gaussian fitting.  We fitted those two Fe \textsc{xiv} emission lines simultaneously with a physical restriction that corresponding components in two emission lines must have the same Doppler velocity and thermal width, which previous EIS analysis on the density diagnostics have not been tried.  The results were listed in Table \ref{tab:dns_cmpl}.
  
The derived electron density for the major component ($n_{\mathrm{Major}}$) and that for EBW component ($n_{\mathrm{EBW}}$) had opposite relationship in their magnitudes at the eastern and western outflow regions.  There are several possibilities which cause the difference in the magnitude relationship between the east and west outflow region as follows.  (1) The major component and EBW in Fe \textsc{xiv} line profiles are not directly related (\textit{e.g.}, superposition of structures along the line of sight).  The electron density of EBW component just reflects the energy input amount.  (2) The eastern outflow regions consist of the footpoints of corona loops extending to the north and connected to the opposite magnetic polarity around $(x,y)=(-170'', -70'')$, while longer coronal loops emanate in the western outflow regions and extend to the north west considering the appearance in Figure \ref{fig:fexiv_1G_map}.  The difference in length may influence the plasma density by the same driving mechanism for the outflow, since it is easier for the upflows in an open structure to flow without condensation than for those in a closed loop.  

We also calculated the column depth for each component ($h_{\mathrm{Major}}$ and $h_{\mathrm{EBW}}$).  In the eastern region, $h_{\mathrm{EBW}}$ was smaller than $h_{\mathrm{Major}}$ by roughly one order of magnitude, which implies that the upflows possess only a small fraction ($\sim 0.1$).  
Considering this implication with the result for the electron density ($n_{\mathrm{EBW}} \ge n_{\mathrm{Major}}$), it leads to a picture that the upflows may play a role in supplying hot plasma ($\logt = 6.2 \text{--} 6.3$) into coronal loops.  On the other hand, in the western outflow region, the upflows have a larger volume by a factor of $5$--$6$ than 
the plasma characterized by the major component, 
from which we consider the western outflow region as a structure composed of extending tubes with unidirectional upflows.


%% file: ndv_sum.tex

We introduced a density diagnostics from a new point of view in Section \ref{chap:ndv}.  Electron density derived in our method is a function of Doppler velocity or wavelength (Equation \ref{eq:ndv_rprsnt}), referred to as $\lambda$-$n_{\mathrm{e}}$ diagram, which was found to be a good indicator of the electron density of minor components in a line profile.  The method has the advantage that it does not depend on any fitting model which might be ill-posed in some cases.  Our aim was to evaluate the electron density of the outflow seen at the edge of the active region, and reinforce the result obtained in Section \ref{sect:dns_results}.  

Using a density-sensitive emission line pair Fe \textsc{xiv} $264.78${\AA}/$274.20${\AA}, we studied $n_{\mathrm{e}} (\lambda)$ by making $\lambda$-$n_{\mathrm{e}}$ diagrams at the active region core and the outflow regions.  The increase in the diagrams was seen on the longer wavelength side for both structures, but we could not ascertain whether that behavior actually implies a physical situation at present.  The diagrams for the active region core were flat around $\log n_{\mathrm{e}} \, [\mathrm{cm}]^{-3} \simeq 9.5$, while those for the outflow regions exhibit some characteristic behaviors at the shorter wavelength side.  They show a small hump around $v = -110 \, \kmpers$ in the eastern region (cuts 1 and 2 in Figure \ref{fig:eis_map_lp_sv}), and a decrease trend from $\log n_{\mathrm{e}} \, [\mathrm{cm}^{-3}]=9.0$ to $\log n_{\mathrm{e}} \, [\mathrm{cm}^{-3}]=8.5$ in a velocity scale of $100 \, \kmpers$ in the western outflow region (cuts 3--5 in Figure \ref{fig:eis_map_lp_sv}) as seen in Figure \ref{fig:eis_ndv_sv}.  Thus we confirmed the results obtained in Section \ref{sect:dns_results} through our new method independent of the double-Gaussian fitting. 

As for the case where intermittent heating is responsible for the outflows, the duration of heating was crudely estimated to be longer than $\tau = 500 \, \mathrm{s}$ for the energy input of $10^{24} \, \mathrm{erg}$ (\textit{i.e.}, nanoflare) so that the density of upflows from the footpoints becomes compatible with that of the observed outflows. The electron density and column depth of the upflows in the eastern and western outflow regions were different, which was considered to be due to the magnetic structure above the outflow regions.  Mass leakage occurs at the western outflow region (small $n_{\mathrm{EBW}}$ and large $h_{\mathrm{EBW}}$).  On the other hand, there is a possibility of the mass supply to active region loops at the eastern outflow region (large $n_{\mathrm{EBW}}$ and small $h_{\mathrm{EBW}}$), which may be related to the coronal heating process.

%% file: dns_app_in.tex

\begin{figure}[!t]
  \centering
  \includegraphics[width=10cm,clip]{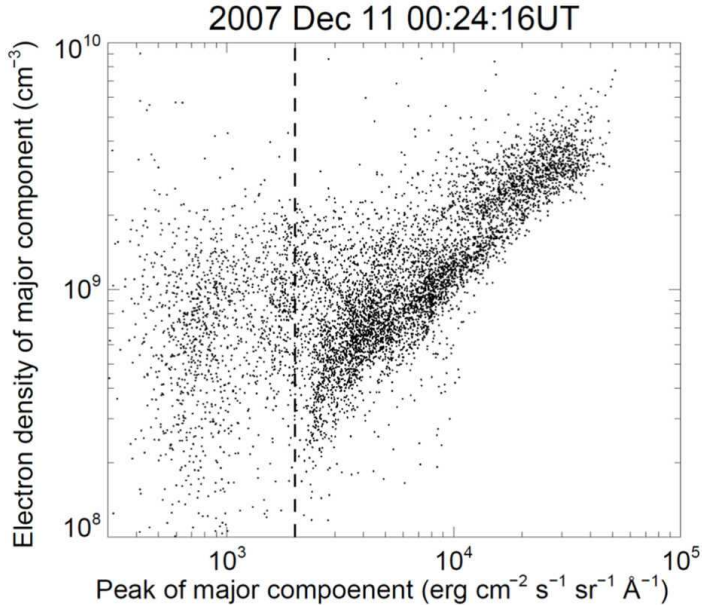}
  \caption{Peak intensity of Fe \textsc{xiv} $264.78${\AA} vs.\ electron density deduced from the major component of Fe \textsc{xiv} $264.78${\AA} and $274.20${\AA} in the double Gaussian fitting.}
  \label{fig:in}
\end{figure}

We can see a clear positive correlation between peak intensity and electron density in the intensity range larger than $I_{\mathrm{Major}} = 2.0 \times 10^{3} \, \mathrm{erg} \, \mathrm{cm}^{-2} \, \mathrm{s}^{-1} \, \mathrm{sr}^{-1} \, \text{\AA}^{-1}$ in Figure \ref{fig:in} (indicated by a \textit{vertical dashed} line) while the plot is more scattered below that intensity.  Not only the photon noise contributes to this large degree of uncertainty, but also unidentified blended emission lines could do so.  Therefore we analyzed the data points with $I_{\mathrm{Major}}$ larger than the value which the \textit{vertical dashed} line indicates. 

%% file: dns_results_sivii.tex

\begin{figure}[!t]
  \centering
  \includegraphics[height=6cm,clip]{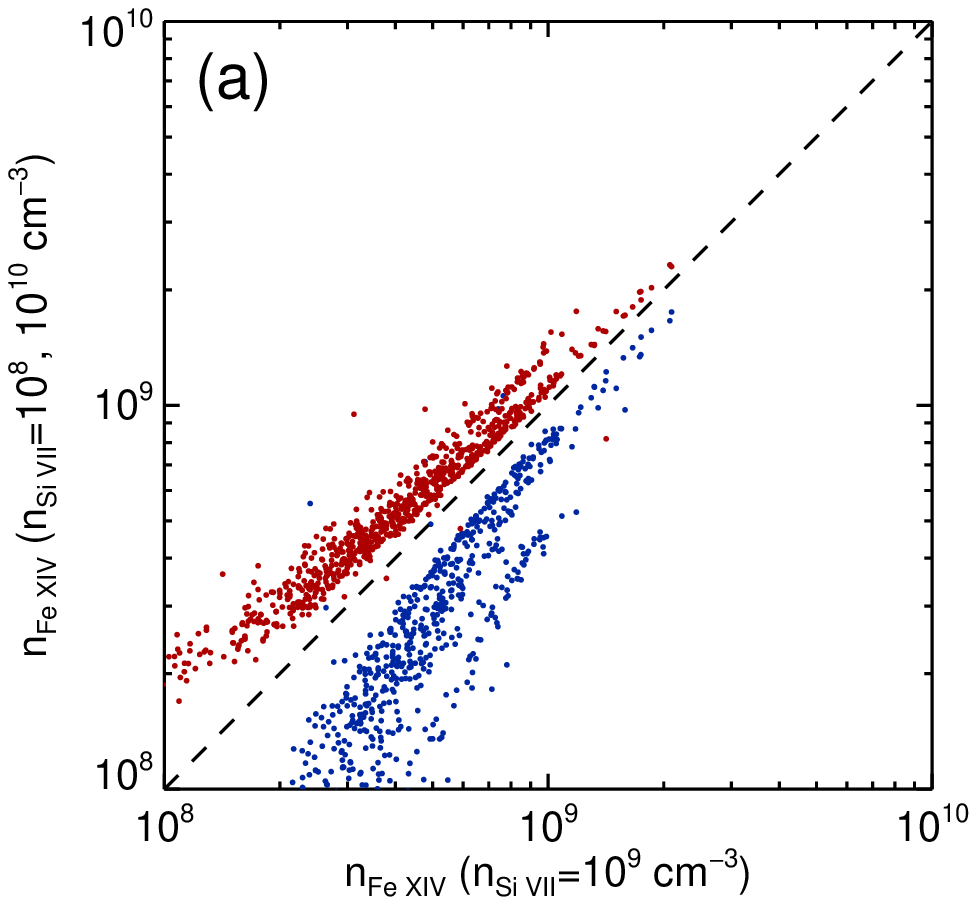}
  \includegraphics[height=6cm,clip]{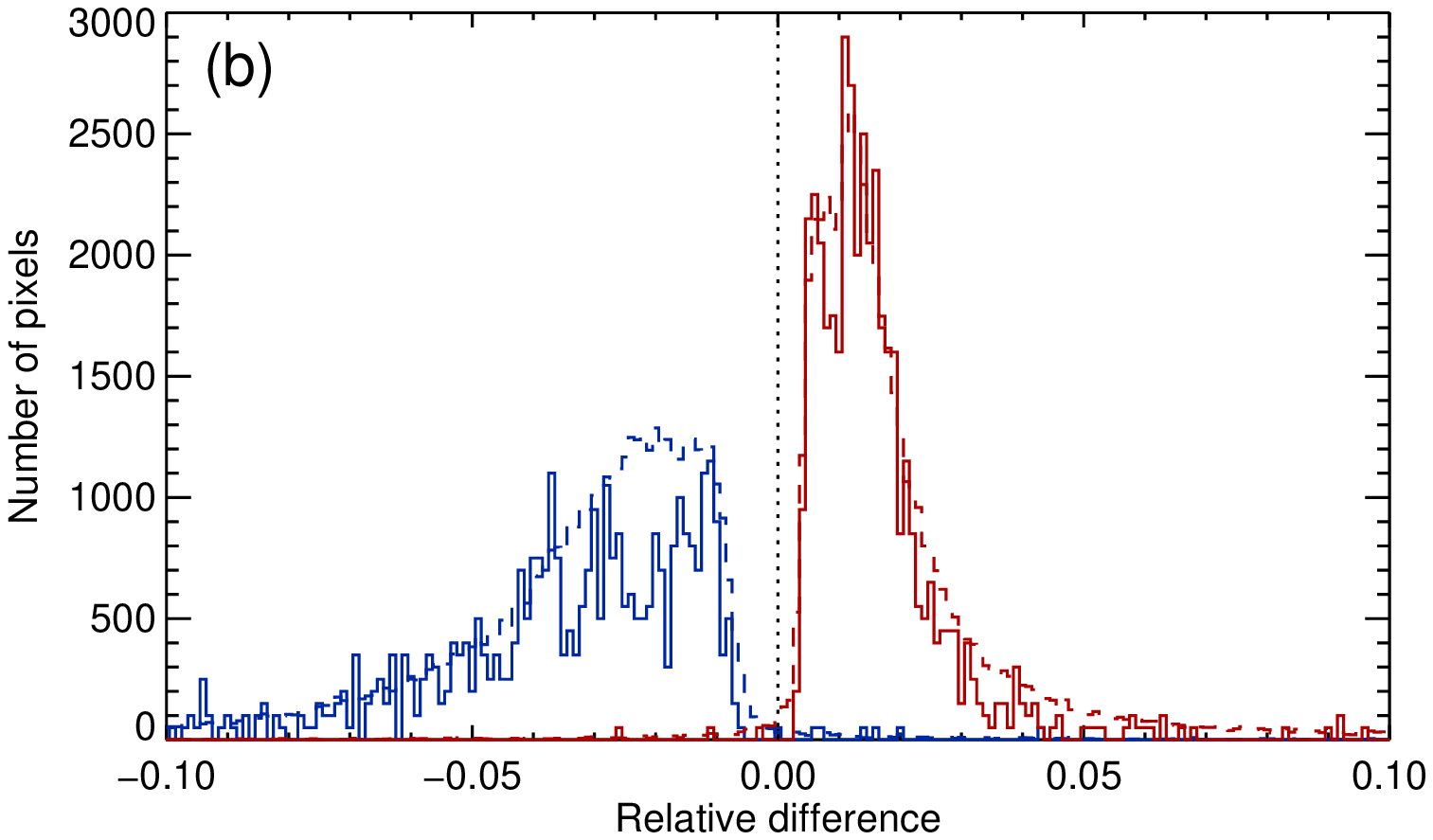}
  \caption{(a) Scatter plots of derived Fe \textsc{xiv} density ($n_{\mathrm{Fe \textsc{xiv}}}$) for different electron density of Si \textsc{vii} ($n_{\mathrm{Si \textsc{vii}}}$).  Horizontal axis indicates $n_{\mathrm{Fe \textsc{xiv}}}$ of EBW component derived by assuming $n_{\mathrm{Si \textsc{vii}}}=10^9 \, \mathrm{cm}^{-3}$.  Vertical axis indicates $n_{\mathrm{Fe \textsc{xiv}}}$ of EBW component derived by assuming $n_{\mathrm{Si \textsc{vii}}}=10^8 \, \mathrm{cm}^{-3}$ (\textit{blue}) and $10^{10} \, \mathrm{cm}^{-3}$ (\textit{red}). (b) The same data as in panel (a) but the \textit{horizontal} axis indicates a relative difference $\Delta n_{\mathrm{Fe \textsc{xiv}}} / n_{\mathrm{Fe \textsc{xiv}}}$, where $\Delta n_{\mathrm{Fe \textsc{xiv}}}$ is a difference of $n_{\mathrm{Fe \textsc{xiv}}}$ for different $n_{\mathrm{Si \textsc{vii}}}$ ($10^{8}$ and $10^{10} \, \mathrm{cm}^{-1}$) measured from the case for $n_{\mathrm{Si \textsc{vii}}}=10^{9} \, \mathrm{cm}^{-3}$.}
  \label{fig:err_sivii}
\end{figure}

Since the electron density is not the same for emission lines with different formation temperature, there is an uncertainty in $n_{\mathrm{Si \textsc{vii}}}$ which cannot be determined from the data used in this analysis.  In order to evaluate the error in the electron density derived for Fe \textsc{xiv} ($n_{\mathrm{Fe \textsc{xiv}}}$) coming from this uncertainty, we remove the blending Si \textsc{vii} at Fe \textsc{xiv} $274.20${\AA} in three cases for $n_{\mathrm{Si \textsc{vii}}}$: $10^{8}$, $10^{9}$, and $10^{10} \, \mathrm{cm}^{-3}$, and derived $n_{\mathrm{Fe \textsc{xiv}}}$ for each case.  Panel (a) in Figure \ref{fig:err_sivii} shows scatter plots for the electron density of EBW component within the entire western outflow region derived for the case $n_{\mathrm{Si \textsc{vii}}}=10^9 \, \mathrm{cm}^{-3}$ vs.\ $10^8 \, \mathrm{cm}^{-3}$ $(10^{10} \, \mathrm{cm}^{-3})$ in \textit{blue} (\textit{red}).  The $n_{\mathrm{Fe \textsc{xiv}}}$ of EBW component derived by assuming $n_{\mathrm{Si \textsc{vii}}}=10^8$ $(10^{10}) \, \mathrm{cm}^{-3}$ becomes smaller (larger).  Panel (b) in Figure \ref{fig:err_sivii} shows those relative differences $\Delta n_{\mathrm{Fe \textsc{xiv}}} / n_{\mathrm{Fe \textsc{xiv}}}$, where $\Delta n_{\mathrm{Fe \textsc{xiv}}}$ is a difference of $n_{\mathrm{Fe \textsc{xiv}}}$ for different $n_{\mathrm{Si \textsc{vii}}}$ ($10^{8}$ and $10^{10} \, \mathrm{cm}^{-1}$) measured from the case for $n_{\mathrm{Si \textsc{vii}}}=10^{9} \, \mathrm{cm}^{-3}$.  Colors (\textit{red} and \textit{blue}) are the same as in panel (a).  \textit{Solid} and \textit{Dashed} histograms indicate the western outflow region (the \textit{white dashed} box in Figure \ref{fig:dens_map}) and for the entire field of view, respectively.  These relative differences were calculated in log scale.  The histograms show that the error coming from the difference of $n_{\mathrm{Si \textsc{vii}}}$ does not exceed $5${\%}.  It means that the error is around $10^{0.4\text{--}0.5}$ at most for the density range $10^{8} \, \mathrm{cm}^{-3} \leq n_{\mathrm{e}} \leq 10^{10} \, \mathrm{cm}^{-3}$, and roughly becomes a factor of $3$ (\textit{i.e.}, comparable to the error originated in the photon noise).